\documentclass[12 pt]{article}
\usepackage{hyperref} 
\usepackage{amssymb} 
\usepackage{GatesHeader}
\usepackage{graphicx}  
\usepackage{tikz}  
\usetikzlibrary{shapes.misc}  
\usepackage{booktabs}
\usepackage[footnotesize,bf,textfont={it},margin=1 cm]{caption}
\usepackage{setspace} 
\usepackage{xcolor}
\usepackage{subfigure}
\usepackage{amsmath}
\allowdisplaybreaks
\definecolor{AdinkraGreen}{rgb}{0.10196079, 0.61176473, 0.21960784 }
\definecolor{AdinkraViolet}{rgb}{0.42352942, 0.15294118, 0.4509804 }
\definecolor{AdinkraOrange}{rgb}{0.89803922, 0.57647061, 0.27450982}
\definecolor{AdinkraRed}{rgb}{0.78431374, 0, 0.12156863}
\def\gD{\mbox{\textcolor{AdinkraGreen}{${\rm D}_1$}}}
\def\vD{\mbox{\textcolor{AdinkraViolet}{${\rm D}_2$}}}
\def\oD{\mbox{\textcolor{AdinkraOrange}{${\rm D}_3$}}}
\def\rD{\mbox{\textcolor{AdinkraRed}{${\rm D}_4$}}}

\def\brL{{\bm {\rm L}}}
\def\brR{{\bm {\rm R}}}

\def\nmSG{{$\not$mSG}}
\def\newSG{$\nu\mbox{SG}$}
\def\newnewSG{$\nu\nu\mbox{SG}$}

\begin{document}
\numberwithin{equation}{section}
\setcounter{equation}{0}
\setcounter{page}{0}

\def\dt#1{\on{\hbox{\rm .}}{#1}}                
\def\Dot#1{\dt{#1}}

\def\gfrac#1#2{\frac {\scriptstyle{#1}}
        {\mbox{\raisebox{-.6ex}{$\scriptstyle{#2}$}}}}
\def\gg{{\hbox{\sc g}}}
\border\headpic {\hbox to\hsize{\today \hfill  
{UMDEPP-012-022} 
}}
\par {$~$ \hfill 
{arXiv:1212.3318 [hep-th]} 
} 
\par 

\setlength{\oddsidemargin}{0.3in}
\setlength{\evensidemargin}{-0.3in}

\begin{center}
\vglue .10in
{\large\rm 4D, ${\cal N}$ = 1 Supergravity Genomics}\\[.5in]

Isaac\,  Chappell\,\footnote{isaac.chappell@gmail.com}, S.\, James Gates, Jr.\footnote{gatess@wam.umd.edu}, 
William D. Linch III\,\footnote{wdlinch3@gmail.com}, 
James Parker\footnote{jp@jamesparker.me}, \\
Stephen Randall\,\footnote{stephenlrandall@gmail.com},  Alexander Ridgway\footnote{alecridgway@gmail.com},~and Kory Stiffler\footnote{kstiffle@gmail.com}
\\[0.3in]
{\it Center for String and Particle Theory\\
Department of Physics, University of Maryland\\
College Park, MD 20742-4111 USA}
\\[0.5in]

{\rm ABSTRACT}\\[.01in]
\end{center}
\begin{quotation}
{\small The off-shell representation theory of 4D, $\mathcal{N}=1$ supermultiplets 
can be categorized in terms of distinct irreducible graphical representations called 
adinkras as part of a larger effort we call supersymmetry `genomics.' Recent evidence has emerged pointing to the existence of three such 
fundamental adinkras associated with distinct equivalence classes of a Coxeter group.
A partial description of these adinkras is given in terms of two types, termed 
cis-and trans-adinkras (the latter being a degenerate doublet) in analogy to cis/trans isomers
in chemistry.  Through a new and simple procedure that uses adinkras, we find the irreducible off-shell adinkra representations of 4D, 
$\mathcal{N}=1$ supergravity, in the old-minimal, non-minimal, and conformal 
formulations. This procedure uncovers what appears to be a selection rule useful to reverse engineer adinkras to higher dimensions. We categorize the supergravity representations in terms of the number of cis-($n_c$) and trans-($n_t$) adinkras in the 
representation and synthesize our new results with our previous supersymmetry genomics results into a group theoretic framework.
}
\endtitle

\setlength{\oddsidemargin}{0.3in}
\setlength{\evensidemargin}{-0.3in}

\setcounter{equation}{0}

\section{Introduction}\label{s:intro}
$~~~~$ We continue to build a comprehensive representation theory for off-shell supersymmetry (SUSY) in a project called supersymmetric `genomics'~\cite{Gates:2009me,Gates:2011aa}.  Our main tools in this venture are adinkras: graphical representations of higher dimensional supersymmetric systems reduced to their one dimensional `shadows' that remain when all spatial dependence has been removed and only time dependence is considered. We refer to this dimensional reduction process as reducing to the \emph{0-brane}. For instance, the 0-brane reduction of the transformation laws for the 4D, $\mathcal{N}=1$ chiral multiplet are \emph{entirely} encoded as the adinkra in Fig.~\ref{f:chiral1} where the colored lines connecting the fields describe supersymmetry transformations. The precise meaning of the lines and nodes in an adinkra such as Fig.~\ref{f:chiral1} is reviewed in Sec.~\ref{s:arev}. 

In the previous genomics works~\cite{Gates:2009me,Gates:2011aa}, we built the adinkra representations for many of the simplest off-shell 4D, $\mathcal{N}=1$ supermultiplets: the chiral, vector, tensor, real scalar, and complex linear supermultiplets. This paper is the natural extension of that work to include supergravity (SUGRA) as one of our ultimate goals is to find the adinkraic representations for \emph{all} 4D, $\mathcal{N}=1$ representations. We study off-shell 4D, ${\mathcal N}=1$ supergravity in components to a depth not previously studied in the literature. For instance, this paper is the only place to the knowledge of the authors that the Lagrangian and transformation laws for linearized  4D, $\mathcal{N}=1$ non-minimal supergravity  are written out explicitly in a Majorana component representation. A main result of this paper is the discovery of what appears to be a selection rule for adinkras that \emph{can not} describe higher dimensional systems. The search for selection rules with SUSY genomics is complementary to the procedures in Refs.~\cite{Faux:2009rz} to `reverse engineer' adinkras to construct the higher dimensional systems in that SUSY genomics does not suffer the  difficulties with gauge theories encountered in Refs.~\cite{Faux:2009rz}. 

\begin{figure}[!ht]\setlength{\unitlength}{.8 mm}
\begin{center}
   \begin{picture}(90,65)(0,0)
\put(0,0){\includegraphics[width = 90\unitlength]{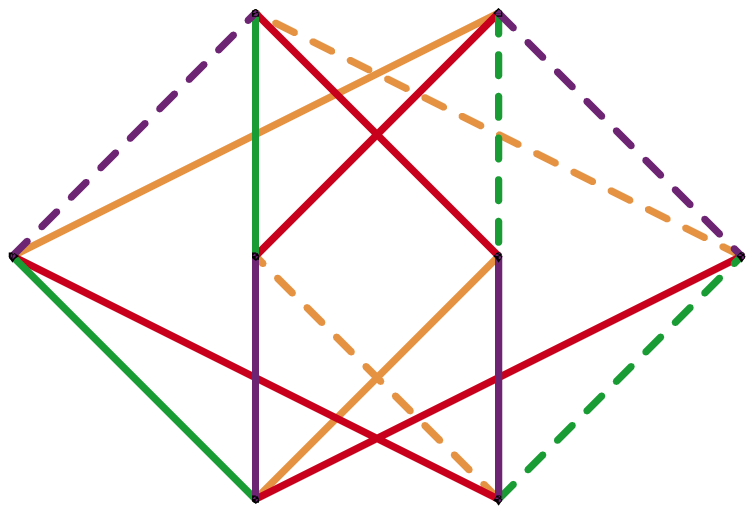}}
\put(29,7){\fcolorbox{black}{white}{$A$}}
\put(54,7){\fcolorbox{black}{white}{$B$}}
\put(1.5,30){\begin{tikzpicture}
 \node[rounded rectangle,draw,fill=white!30]{$\psi_1$};
 \end{tikzpicture}}
\put(27,30){\begin{tikzpicture}
 \node[rounded rectangle,draw,fill=white!30]{$\psi_2$};
 \end{tikzpicture}}
\put(52,30){\begin{tikzpicture}
 \node[rounded rectangle,draw,fill=white!30]{$\psi_3$};
 \end{tikzpicture}}
\put(74.5,30){\begin{tikzpicture}
 \node[rounded rectangle,draw,fill=white!30]{$\psi_4$};
 \end{tikzpicture}}
\put(29,57){\fcolorbox{black}{white}{$F$}}
\put(54,57){\fcolorbox{black}{white}{$G$}}
   \end{picture}
   \end{center}
  \vspace{-25 pt}
\caption{An adinkra for the chiral multiplet.}
\label{f:chiral1}
\end{figure}

Supersymmetric theories are often better understood on-shell than they are off-shell. Some of the most famous examples are 10D and 11D SUGRA and 4D, $\mathcal{N}=4$ super Yang-Mills theory (SYM). Off-shell representation theory is important for many reasons. Practically, it allows for many applications such as model-independent closure of the supersymmetry algebra, solution of the superspace constraints in terms of unconstrained variables and the consequent straightforward quantization, coupling to off-shell supergravity and the consequent description of field theories in rigid curved backgrounds, and the description of spontaneous supersymmetry breaking.  Theoretically, we would like as complete a classification of supersymmetric representations as we have for Lie algebras. So far we achieved this only for the on-shell representations and large families of off-shell representations with no more than eight supercharges.

There is a well-established no-go theorem to the effect that requiring a finite number of auxiliary fields is only consistent with off-shell supersymmetry for generic representations if the number of supercharges does not exceed four \cite{Siegel:1981dx}. In the case of eight supercharges, the harmonic and projective superspaces classify generic representations using an infinite number of auxiliary superfields. The generic classification with more than eight supercharges is unknown. Also unclassified are the non-generic representations. Well-known examples of these are the $\mathcal N=2$ tensor multiplet and (relaxed-)hypermultiplets. We would like to classify with adinkras the theories that lie outside of the no-go theorem of Ref.~\cite{Siegel:1981dx}. A completed off-shell representation theory would clarify these issues and we will explain in this paper how we are doing this with adinkras.

 The possibility that lower dimensional supersymmetric systems hold information about higher dimensional supersymmetric systems was motivation for Ref.~\cite{Gates:2002bc} to use lower dimensional systems to classify more relevant higher dimensional systems.  As adinkras are one dimensional `shadow' representations of higher dimensional systems, they can be reverse engineered to find \emph{new} representations of supersymmetry in higher dimensions. In Ref.~\cite{Doran:2007bx} for instance, a previously unknown 4D, ${\mathcal N} =2$ off-shell multiplet dubbed the `relaxed extended' tensor multiplet was discovered with the aid of adinkras. This multiplet, inspired by attempts to realize an off-shell hypermultiplet, is an extended version of the 4D, ${\mathcal N} =2$ tensor multiplet in the sense that it has more fields, and is relaxed in the sense that the divergence constraint on the vector field is removed. 
 
  As there are more supersymmetric systems in one dimension than there are in higher dimensions, a difficulty in reverse engineering adinkras is in knowing which adinkras are related to higher dimensional systems. A rigorous approach to reverse engineering adinkras to higher dimensions, termed `dimensional enhancement,' was begun by Faux, Iga, and Landweber in Ref.~\cite{Faux:2009rz}. Some progress was made in Ref.~\cite{Faux:2009rz} toward understanding the way in which the representation theory for the dimensionally enhanced system is encoded in its `shadow' adinkra, though the work with gauge multiplets is incomplete. More recently, dimensional enhancement to the world-sheet was investigated in Ref.~\cite{Gates:2011mu}.
  
 Our approach to dimensional enhancement in this paper is to categorize dimensional reductions from higher dimensions to lower dimensions and to look for patterns that give us selection rules as to which adinkras can be extended to higher dimensions. This approach works well for both gauge and non-gauge multiplets because there is a straightforward process with which to dimensionally reduce a system even if it contains gauge degrees of freedom.  In our data driven approach, we use adinkras as the building blocks of the higher dimensional representations analogous to how quarks are the building blocks of $SU(3)$ representations. As 4D, $\mathcal{N}=1$ supersymmetry is well known, we have been focusing on this area before moving on to lesser known supersymmetries, such as the aforementioned SYM and the 10D and 11D SUGRA's. Working first in 4D, $\mathcal{N}=1$ is also useful as in the complementary approach of Ref.~\cite{Faux:2009rz} 4D, $\mathcal{N}=1$ was specifically discussed. In another paper of two of the current authors that is yet in progress, we are extending our adinkra analysis to 4D, $\mathcal{N}=2$ where we  find adinkraic criteria based on the underlying $\mathcal{N}=1$ superfield content that seems to encode information about the $\mathcal{N}=2$ theories to which they can extend~\cite{Gates:2013xx}.

In this paper, we add supergravity to the catalog of adinkraic representations and uncover what appears to be a selection rule that encodes which adinkras \emph{can not} be dimensionally enhanced. For the data we have seen thus far, this selection rule works for both gauge, or diffeomorphisms in the case of supergravity, and non-gauge multiplets. There are two fundamental adinkras that exist with which we are building the representation theory for all 4D, $\mathcal{N}=1$ off-shell representations. Fig.~\ref{f:cvtvintro} shows the two fundamental adinkras, the cis- and trans-adinkras, as they describe the chiral and vector off-shell multiplets, respectively. The names cis and trans come from an analogy with cis/trans isomers in chemistry which have the same chemical form and are connected in the same way, but occur in two different spatial forms~\cite{Wiki:Isomer}. As cis- and trans-isomers in chemistry are distinct within certain classes of spatial transformations, typically rotations, the cis- and trans-adinkras are distinct with respect to `adinkra transformations' that involve rearrangements of the nodes. This will be thoroughly discussed in Sec.~\ref{s:chromo}. For now it is instructive to note that the cis- and trans-adinkras in Fig.~\ref{f:cvtvintro} are related via switching all orange solid lines to orange dotted lines and vice versa. This particular transformation is outside of the set of adinkra transformations and so the cis- and trans-adinkras are considered distinct.

\begin{figure}[!ht]\setlength{\unitlength}{.8 mm}
\begin{center}
   \begin{picture}(200,80)(0,0)
\put(0,40){\includegraphics[width = 90\unitlength]{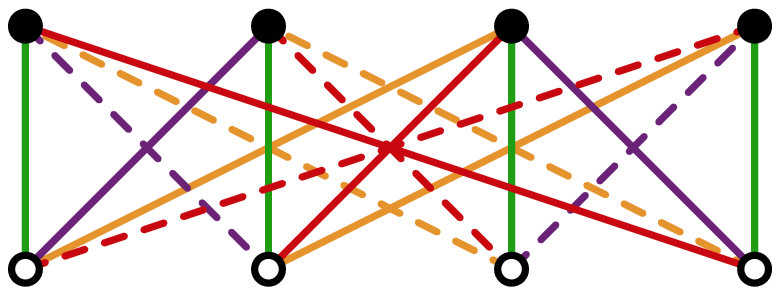}}
\put(3,47){\fcolorbox{black}{white}{$A$}}
\put(25,47){\fcolorbox{black}{white}{$\int dt F$}}
\put(47,47){\fcolorbox{black}{white}{$-\int dt G$}}
\put(79,47){\fcolorbox{black}{white}{$B$}}
\put(1.5,70){\begin{tikzpicture}
 \node[rounded rectangle,draw,fill=white!30]{$\psi_1$};
 \end{tikzpicture}}
\put(27,70){\begin{tikzpicture}
 \node[rounded rectangle,draw,fill=white!30]{$\psi_2$};
 \end{tikzpicture}}
\put(52,70){\begin{tikzpicture}
 \node[rounded rectangle,draw,fill=white!30]{$\psi_3$};
 \end{tikzpicture}}
\put(74,70){\begin{tikzpicture}
 \node[rounded rectangle,draw,fill=white!30]{$-\psi_4$};
 \end{tikzpicture}}
\quad
\put(100,40){\includegraphics[width = 90\unitlength]{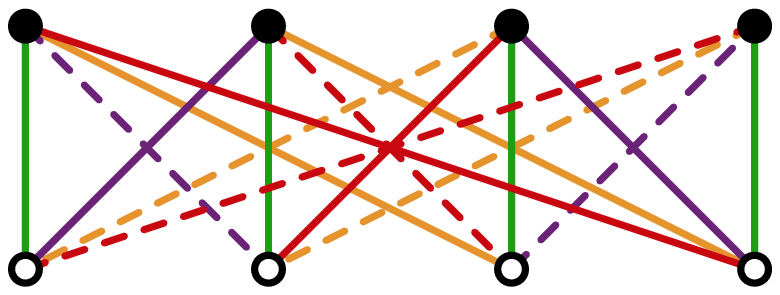}}
\put(103,47){\fcolorbox{black}{white}{$A_3$}}
\put(126,47){\fcolorbox{black}{white}{$-A_1$}}
\put(150,47){\fcolorbox{black}{white}{$\int dt~{\rm d}$}}
\put(176,47){\fcolorbox{black}{white}{$-A_2$}}
\put(102,70){\begin{tikzpicture}
 \node[rounded rectangle,draw,fill=white!30]{$\l_1$};
 \end{tikzpicture}}
\put(125,70){\begin{tikzpicture}
 \node[rounded rectangle,draw,fill=white!30]{$-\l_2$};
 \end{tikzpicture}}
\put(150,70){\begin{tikzpicture}
 \node[rounded rectangle,draw,fill=white!30]{$-\l_3$};
 \end{tikzpicture}}
\put(178,70){\begin{tikzpicture}
 \node[rounded rectangle,draw,fill=white!30]{$\l_4$};
 \end{tikzpicture}}
\put(18,0){\includegraphics[scale=.25]{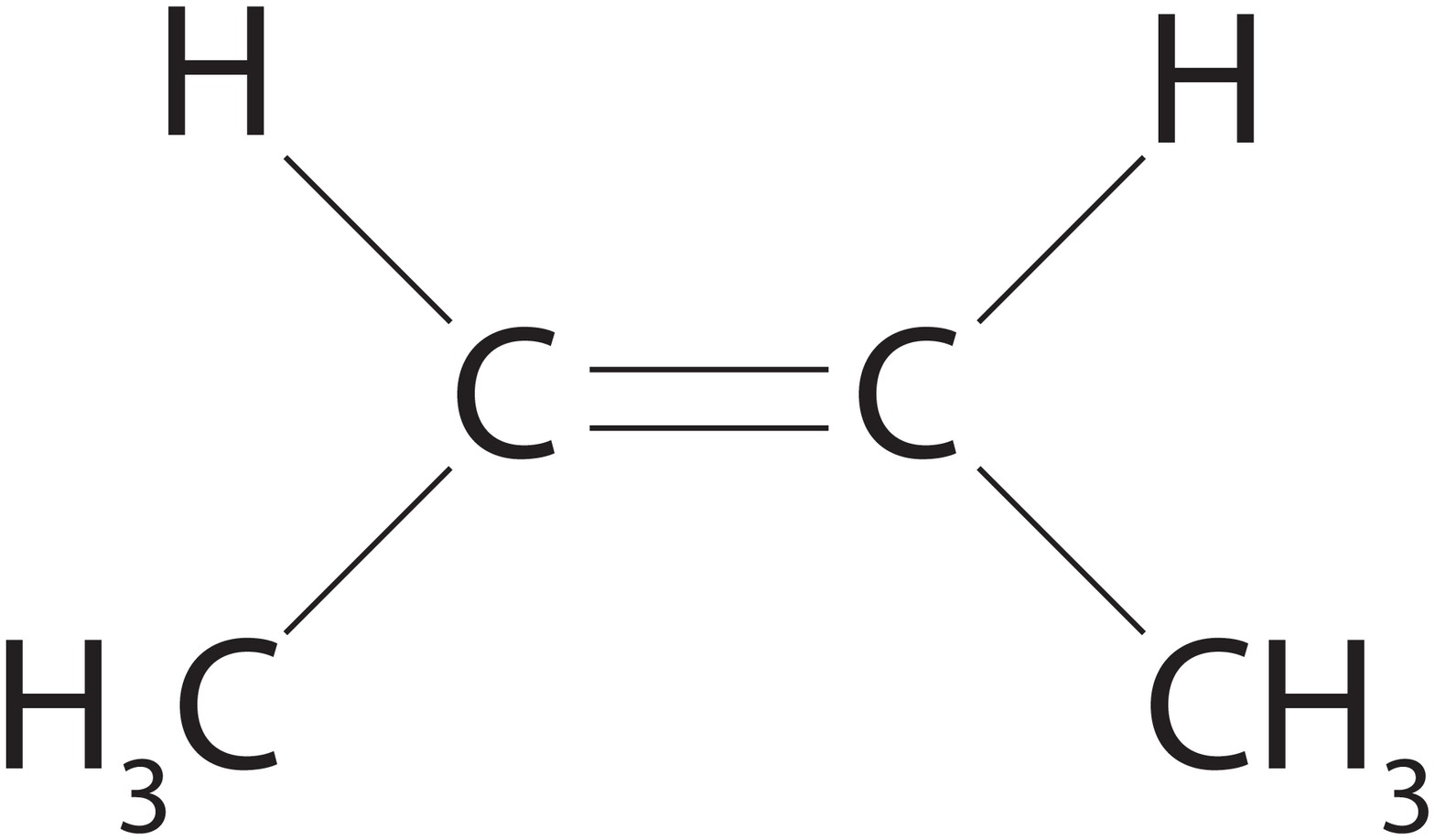}}
\put(116,0){\includegraphics[scale=.25]{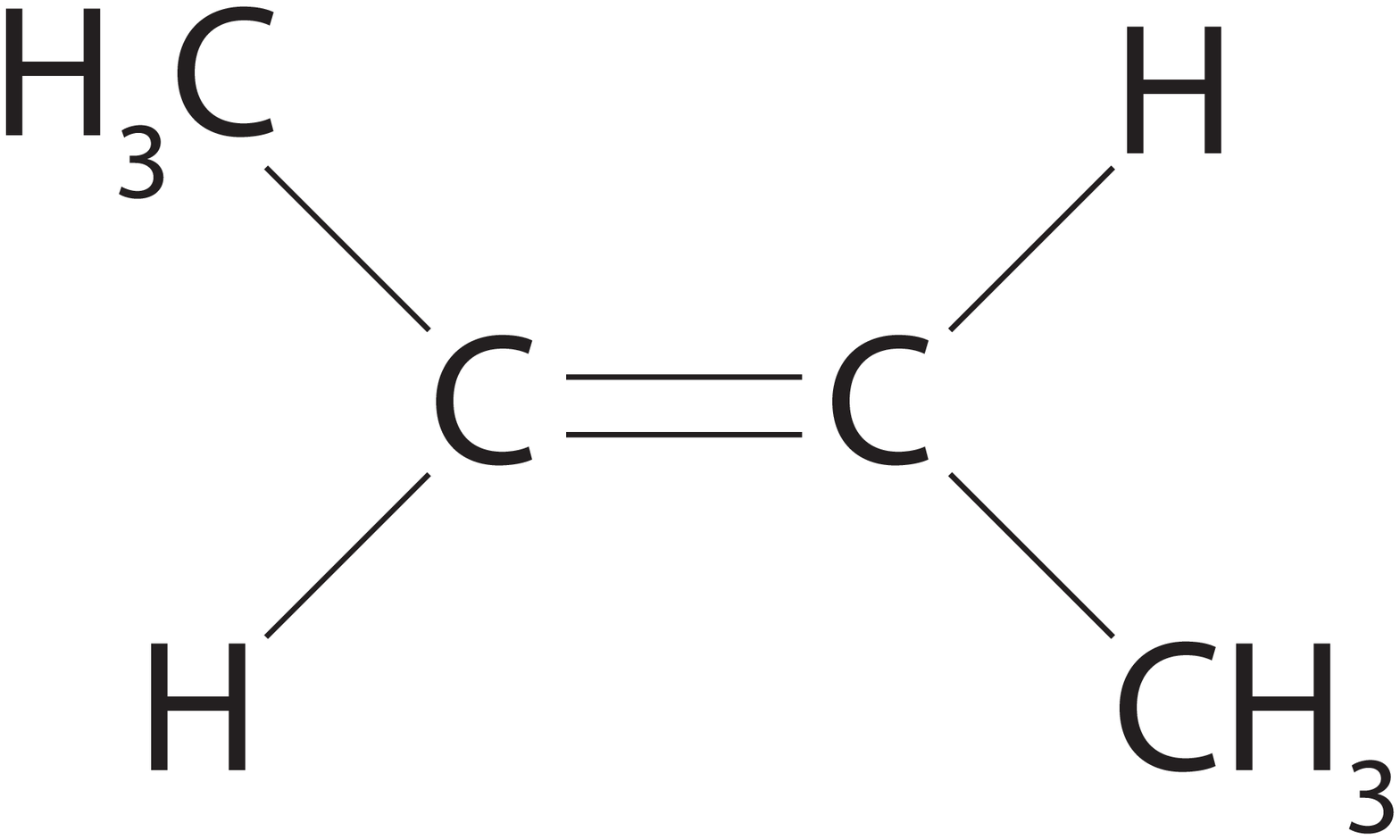}}
   \end{picture}
   \end{center}
   \vspace{-20 pt}
\caption{The cis- (top left) is the adinkraic representation of the chiral multiplet. The trans- (top right) adinkra is the adinkraic representation of both the vector and tensor multiplets. Examples of cis/trans isomers in chemistry are cis-2-butene (bottom left) and trans-2-butene (bottom right)~\cite{Wiki:Isomer}. }
\label{f:cvtvintro}
\end{figure} 

The cis- and trans-adinkras each describe four supersymmetries between four bosons and four fermions. More generally, adinkras have matrix representations that form the algebra of $N$ general real $d_b \times d_f$  matrices that describe $N$ supersymmetries between $d_b$ bosons and $d_f$ fermions: the so-called `garden algebra' which we denote by ${\mathcal{G R}}$($d_b$, $d_f$,$N)$. For off-shell SUSY, we have $d_b  = d_f = d$ and so we will denote off-shell garden algebras by  ${\mathcal{G R}}$($d$,$N)$. Where the cis-adinkra represents the chiral multiplet, the trans-adinkra can represent either the tensor or the vector multiplets. At the time Refs.~\cite{Gates:2009me,Gates:2011aa} were completed, all 
evidence pointed toward there being one cis-  and one trans-adinkra representation. These works 
found no adinkraic distinction between the tensor and vector multiplets. The degeneracy in the trans-adinkra is reminiscent of the historical story of the degeneracy between the anti-up-quark and anti-down-quark that is present if one is \emph{not} aware of the $su(2)$ isospin operator. Once one becomes aware of the $su$(2) isospin operator and its action upon 
the various quarks, the anti-up-quark and anti-down-quark can be seen to be distinct.  More recently Ref.~\cite{CxteR} revealed the existence of a 
${\cal {GR}}$($d$,$N$) operator (roughly analogous to an $su$(2) isospin operator) that shows the 
vector and tensor supermultiplets in the space of adinkras may be regarded as 
distinct. This evidence is embedded in a definition of 
equivalence classes of the ${\mathcal{G R}}$($d$,$N)$ algebras and is related to 
permutation elements. While we are aware of this distinction, we will not use 
this finer definition in the analysis of this paper. Work on the understanding of this additional structure was begun in Ref.~\cite{SUSYHolo}.

Whereas dimensional reduction of a multiplet is trivial, representing this reduction with adinkras generally is not. The 4D, $\mathcal{N}=1$ off-shell chiral, vector, and tensor multiplets are trivially represented as adinkras as they have only four bosonic and four fermionic degrees of freedom $(4|4)$ once gauge degrees of freedom have been removed. Generally speaking, 4D, $\mathcal{N}=1$ off-shell multiplets will have $(4k|4k)$ bosonic and fermionic degrees of freedom, describable in terms of $k$ adinkras, where $k$ is the number of bosonic (or fermionic) off-shell degrees of freedom divided by four
\begin{equation}
	k \equiv d/4 = n_c + n_t
\end{equation} 
with $n_c$ and $n_t$ the number of cis- and trans-adinkras composing the representation, respectively. We therefore use the cis- and trans-adinkras as the building blocks of 4D, $\mathcal{N}=1$ representations. It is important to note that at first glance it is not at all obvious that it is possible to uniquely decompose a 4D, $\mathcal{N}=1$ representation in terms of $n_c$ cis- and $n_t$ trans-adinkras, though remarkably we have always found this to be the case.  For brevity, we refer to $n_c$ and $n_t$ collectively as the SUSY `isomer' numbers and cis- and trans-adinkras collectively as `isomer' adinkras throughout the rest of the paper.\footnote{In Ref.~\cite{Gates:2011aa}, we referred to $n_c$ and $n_t$ as `enantiomer' numbers. We switch here to the overarching word `isomer' as it is both shorter and technically a better analogy between cis/trans isomers in chemistry.} We will also reserve the letter $k$ to refer to the number of isomer adinkras that are necessary to compose a given off-shell representation of size $d=4k$.

For systems with $k>1$, the stumbling point has been in finding the field redefinitions that are generally necessary to express the system in terms of isomer adinkras. When such field redefinitions are necessary, the system has been termed \emph{non-adinkraic} in some of the literature (c.f. Ref.~\cite{Gates:2012zr} for the complex linear supermultiplet). We will not make such distinctions in this paper as we term any multiplet that can be expressed in terms of isomers as `adinkraic' even if it requires field redefinitions. In fact, a main result of this paper is the unveiling of a new methodical procedure that utilizes cis- and trans-adinkras to easily find these field redefinitions. 
This is interesting and quite useful as Ref.~\cite{Gates:2012zr} pointed out that adinkra studies that are restricted to systems that do \emph{not} require field definitions will most likely miss many possibilities, the complex linear supermultiplet being one of them.

 The aforementioned selection rule that has been consistent with all SUSY genomics data to date is that there are only four possible pairs of isomer numbers $n_c$ and $n_t$ that a multiplet of size $k =n_c + n_t$ can have:
 \begin{equation}\label{e:sr}
 (n_c, n_t) =(0,k),~ (1,k-1),~(k,0),~\text{or}~(k-1,0). 
 \end{equation}
This implied selection rule may also contain information about the representation theory of the dimensionally enhanced system. There are some interesting patterns that we see in the field redefinitions that the new adinkranization procedure in this paper brings to light for the first time. While we are curious that things like superhelicity may indeed be embedded in the field redefinition patters, we hold off on this analysis until a later time when we have compiled more data from other multiplets.

We see the selection rule emerge when we build the adinkraic representations for supergravity.
The subject of off-shell 4D, ${\mathcal N} = 1$ supergravity continues 
to be one that reveals surprises even today.  Among the earliest off-shell 
formulations of 4D, ${\mathcal N} = 1$ supergravity were those in Refs.~\cite{Ogievet2,Breit}.
Shortly thereafter the topic received much more attention due to the near simultaneous work in 
Refs.~\cite{Stelle:1978ye,Ferrara:1978em}. 
Recently, the auxiliary field structure of the various off-shell supergravity theories~\cite{ModSG} has been exploited to yield new insights into supersymmetric field theories in rigid, curved backgrounds \cite{Festuccia:2011ws, Dumitrescu:2012ha, Klare:2012gn}. 

A long time ago, 4D, ${\mathcal N} = 1$ Poincar\'e 
supergravity, in its minimal and non-minimal off-shell representations, was
succinctly written in terms of a real parameter $n$ that originated in a superspace 
prepotential analysis~\cite{Siegel:1978mj}. There the discussion was 
given using superspace in terms of non-Riemannian geometry, with torsion.  
 This parameter takes the value $n= - \tfrac13$ for the old-minimal formulation (mSG), $n=0$ for the so-called new-minimal supergravity ($\nu$SG) and any $n\ne -\tfrac13,0$ for the non-minimal representation ($\not$mSG).\footnote{There is, in fact, yet another theory at the linearized level discovered in Ref.~\cite{Buchbinder:2002gh} and explained in Ref.~\cite{Gates:2003cz}. This ``new-new-minimal'' theory ($\nu\nu$SG) is minimal in that it is described by $(12|12)$ degrees of freedom and is very similar in structure to the new-minimal formulation.} We use the same parameter $n$ in this paper. Besides the aforementioned use of these various formulations to describe rigid curved backgrounds, the non-minimal generalizations have certain roles to play in other areas. In the covariant formulation of the $\mathcal{N}=1$ heterotic string in four dimensions, for example, the explicit form of the graviton vertex operator singles out the $n=-1$ non-minimal theory. This same theory was also shown in Ref.~\cite{Gates:1996xs} to belong to a family of $\mathcal{N}=1$ higher-spin theories for which it is possible to find a non-linearly realized $\mathcal{N}=2$ extension. 

In Ref.~\cite{Gates:2003cz}, 4D, $\mathcal{N} =1$ off-shell linearized supergravity was investigated in terms of superprojectors. There it was shown that all $(12|12)$ minimal models (mSG, \newSG, and new-new-minimal (\newnewSG) supergravity) could be formulated in terms of two superprojectors. All models with three superprojectors led to reducible $(16|16)$ models, and models with four superprojectors were the non-minimal models, parameterized by a real number $n$.

Stated another way, the various supergravity representations can be thought of in terms of which compensator is added to the base $(8|8)$ 4D, $\mathcal{N} =1$ conformal supergravity (cSG). For mSG, the compensator is a chiral superfield, $\sigma$. For \newSG ~or \newnewSG, the compensator is a real linear superfield $\mathcal{U}$ or $U$, respectively, where the difference is in the gauge transformation. For \nmSG, the compensator is a complex linear superfield $\Sigma = a \sigma + b~\mathcal{U} + c~U$ with $a$, $b$, and $c$ constants. We can thus succinctly denote the representation sizes as $(4k|4k)$ bosonic and fermionic degrees of freedom, with $k=2$ cSG, $k=3$ mSG, \newSG, or \newnewSG, $k=4$ reducible,  and $k=5$ \nmSG. This leads to $k$ adinkras for a representation: The question is how many cis- and trans-adinkras are there and which component fields are part of which irreducible representation. 

This paper is organized as follows. In Sec.~\ref{s:arev} we review the previous two genomics Refs.~\cite{Gates:2009me,Gates:2011aa}. We review the construction of all adinkras for the off-shell systems presented in those works. We also review various terminology which has become standard in our analyses and that we will use throughout the rest of the paper.  In Sec.~\ref{s:MSG} we show how mSG is decomposed into an adinkra. During this decomposition, we reveal our new and simple procedure that will always produce the adinkra for the system if it exists. In Secs.~\ref{s:MSG} and~\ref{s:CSG} we use this procedure to uncover the adinkra for \nmSG ~and cSG, respectively. Throughout Secs.~\ref{s:MSG}, \ref{s:NMSG}, and \ref{s:CSG} we will show how the adinkras for the compensator field and cSG emerge to compose the full representations. In Sec.~\ref{s:synth}, we explain the pattern from $SO(4)$ representation theory. This includes putting all adinkra data from the previous genomics works together with the supergravity adinkra data to form a cohesive framework based on group theory. We find that adinkras are graphical depictions of characters of the $Spin(4)_R$ group that remains after dimensional reduction.  Section~\ref{s:conc} is the conclusion where we analyze all cis- and trans-adinkra data that has been compiled to date and explain the selection rule that is  consistent with this data and that tells us which adinkras can not be dimensionally enhanced. Our conventions for gamma matrices are shown in Appendix~\ref{a:conv}. The conventions are the same as in Refs.~\cite{Gates:2009me,Gates:2011aa}.

\section{Review of Genomics I and II}\label{s:arev}
$~~~~$ In this section we review the construction of adinkras from dimensional reduction of various 4D, $\mathcal{N}=1$ supersymmetric representations as was done in Refs~\cite{Gates:2009me,Gates:2011aa}. Along the way, this section reviews some of the nomenclature that have become standard in our adinkra analyses such as valises and chromocharacters.

\subsection{The Chiral Multiplet}\label{s:crev}
$~~~~$ The 4D, $\cal N$ = 1 off-shell chiral multiplet (CM) is very well known to consist of a scalar $A$,
a pseudoscalar $B$, a Majorana fermion $\psi_a$, a scalar auxiliary field $F$, and a
pseudoscalar auxiliary field $G$.  The Lagrangian for this system which is supersymmetric with respect to the transformation laws investigated in~\cite{Gates:2009me} is:
\be\label{eq:CMLagrangian}\eqalign{
   {\mathcal L}_{\text{CM}} = & -\frac{1}{2}(\partial_{\mu}A)(\partial^{\mu}A) -\frac{1}{2}(\partial_{\mu} B)(\partial^{\mu}B)+i\frac{1}{2}(\gamma^{\mu})^{ab}\psi_{a}\partial_{\mu}\psi_{b} +\frac{1}{2} F^{2}+\frac{1}{2} G^{2} 
}\ee
This Lagrangian is invariant with respect to the supersymmetric transformation laws
\be\label{e:LCM4D} \eqalign{
{\rm D}_a A ~&=~ \psi_a  ~~~, \cr
{\rm D}_a B ~&=~ i \, (\gamma^5){}_a{}^b \, \psi_b  ~~~, \cr
{\rm D}_a \psi_b ~&=~ i\, (\gamma^\mu){}_{a \,b}\,  \partial_\mu A 
~-~  (\gamma^5\gamma^\mu){}_{a \,b} \, \partial_\mu B ~-~ i \, C_{a\, b} 
\,F  ~+~  (\gamma^5){}_{ a \, b} G  ~~, \cr
{\rm D}_a F ~&=~  (\gamma^\mu){}_a{}^b \, \partial_\mu \, \psi_b   ~~~, \cr
{\rm D}_a G ~&=~ i \,(\gamma^5\gamma^\mu){}_a{}^b \, \partial_\mu \,  
\psi_b  ~~~.
}
\ee
The transformation laws satisfy the algebra
\begin{align}\label{e:DcloseCM}
	\{ {\rm D}_a , {\rm D}_b \} = 2 i (\g^\m_{ab})\partial_\mu~~~.
\end{align}
The 0-brane reduction of the chiral multiplet Lagrangian is acquired by assuming only time dependence of the fields:
\be\label{eq:CMLagrangian0}\eqalign{
   {\mathcal L}_{\text{CM}}^{(0)}= & \frac{1}{2}\dot{A}^2 +\frac{1}{2}\dot{B}^2 +i\frac{1}{2}\delta^{ab}\psi_{a}\dot{\psi}_{b} +\frac{1}{2} F^{2}+\frac{1}{2} G^{2} ~~~,
}\ee
where here and throughout the rest of the paper we denote time derivatives with a dot over the field:
\begin{align}\label{e:dotdef}
	\dot{A} \equiv \partial_0 A~~~,~~~\dot{B} \equiv \partial_0 B~~~,~~~\dot{\psi}_a \equiv \partial_0 \psi_a~~~.
\end{align}
Under this 0-brane reduction, the transformations reduce to those in Tab.~\ref{t:cmred}.
\begin{table}[!ht]
\renewcommand{\arraystretch}{1.3}
\centering
\caption{The 0-brane reduced transformation laws for the 4D, $\mathcal{N}=1$ chiral multiplet.}
\label{t:cmred}
\begin{tabular}{c|cc|cccc|cc}
& $A$ & $B$ & $\psi_1$ & $\psi_2$ & $\psi_3$ & $\psi_4$ & $F$ & $G$ \\
\hline
\gD & $\psi_1$ & $-\psi_4$ & $i \dot{A}$ & $iF$ & $-iG$ & $-i \dot{B}$ & $\dot{\psi}_2$ & $-\dot{\psi}_3$ \\
\hline
\vD & $\psi_2$ & $\psi_3$ & $-iF$ & $i \dot{A}$ & $i\dot{B}$ & $-iG$ & $-\dot{\psi}_1$ & $-\dot{\psi}_4$ \\
\hline
\oD & $\psi_3$ & $-\psi_2$ & $iG$ & $-i\dot{B}$ & $i\dot{A}$ & $-iF$ & $-\dot{\psi}_4$ & $\dot{\psi}_1$ \\
\hline
\rD & $\psi_4$ & $\psi_1$ & $i\dot{B}$ & $iG$ & $iF$ & $i\dot{A}$ & $\dot{\psi}_3$ & $\dot{\psi}_2$
\end{tabular}
\end{table}

The rules for constructing an adinkra from the 0-brane transformation laws are as follows:

\begin{enumerate}
	\item Put bosonic fields in square nodes and fermionic fields in round nodes and arrange them in rows according to their engineering dimensions, increasing from bottom to top. We use the standard shorthand of brackets
\begin{equation}
	[field] = \text{mass dimension of}~field.
\end{equation}
For instance, the dimensions of the fields in the chiral multiplet can be easily seen from the Lagrangian~(\ref{e:LCM4D}) to be: $[A] = [B] = 1$, $[\psi_a] = 3/2$, and $[F]=[G] = 2$.
	\item Draw color coded lines connecting the fields that have supersymmetry transformations between them. The color coding we use throughout this paper is: green for \gD, violet for \vD, orange for \oD, and red for \rD.
	\item A dashed line encodes an overall minus sign in the transformation law. For instance, in the chiral multiplet we see from Tab.~\ref{t:cmred} that $\gD B = -\psi_4$ and $\gD \psi_4 = -i \dot{B}$ so there will be a dashed green line connecting these two fields, as shown in Fig.~\ref{f:CMExt}.
	\item If the transformation laws from boson to fermion have a factor of $i$ multiplying the fermion, place a factor of $i$ in the fermion node. 
\end{enumerate}
\noindent Following these rules we encode the chiral multiplet's 0-brane reduced transformation rules shown in Tab.~\ref{t:cmred} as the adinkra in Fig.~\ref{f:CMExt}. Notice that transformations from upper nodes to lower nodes encode time derivatives acting on the lower nodes. One can see that this must always be true as engineering dimensions decrease from top to bottom. Also, notice that rule four regarding the imaginary number $i$ is not needed for the chiral multiplet as boson to fermion transformations do not contain factors of $i$. We will need to invoke this rule later in this section when we make a field redefinition that absorbs a factor of $i$ into the fermions.

\begin{figure}[!ht]\setlength{\unitlength}{.8 mm}
\begin{center}
   \begin{picture}(90,65)(0,0)
\put(0,0){\includegraphics[width = 90\unitlength]{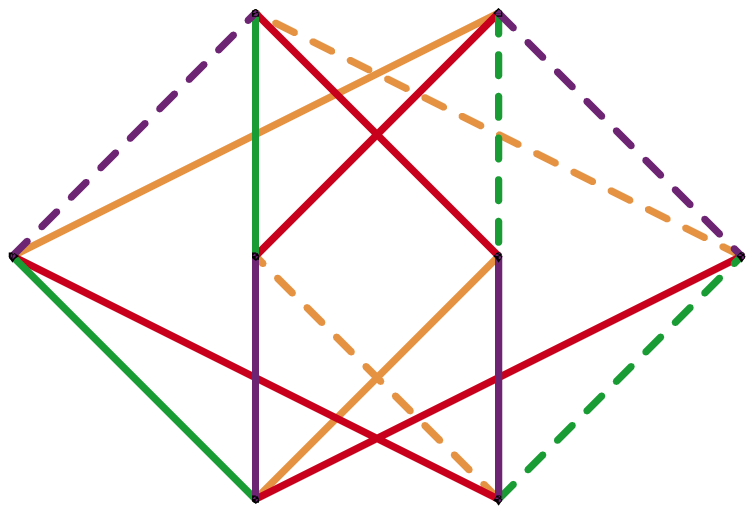}}
\put(29,7){\fcolorbox{black}{white}{$A$}}
\put(54,7){\fcolorbox{black}{white}{$B$}}
\put(1.5,30){\begin{tikzpicture}
 \node[rounded rectangle,draw,fill=white!30]{$\psi_1$};
 \end{tikzpicture}}
\put(27,30){\begin{tikzpicture}
 \node[rounded rectangle,draw,fill=white!30]{$\psi_2$};
 \end{tikzpicture}}
\put(52,30){\begin{tikzpicture}
 \node[rounded rectangle,draw,fill=white!30]{$\psi_3$};
 \end{tikzpicture}}
\put(74.5,30){\begin{tikzpicture}
 \node[rounded rectangle,draw,fill=white!30]{$\psi_4$};
 \end{tikzpicture}}
\put(29,57){\fcolorbox{black}{white}{$F$}}
\put(54,57){\fcolorbox{black}{white}{$G$}}
   \end{picture}
   \end{center}
   \vspace{-25 pt}
\caption{An adinkra for the chiral multiplet. The 0-brane transformation laws in Tab.~\ref{t:cmred} are completely encoded by the colored lines with the identifications \gD, \vD, \oD, and \rD. }
\label{f:CMExt}
\end{figure}

From the adinkra in Fig.~\ref{f:CMExt}, we can lower the $F$ and $G$ nodes to the $A$ and $B$ level by redefining the $F$ and $G$ nodes with an integral, as shown in Fig.~\ref{f:CMVal1}, so that the engineering dimensions of the integrated fields $\int dt F$ and $\int dt G$ are the same as for the non-integrated fields $A$  and $B$. 

\begin{figure}[!ht]\setlength{\unitlength}{.8 mm}
\begin{center}
   \begin{picture}(90,40)(0,0)
\put(0,0){\includegraphics[width = 90\unitlength]{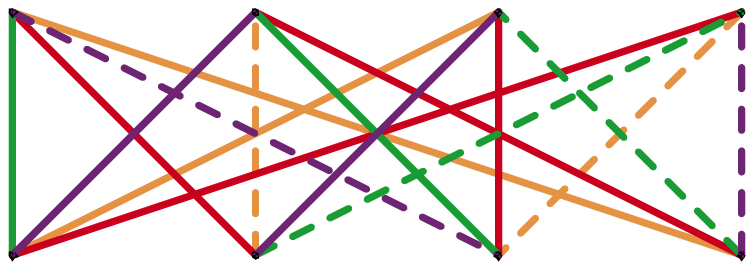}}
\put(3,7){\fcolorbox{black}{white}{$A$}}
\put(28,7){\fcolorbox{black}{white}{$B$}}
\put(49,7){\fcolorbox{black}{white}{$\int dt F$}}
\put(75,7){\fcolorbox{black}{white}{$\int dt G$}}
\put(1.5,30){\begin{tikzpicture}
 \node[rounded rectangle,draw,fill=white!30]{$\psi_1$};
 \end{tikzpicture}}
\put(27,30){\begin{tikzpicture}
 \node[rounded rectangle,draw,fill=white!30]{$\psi_2$};
 \end{tikzpicture}}
\put(52,30){\begin{tikzpicture}
 \node[rounded rectangle,draw,fill=white!30]{$\psi_3$};
 \end{tikzpicture}}
\put(78,30){\begin{tikzpicture}
 \node[rounded rectangle,draw,fill=white!30]{$\psi_4$};
 \end{tikzpicture}}
   \end{picture}
   \end{center}
   \vspace{-25 pt}
\caption{A valise adinkra for the chiral multiplet. The nodes for the $F$ and $G$ fields have been moved down to the $A$ and $B$ level via integration.}
\label{f:CMVal1}
\end{figure} 
\noindent Adinkras such as Fig.~
\ref{f:CMVal1} that contain field nodes of only two different engineering dimensions so that they have been packed into their most compact form, as in a suitcase or valise, are known as \emph{valise} adinkras. We can translate the valise adinkra into four adinkra matrices \textcolor{AdinkraGreen}{$\brL_{\rm 1}$}, \textcolor{AdinkraViolet}{$\brL_{\rm 2}$}, \textcolor{AdinkraOrange}{$\brL_{\rm 3}$}, and \textcolor{AdinkraRed}{$\brL_{\rm 4}$} corresponding to each color link encoding \gD, \vD, \oD, and \rD ~respectively, where the rows correspond to the bosons in the adinkra, left to right, and the columns correspond to the fermions in the adinkra, left to right.  A plus (minus) one means there is a solid (dashed) link of that color between the boson and fermion. A zero means there is not a link of that color between the boson and fermion. The adinkra matrices corresponding to the valise adinkra for the chiral multiplet in Fig.~\ref{f:CMVal1} are:  
\begin{align}
\mbox{\textcolor{AdinkraGreen}
{
$\brL_1 ~=
\left[\begin{array}{cccc}
~1 & ~~0 &  ~~0  &  ~~0 \\
~0 & ~~0 &  ~~0  &  ~-\, 1 \\
~0 & ~~1 &  ~~0  &  ~~0 \\
~0 & ~~0 &  ~-\, 1  &  ~~0 \\
\end{array}\right]$
}
}
 &~~~,~~~
\mbox{\textcolor{AdinkraViolet}{
$\brL_2 ~=\left[\begin{array}{cccc}
~0 & ~~1 &  ~~0  &  ~ \, \, 0 \\
~0 & ~~ 0 &  ~~1  &  ~~0 \\
-\, 1 & ~~ 0 &  ~~0  &  ~~0 \\
~ 0 & ~~~0 &  ~~0  &   -\, 1 \\
\end{array}\right]$
}
}  ~~~
, 
\cr
\mbox{\textcolor{AdinkraOrange}
{
$\brL_3  ~=\left[\begin{array}{cccc}
~0 & ~~0 &  ~~1  &  ~~0 \\
~0 & ~- \, 1 &  ~~0  &  ~~0 \\
~0 & ~~0 &  ~~0  &  -\, 1 \\
~1 & ~~0 &  ~~0  &  ~~0 \\
\end{array}\right]$
}
}
 &~~~,~~~
 \mbox{\textcolor{AdinkraRed}
{
$\brL_4  ~=~
\left[\begin{array}{cccc}
~0 & ~~0 &  ~~0  &  ~ \, \, 1 \\
~1 & ~~ 0 &  ~~0  &  ~~0 \\
~0 & ~~ 0 &  ~~1  &  ~~0 \\
~ 0 & ~~~1 &  ~~0  &   ~~0  \\
\end{array}\right]
$
}
}  ~~~.
 \label{chiD0F}
 \end{align}
 Similarly, we can define four adinkra matrices \textcolor{AdinkraGreen}{$\brR_{\rm 1}$}, \textcolor{AdinkraViolet}{$\brR_{\rm 2}$}, \textcolor{AdinkraOrange}{$\brR_{\rm 3}$}, and \textcolor{AdinkraRed}{$\brR_{\rm 4}$} that are the transposes of the $\brL_{\rm I}$ adinkra matrices such that the rows correspond to \emph{fermions} and the columns correspond to \emph{bosons}:
 \begin{align}\label{e:Rdef}
 	\brR_{\rm I} \equiv \left( \brL_{\rm I} \right)^t = \left( \brL_{\rm I} \right)^{-1}~~~.
 \end{align}
  The ${\mathcal{G R}}$($d$,$N$) algebra, or garden algebra, is the algebra of $d \times d$ real matrices describing $N$ supersymmetries
  \begin{subequations}\label{e:GRdN}
 \begin{align}
 	\brL_{\rm I} \brR_{\rm J} + \brL_{\rm J} \brR_{\rm I} = & 2 \delta_{\rm IJ} {\bm{\rm I}}_d \\
 	\brR_{\rm I} \brL_{\rm J} + \brR_{\rm J} \brL_{\rm I} = & 2 \delta_{\rm IJ} {\bm{\rm I}}_d
 \end{align}
 \end{subequations}
 with $\bm{\rm I}_d$ the $d \times d$ identity matrix. The adinkra matrices \ref{chiD0F} and ~\ref{e:Rdef} for the chiral multiplet satisfy the garden algebra for $d=4$ and $N=4$, i.e., they satisfy the ${\mathcal{G R}}$(4,4) algebra. In fact, any valise adinkra that describes off-shell supersymmetry will have a matrix description that satisfies the corresponding garden algebra ${\mathcal{G R}}$($d$,$N$). The garden algebra is the one-dimensional remnant of the closure relation that the ${\rm D}$-operators satisfy in higher dimensions. For the chiral multiplet, the closure relation in four dimensions was simply Eq.~(\ref{e:DcloseCM}).
 
In the first two genomics works, we lumped the $d$ bosonic and $d$ fermionic nodes of a valise adinkra into $d$-plets, which for the chiral multiplet valise adinkra in Fig.~\ref{f:CMVal1} are the 4-plets:
\begin{align}\label{e:chiPhiPsi}
	\Phi \equiv & \left( 
				\begin{array}{c}
	                	A \\
	            	    B \\
	                	\int dt F \\
	                	\int dt G\\
	            \end{array}	
	        \right) 
	~~~,~~~
	i \Psi \equiv \left(
				\begin{array}{c}
					\psi_1 \\
					\psi_2 \\
					\psi_3 \\
					\psi_4
				\end{array}
			\right)~~~.
\end{align}
Throughout this paper, $\Phi$ and $\Psi$ will always be implicitly defined this way off of a valise adinkra, left to right in the adinkra corresponding to the one to $d$ (up to down) components $\Phi$ and $i\Psi$, so we will always have the succinct relation between the supersymmetry transformation and the ${\bm {\rm L}}_{\rm I}$ and ${\bm {\rm R}}_{\rm I}$ matrices as 
\begin{align}\label{e:LRdef}
	{\rm D}_{\rm I} \Phi = i {\bm {\rm L}}_{\rm I}\Psi~~~,~~~{\rm D}_{\rm I} \Psi = {\bm {\rm R}}_{\rm I} \dot{\Phi}~~~.
\end{align}

Notice that we have absorbed a factor of $i$ into the field redefinitions for the fermions in Eq.~(\ref{e:chiPhiPsi}). The consequence of this is shown in Fig.~\ref{f:CMValPhiPsi} where rule four from the adinkra construction rules listed above has made us place a factor of $i$ in the fermionic nodes. Field redefinitions of this sort have become standard in our analysis of and we will continue to use them in this paper. The next two subsections reviewing the vector and tensor multiplets are simple enough that explicit use of these field redefinitions are not necessary.

\begin{figure}[!ht]\setlength{\unitlength}{.8 mm}
\begin{center}
   \begin{picture}(90,40)(0,0)
\put(0,0){\includegraphics[width = 90\unitlength]{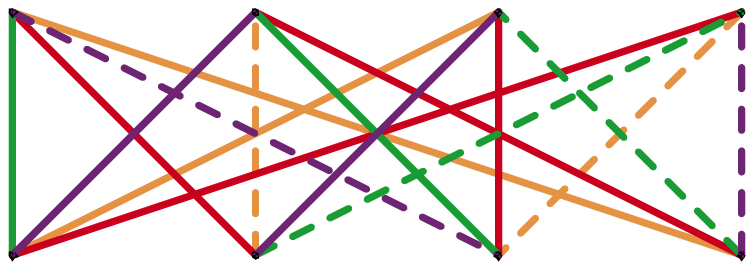}}
\put(2.5,7){\fcolorbox{black}{white}{$\Phi_1$}}
\put(28,7){\fcolorbox{black}{white}{$\Phi_2$}}
\put(53,7){\fcolorbox{black}{white}{$\Phi_3$}}
\put(78,7){\fcolorbox{black}{white}{$\Phi_4$}}
\put(1,30){\begin{tikzpicture}
 \node[rounded rectangle,draw,fill=white!30]{$i\Psi_1$};
 \end{tikzpicture}}
\put(26,30){\begin{tikzpicture}
 \node[rounded rectangle,draw,fill=white!30]{$i\Psi_2$};
 \end{tikzpicture}}
\put(51,30){\begin{tikzpicture}
 \node[rounded rectangle,draw,fill=white!30]{$i\Psi_3$};
 \end{tikzpicture}}
\put(76,30){\begin{tikzpicture}
 \node[rounded rectangle,draw,fill=white!30]{$i\Psi_4$};
 \end{tikzpicture}}
   \end{picture}
   \end{center}
   \vspace{-25 pt}
\caption{A valise adinkra for the chiral multiplet under the field redefinitions in Eq.~(\ref{e:chiPhiPsi}). Notice the factors of $i$ in the fermion nodes that are necessary to be consistent with rule four of the adinkra construction rules.}
\label{f:CMValPhiPsi}
\end{figure}

\subsection{The Vector Multiplet}
$~~~~$ The 4D, $\cal N$ = 1 off-shell vector multiplet (VM)  is described by a vector
 $A{}_{\mu}$, a Majorana fermion $\l_a$, and a pseudoscalar auxiliary field
 d.  The Lagrangian for the vector multiplet which is supersymmetric with respect to the transformation laws investigated in~\cite{Gates:2009me} is:
\be\label{eq:VMLagrangian}\eqalign{
   {\mathcal L}_{\text{VM}} = &-\frac{1}{4}F_{\mu\nu}F^{\mu\nu} +\frac{1}{2}i(\gamma^{\mu})^{ab}\lambda_{a}\partial_{\mu}\lambda_{b}+\frac{1}{2}{\rm d}^2
}\ee 
where the gauge field strength is the canonical
\begin{align}
	F_{\mu\nu} = \partial_\mu A_\nu - \partial_\nu A_\mu~~~.
\end{align}
The 0-brane reduction is acquired by assuming only time dependence of the fields and selecting the temporal gauge
\begin{equation}
	A_0 = 0~~~.
\end{equation}  
The resulting 0-brane Lagrangian is:
\be\label{eq:VMLagrangian0}\eqalign{
   {\mathcal L}_{VM}^{(0)} = &\frac{1}{2}(\dot{A}_1^2 + \dot{A}_2^2 + \dot{A}_3^2) +\frac{1}{2}i \delta^{ab}\lambda_{a}\dot{\lambda}_{b}+\frac{1}{2}{\rm d}^2
}\ee 
The 0-brane reduced SUSY transformation laws which are a symmetry of this Lagrangian were shown in Ref.~\cite{Gates:2009me}. They can be succinctly displayed as the adinkra in Fig.~\ref{f:VMExt}.

\begin{figure}[!ht]\setlength{\unitlength}{.8 mm}
\begin{center}
   \begin{picture}(90,65)(0,0)
\put(0,0){\includegraphics[width = 90\unitlength]{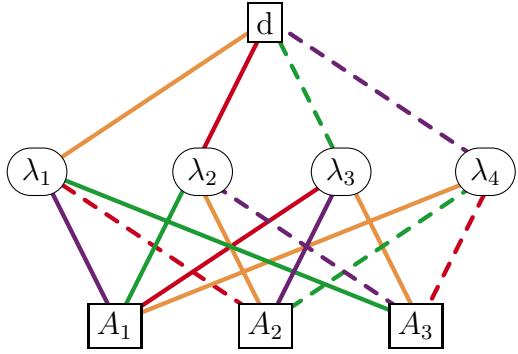}}
\put(15,7){\fcolorbox{black}{white}{$A_1$}}
\put(40,7){\fcolorbox{black}{white}{$A_2$}}
\put(65,7){\fcolorbox{black}{white}{$A_3$}}
\put(41.5,57){\fcolorbox{black}{white}{${\rm d}$}}
\put(1.5,30){\begin{tikzpicture}
 \node[rounded rectangle,draw,fill=white!30]{$\l_1$};
 \end{tikzpicture}}
\put(29,30){\begin{tikzpicture}
 \node[rounded rectangle,draw,fill=white!30]{$\l_2$};
 \end{tikzpicture}}
\put(52,30){\begin{tikzpicture}
 \node[rounded rectangle,draw,fill=white!30]{$\l_3$};
 \end{tikzpicture}}
\put(76,30){\begin{tikzpicture}
 \node[rounded rectangle,draw,fill=white!30]{$\l_4$};
 \end{tikzpicture}}
   \end{picture}
   \end{center}
   \vspace{-20 pt}
\caption{An adinkra for the vector multiplet.}
\label{f:VMExt}
\end{figure} 
Lowering the $\rm d$ node in this adinkra leads to a valise adinkra description as shown in Fig.~\ref{f:VMVal1}.

\begin{figure}[!ht]\setlength{\unitlength}{.8 mm}
\begin{center}
   \begin{picture}(90,40)(0,0)
\put(0,0){\includegraphics[width = 90\unitlength]{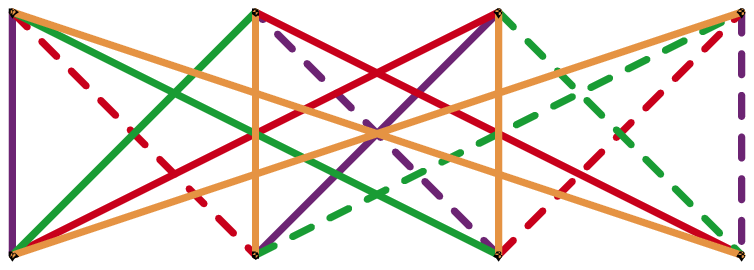}}
\put(3,7){\fcolorbox{black}{white}{$A_1$}}
\put(28,7){\fcolorbox{black}{white}{$A_2$}}
\put(53,7){\fcolorbox{black}{white}{$A_3$}}
\put(74,7){\fcolorbox{black}{white}{$\int dt~ {\rm d}$}}
\put(1.5,30){\begin{tikzpicture}
 \node[rounded rectangle,draw,fill=white!30]{$\l_1$};
 \end{tikzpicture}}
\put(27,30){\begin{tikzpicture}
 \node[rounded rectangle,draw,fill=white!30]{$\l_2$};
 \end{tikzpicture}}
\put(52,30){\begin{tikzpicture}
 \node[rounded rectangle,draw,fill=white!30]{$\l_3$};
 \end{tikzpicture}}
\put(78,30){\begin{tikzpicture}
 \node[rounded rectangle,draw,fill=white!30]{$\l_4$};
 \end{tikzpicture}}
   \end{picture}
   \end{center}
   \vspace{-20 pt}
\caption{A valise adinkra for the vector multiplet.}
\label{f:VMVal1}
\end{figure} 
In the same way that we defined adinkra matrices for the chiral multiplet in Sec.~\ref{s:crev}, the adinkra matrices can be read directly off of Fig.~\ref{f:VMVal1} for the vector multiplet:
\begin{align}
\mbox{\textcolor{AdinkraGreen}
{
$\brL_1 ~=
\left[\begin{array}{cccc}
~0 & ~1 &  ~ 0  &  ~ 0 \\
~0 & ~0 &  ~0  &  -\,1 \\
~1 & ~0 &  ~ 0  &  ~0 \\
~0 & ~0 &  -\, 1  &  ~0 \\
\end{array}\right]$
}
}
 &~~~,~~~
\mbox{\textcolor{AdinkraViolet}{
$\brL_2 ~=\left[\begin{array}{cccc}
~1 & ~ 0 &  ~0  &  ~ 0 \\
~0 & ~ 0 &  ~1  &  ~ 0 \\
 ~0 & - \, 1 &  ~0  &   ~ 0 \\
~0 & ~0 &  ~0  &  -\, 1 \\
\end{array}\right]$
}
}  ~~~
, 
\cr
\mbox{\textcolor{AdinkraOrange}
{
$\brL_3  ~=\left[\begin{array}{cccc}
~0 & ~0 &  ~ 0  &  ~ 1 \\
~0 & ~1 &  ~0  &   ~0 \\
~0 & ~0 &  ~ 1  &  ~0 \\
~1 & ~0 &  ~0  &  ~0 \\
\end{array}\right]$
}
}
 &~~~,~~~
 \mbox{\textcolor{AdinkraRed}
{
$\brL_4  ~=~
\left[\begin{array}{cccc}
~0 & ~0 &  ~1  &  ~ 0 \\
-\,1 & ~ 0 &  ~0  &  ~ 0 \\
 ~0 & ~0 &  ~0  &   - \, 1 \\
~0 & ~1 &  ~0  &  ~  0 \\
\end{array}\right]
$
}
}  ~~~.
 \label{e:VMval}
 \end{align}
 As before, the $\brR_{\rm I}$ matrices are defined as the transposes of the $\brL_{\rm I}$ matrices, Eq.~\ref{e:Rdef}, and together they satisfy the $\mathcal{GR}(4,4)$ algebra~(\ref{e:GRdN}) with $d=4$.
 
\subsection{The Tensor Multiplet}
$~~~~$ The 4D, $\cal N$ = 1 off-shell tensor multiplet (TM) consists of a scalar $\varphi$, a second-rank
skew symmetric tensor, $B{}_{\mu \, \nu }$, and a Majorana fermion $\chi_a$.  The Lagrangian for the tensor multiplet which is supersymmetric with respect to the transformation laws investigated in~\cite{Gates:2009me} is:
\be\eqalign{
 \mathcal{L}_{\text{TM}} = &  - \frac{1}{3}H_{\mu\nu\alpha}H^{\mu\nu\alpha}- \frac{1}2 \partial_\mu \varphi \partial^\mu \varphi   + \frac{1}{2} i (\gamma^\mu)^{bc} \chi_b \partial_\mu \chi_c  
}\ee
where
\be
  H_{\mu\nu\alpha} \equiv \partial_\m B_{\n\a} + \partial_\n B_{\a\m} + \partial_\a B_{\m\n}  ~~~.
\ee
Choosing temporal gauge
\begin{align}
	B_{0\mu} = 0
\end{align}
the 0-brane reduced Lagrangian is:
\begin{align}
 \mathcal{L}^{(0)}_{\text{TM}} = & 2\left( \dot{B}_{12}^2 + \dot{B}_{23}^2 + \dot{B}_{31}^2\right) + \frac{1}{2} \dot{\varphi}^2 + \frac{1}{2} i \delta^{bc} \chi_b \dot{\chi}_c  ~~~.
\end{align}
The 0-brane reduced SUSY transformation laws which are a symmetry of this Lagrangian were shown in Ref.~\cite{Gates:2009me}. They can be succinctly displayed as the valise adinkra in Fig.~\ref{f:TMVal1} without any need to lower nodes. 
\begin{figure}[!ht]\setlength{\unitlength}{.8 mm}
\begin{center}
   \begin{picture}(90,40)(0,0)
\put(0,0){\includegraphics[width = 90\unitlength]{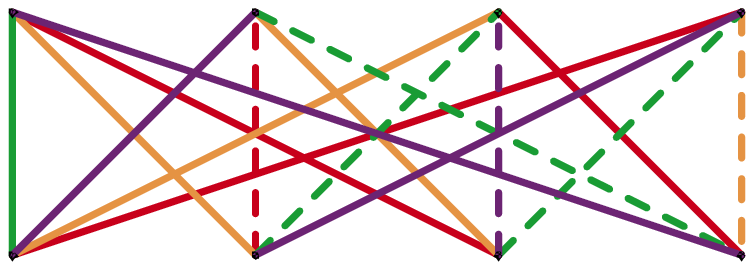}}
\put(3,7){\fcolorbox{black}{white}{$\varphi$}}
\put(26,7){\fcolorbox{black}{white}{$2B_{12}$}}
\put(51,7){\fcolorbox{black}{white}{$2B_{23}$}}
\put(76,7){\fcolorbox{black}{white}{$2B_{31}$}}
\put(1.5,30){\begin{tikzpicture}
 \node[rounded rectangle,draw,fill=white!30]{$\chi_1$};
 \end{tikzpicture}}
\put(27,30){\begin{tikzpicture}
 \node[rounded rectangle,draw,fill=white!30]{$\chi_2$};
 \end{tikzpicture}}
\put(52,30){\begin{tikzpicture}
 \node[rounded rectangle,draw,fill=white!30]{$\chi_3$};
 \end{tikzpicture}}
\put(78,30){\begin{tikzpicture}
 \node[rounded rectangle,draw,fill=white!30]{$\chi_4$};
 \end{tikzpicture}}
   \end{picture}
   \end{center}
   \vspace{-20 pt}
\caption{A valise adinkra for the tensor multiplet.}
\label{f:TMVal1}
\end{figure}
As we did for the vector and chiral multiplets in the previous two subsections, here we define the adinkra matrices for the tensor multiplet directly off of its valise adinkra in Fig.~\ref{f:TMVal1}:
\begin{align}
\mbox{\textcolor{AdinkraGreen}
{
$\brL_1 ~=
\left[\begin{array}{cccc}
~1 & ~0 &  ~0  &  ~0 \\
~0 & ~0 &  -\, 1  &  ~ 0 \\
~0 & ~0 &  ~0  &  -\,1 \\
~0 & -\,1 &  ~ 0  &  ~0 \\
\end{array}\right]$
}
}
 &~~~,~~~
\mbox{\textcolor{AdinkraViolet}{
$\brL_2 ~=\left[\begin{array}{cccc}
~0 & ~1 &  ~0  &  ~  0 \\
~0 & ~ 0 &  ~0  &  ~ 1 \\
~0 & ~ 0 &  -\,1  &  ~ 0 \\
 ~ 1 & ~0 &  ~0  &   ~ 0 \\
\end{array}\right]$
}
}  ~~~
, 
\cr
\mbox{\textcolor{AdinkraOrange}
{
$\brL_3  ~=\left[\begin{array}{cccc}
~0 & ~0 &  ~1  &  ~0 \\
~1 & ~0 &  ~ 0  &  ~ 0 \\
~0 & ~1 &  ~0  &   ~0 \\
~0 & ~0 &  ~ 0  &  -\, 1 \\
\end{array}\right]$
}
}
 &~~~,~~~
 \mbox{\textcolor{AdinkraRed}
{
$\brL_4  ~=~
\left[\begin{array}{cccc}
~0 & ~0 &  ~0  &  ~  1 \\
~0 & -\, 1 &  ~0  &  ~ 0 \\
~1 & ~ 0 &  ~0  &  ~ 0 \\
 ~0 & ~0 &  ~1  &   ~ 0 \\
\end{array}\right]
$
}
}  ~~~.
 \label{e:TML}
 \end{align}
 As before, the $\brR_{\rm I}$ matrices are defined as the transposes of the $\brL_{\rm I}$ matrices, Eq.~\ref{e:Rdef}, and together they satisfy the $\mathcal{GR}(4,4)$ algebra~(\ref{e:GRdN}) with $d=4$.
 
\subsection{Adinkra Transformations, Cis- and Trans-Adinkras, and Chromocharacters}\label{s:chromo}
$~~~~$ Here we review the main result of Genomics I~\cite{Gates:2009me} pertinent to the discussion of the current paper: adinkra transformations, cis- and trans-adinkras, and chromocharacters which are the mathematical formulas that pull out the isomer numbers $n_c$ and $n_t$ from the $\brL_{\rm I}$ and $\brR_{\rm I}$ matrices. Given a valise adinkra, moving the nodes around left and right and redefining nodes on the same level will leave us with an equivalent valise adinkra that holds the same 0-brane information about the higher dimensional system as before. We refer to nodal rearrangements and redefinitions such as these as \emph{adinkra transformations}. To use valise adinkras to classify higher dimensional systems, it is therefore necessary to find which valise adinkras are equivalent to other valise adinkras under adinkra transformations. For this purpose, it is useful to rearrange the nodes in the canonical valises for the chiral, vector, and tensor multiplets as shown in Figs.~\ref{f:CMValReorder}, \ref{f:VMValReorder}, and~\ref{f:TMValReorder}. Comparing the transformed adinkras on the right in each of these, we notice that that the connections are all the same if we disregard the dashings. In fact, the rightmost valise for the vector and tensor multiplets are identical and the rightmost valise for the chiral multiplet differs from the vector and tensor only by a minus sign in the orange transformation \oD. Such color transformations that swap the sign of a D-transformation are outside of what we classify as adinkra transformations. As such, no adinkra transformation exists that can make the chiral multiplet's valise exactly equivalent to the vector and tensor multiplets valise: the 1D shadow adinkras `know' that a difference exists between the chiral and vector multiplets and between the chiral and tensor multiplets in the higher dimensional representation theory! In fact, \emph{all} $d=4$, $N=4$ valise adinkras can be transformed to one and only one of these two valise adinkras via adinkra transformations. We therefore classify $d=4$, $N=4$ valise adinkras into two types: cis for the chiral multiplet and trans for the vector and tensor multiples~\cite{Gates:2009me}. Throughout this paper, the terms cis- and trans-adinkra refer to the rightmost adinkras in Figs.~\ref{f:CMValReorder} and ~\ref{f:VMValReorder} (or \ref{f:TMValReorder}), respectively. Since when we refer to the cis- and trans-adinkra we are always referring to a valise, we will often drop the adjective valise throughout this paper.

\begin{figure}[!ht]
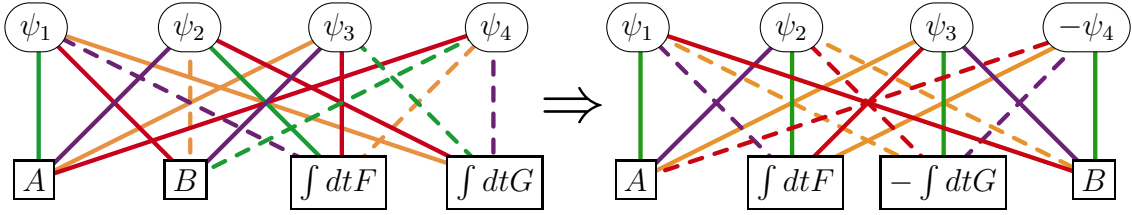
\setlength{\unitlength}{.8 mm}
\begin{center}
   \begin{picture}(200,40)(0,0)
\put(0,0){\includegraphics[width = 90\unitlength]{CMVal1}}
\put(3,7){\fcolorbox{black}{white}{$A$}}
\put(28,7){\fcolorbox{black}{white}{$B$}}
\put(49,7){\fcolorbox{black}{white}{$\int dt F$}}
\put(75,7){\fcolorbox{black}{white}{$\int dt G$}}
\put(1.5,30){\begin{tikzpicture}
 \node[rounded rectangle,draw,fill=white!30]{$\psi_1$};
 \end{tikzpicture}}
\put(27,30){\begin{tikzpicture}
 \node[rounded rectangle,draw,fill=white!30]{$\psi_2$};
 \end{tikzpicture}}
\put(52,30){\begin{tikzpicture}
 \node[rounded rectangle,draw,fill=white!30]{$\psi_3$};
 \end{tikzpicture}}
\put(78,30){\begin{tikzpicture}
 \node[rounded rectangle,draw,fill=white!30]{$\psi_4$};
 \end{tikzpicture}}
\put(90.5,18.5){\Huge $\Rightarrow$}
\quad
\put(100,0){\includegraphics[width = 90\unitlength]{CisValise}}
\put(103,7){\fcolorbox{black}{white}{$A$}}
\put(125,7){\fcolorbox{black}{white}{$\int dt F$}}
\put(147,7){\fcolorbox{black}{white}{$-\int dt G$}}
\put(179,7){\fcolorbox{black}{white}{$B$}}
\put(101.5,30){\begin{tikzpicture}
 \node[rounded rectangle,draw,fill=white!30]{$\psi_1$};
 \end{tikzpicture}}
\put(127,30){\begin{tikzpicture}
 \node[rounded rectangle,draw,fill=white!30]{$\psi_2$};
 \end{tikzpicture}}
\put(152,30){\begin{tikzpicture}
 \node[rounded rectangle,draw,fill=white!30]{$\psi_3$};
 \end{tikzpicture}}
\put(174,30){\begin{tikzpicture}
 \node[rounded rectangle,draw,fill=white!30]{$-\psi_4$};
 \end{tikzpicture}}
   \end{picture}
   \end{center}
   \vspace{-20 pt}
\caption{Transformation of the canonical chiral multiplet valise (Fig.~\ref{f:CMVal1}) into the cis-valise representation. }
\label{f:CMValReorder}
\end{figure} 

\begin{figure}[!ht]\setlength{\unitlength}{.8 mm}
\begin{center}
   \begin{picture}(200,40)(0,0)
\put(0,0){\includegraphics[width = 90\unitlength]{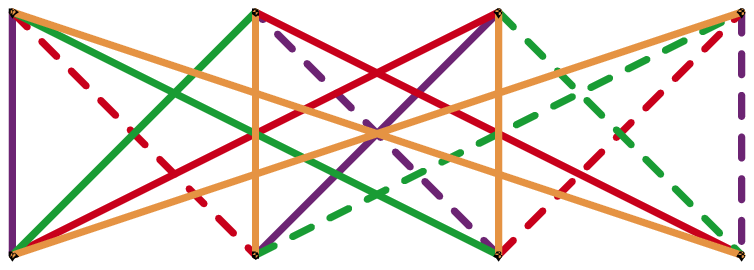}}
\put(3,7){\fcolorbox{black}{white}{$A_1$}}
\put(28,7){\fcolorbox{black}{white}{$A_2$}}
\put(53,7){\fcolorbox{black}{white}{$A_3$}}
\put(74,7){\fcolorbox{black}{white}{$\int dt~ {\rm d}$}}
\put(1.5,30){\begin{tikzpicture}
 \node[rounded rectangle,draw,fill=white!30]{$\l_1$};
 \end{tikzpicture}}
\put(27,30){\begin{tikzpicture}
 \node[rounded rectangle,draw,fill=white!30]{$\l_2$};
 \end{tikzpicture}}
\put(52,30){\begin{tikzpicture}
 \node[rounded rectangle,draw,fill=white!30]{$\l_3$};
 \end{tikzpicture}}
\put(78,30){\begin{tikzpicture}
 \node[rounded rectangle,draw,fill=white!30]{$\l_4$};
 \end{tikzpicture}}
\put(90.5,18.5){\Huge $\Rightarrow$}
\quad
\put(100,0){\includegraphics[width = 90\unitlength]{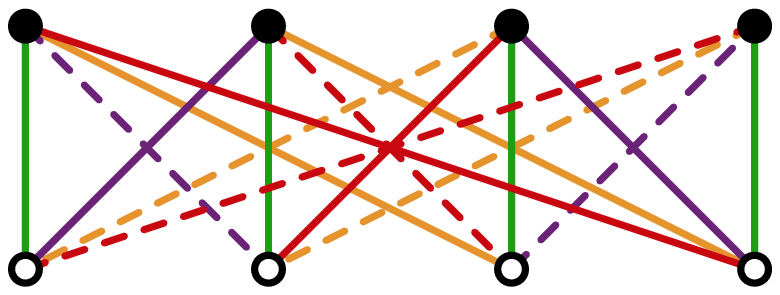}}
\put(103,7){\fcolorbox{black}{white}{$A_3$}}
\put(126,7){\fcolorbox{black}{white}{$-A_1$}}
\put(150,7){\fcolorbox{black}{white}{$\int dt~{\rm d}$}}
\put(176,7){\fcolorbox{black}{white}{$-A_2$}}
\put(102,30){\begin{tikzpicture}
 \node[rounded rectangle,draw,fill=white!30]{$\l_1$};
 \end{tikzpicture}}
\put(125,30){\begin{tikzpicture}
 \node[rounded rectangle,draw,fill=white!30]{$-\l_2$};
 \end{tikzpicture}}
\put(150,30){\begin{tikzpicture}
 \node[rounded rectangle,draw,fill=white!30]{$-\l_3$};
 \end{tikzpicture}}
\put(178,30){\begin{tikzpicture}
 \node[rounded rectangle,draw,fill=white!30]{$\l_4$};
 \end{tikzpicture}}
   \end{picture}
   \end{center}
   \vspace{-20 pt}
\caption{Transformation of the canonical vector multiplet valise (Fig.~\ref{f:VMVal1}) into the trans-adinkra representation. }
\label{f:VMValReorder}
\end{figure}

\begin{figure}[!ht]\setlength{\unitlength}{.8 mm}
\begin{center}
   \begin{picture}(200,40)(0,0)
\put(0,0){\includegraphics[width = 90\unitlength]{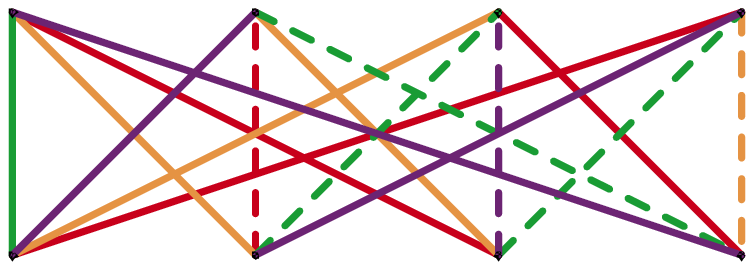}}
\put(3,7){\fcolorbox{black}{white}{$\varphi$}}
\put(26,7){\fcolorbox{black}{white}{$2B_{12}$}}
\put(51,7){\fcolorbox{black}{white}{$2B_{23}$}}
\put(76,7){\fcolorbox{black}{white}{$2B_{31}$}}
\put(1.5,30){\begin{tikzpicture}
 \node[rounded rectangle,draw,fill=white!30]{$\chi_1$};
 \end{tikzpicture}}
\put(27,30){\begin{tikzpicture}
 \node[rounded rectangle,draw,fill=white!30]{$\chi_2$};
 \end{tikzpicture}}
\put(52,30){\begin{tikzpicture}
 \node[rounded rectangle,draw,fill=white!30]{$\chi_3$};
 \end{tikzpicture}}
\put(78,30){\begin{tikzpicture}
 \node[rounded rectangle,draw,fill=white!30]{$\chi_4$};
 \end{tikzpicture}}
\put(90.5,18.5){\Huge $\Rightarrow$}
\quad
\put(100,0){\includegraphics[width = 90\unitlength]{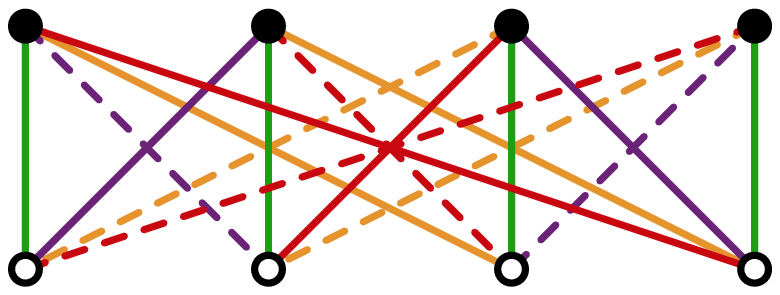}}
\put(104,7){\fcolorbox{black}{white}{$\varphi$}}
\put(123,7){\fcolorbox{black}{white}{$-2B_{31}$}}
\put(151,7){\fcolorbox{black}{white}{$2B_{12}$}}
\put(176,7){\fcolorbox{black}{white}{$2B_{23}$}}
\put(102.5,30){\begin{tikzpicture}
 \node[rounded rectangle,draw,fill=white!30]{$\chi_1$};
 \end{tikzpicture}}
\put(126,30){\begin{tikzpicture}
 \node[rounded rectangle,draw,fill=white!30]{$\chi_2$};
 \end{tikzpicture}}
\put(150,30){\begin{tikzpicture}
 \node[rounded rectangle,draw,fill=white!30]{$-\chi_3$};
 \end{tikzpicture}}
\put(176,30){\begin{tikzpicture}
 \node[rounded rectangle,draw,fill=white!30]{$-\chi_4$};
 \end{tikzpicture}}
   \end{picture}
   \end{center}
   \vspace{-20 pt}
\caption{Transformation of the canonical tensor multiplet valise (Fig.~\ref{f:TMVal1}) into the trans-adinkra representation. }
\label{f:TMValReorder}
\end{figure} 
It is useful to introduce the $SO(4)$ generators
\be\label{eq:SO4Generators}
\begin{array}{lll}
    i{\bm \alpha}_1 = i{\bm \sigma}^2 \otimes {\bm \sigma}^1~~, & i{\bm \alpha}_2 = i{\bm {\rm I}}_2 \otimes {\bm \sigma}^2~~, & i{\bm \alpha}_3 = i{\bm \sigma}^2 \otimes {\bm \sigma}^3~~, \\
     i{\bm \beta}_1 = i{\bm \sigma}^1 \otimes {\bm \sigma}^2~~, & i{\bm \beta}_2 = i{\bm \sigma}^2 \otimes {\bm {\rm I}}_2 ~~, & i{\bm \beta}_3 = i{\bm \sigma}^3 \otimes {\bm \sigma}^2
   \end{array}~~~,
\ee
and the parameter
\begin{equation}\label{e:chi0}
	\chi_0 = \left\{ \begin{array}{ll}
						+1 & \text{cis} \\
						-1 & \text{trans}
					\end{array}
					\right.~~~,
\end{equation}
so that we may write the cis- and trans-adinkra matrices simultaneously as:
\be\label{e:ctval}
\begin{array}{llll}
\textcolor{AdinkraGreen}{\brL_1 =  {\bm {\rm I}}_4} & ,& \textcolor{AdinkraViolet}{\brL_2 = i {\bm {\rm \beta}}_3} &, \\
\textcolor{AdinkraOrange}{\brL_3 =}  \textcolor{AdinkraOrange}{\chi_0 i  {\bm {\rm \beta}}_2} &, & \textcolor{AdinkraRed}{\brL_4 = - i {\bm {\rm \beta}}_1} & .
\end{array}
\ee

The adinkra transformations in Figs.~\ref{f:CMValReorder}, ~\ref{f:VMValReorder}, and~\ref{f:TMValReorder} can be realized on their corresponding matrix representations as:
\begin{subequations}
\label{e:XLY}
\begin{align}
	\brL_{\rm I}^{(cis)} =& \mathcal{X}^{(\text{CM})} \brL_{\rm I}^{(\text{CM})} \mathcal{Y}^{(\text{CM})}~~~,\\
	\brL_{\rm I}^{(trans)} =& \mathcal{X}^{(\text{VM})} \brL_{\rm I}^{(\text{VM})} \mathcal{Y}^{(\text{VM})} \cr
	=& \mathcal{X}^{(\text{TM})} \brL_{\rm I}^{(\text{TM})} \mathcal{Y}^{(\text{TM})}  ~~~.
\end{align}
\end{subequations}
The matrices $\mathcal{X}$ and $\mathcal{Y}$ for each representation are just a reordering of the $\Phi$ and $\Psi$ node $4$-plets
\begin{equation}
	\Phi \to \mathcal{X} \Phi~~~,~~~\Psi \to \mathcal{Y}^t\Psi~~~
\end{equation}
where $\Phi$ and $\Psi$ for the vector and tensor multiplets are defined canonically off their valise adinkras in the same was as for the chiral multiplet in Eq.~(\ref{e:chiPhiPsi}): the left to right nodes correspond to the one to four (up to down) components of $\Phi$ and $i\Psi$. 
The matrices $\mathcal{X}$ and $\mathcal{Y}$ can be easily read off Figs.~\ref{f:CMValReorder}, \ref{f:VMValReorder}, and~\ref{f:TMValReorder} for each representation:
\begin{align}\label{e:CMXY}
	\mathcal{X}^{(\text{CM})} = \left(
					\begin{array}{cccc}
					    1 & 0 & 0 & 0 \\
					    0 & 0 & 1 & 0 \\
					    0 & 0 & 0 & -1 \\
					    0 & 1 & 0 & 0 
					\end{array}
					\right)~~~&,~~~\mathcal{Y}^{(\text{CM})}  
					=\left(
					\begin{array}{cccc}
						1 & 0 & 0 & 0 \\
						0 & 1 & 0 & 0 \\
						0 & 0 & 1 & 0 \\
						0 & 0 & 0 & -1
					\end{array}
					\right) ~~~,
	\\
	\label{e:VMXY}
	\mathcal{X}^{(\text{VM})} = \left(
					\begin{array}{cccc}
					    0 & 0 & 1 & 0 \\
					    -1 & 0 & 0 & 0 \\
					    0 & 0 & 0 & 1 \\
					    0 & -1 & 0 & 0 
					\end{array}
					\right)~~~&,~~~\mathcal{Y}^{(\text{VM})}  
					=\left(
					\begin{array}{cccc}
						1 & 0 & 0 & 0 \\
						0 & -1 & 0 & 0 \\
						0 & 0 & -1 & 0 \\
						0 & 0 & 0 & 1
					\end{array}
					\right)~~~,
	\\
	\label{e:TMXY}
	\mathcal{X}^{(\text{TM})} = \left(
					\begin{array}{cccc}
					    1 & 0 & 0 & 0 \\
					    0 & 0 & 0 & -1 \\
					    0 & 1 & 0 & 0 \\
					    0 & 0 & 1 & 0 
					\end{array}
					\right)~~~&,~~~\mathcal{Y}^{(\text{TM})}  
					=\left(
					\begin{array}{cccc}
						1 & 0 & 0 & 0 \\
						0 & 1 & 0 & 0 \\
						0 & 0 & -1 & 0 \\
						0 & 0 & 0 & -1
					\end{array}
					\right) ~~~.
\end{align}

The first and second order chromocharacters, $\varphi^{(1)}_{\rm IJ}$ and $\varphi^{(2)}_{\rm IJKL}$ in  Eqs.~\ref{e:1stChromo} and~\ref{e:2ndChromo} respectively, were defined in Genomics I~\cite{Gates:2009me}, to give a precise mathematical formula that determines whether an arbitrary $d=4$, $N=4$ valise adinkra is equivalent to the cis- or the trans-adinkra
\begin{subequations}\label{eq:TraceConjecture}
\begin{align}\label{e:1stChromo}
   \varphi^{(1)}_{\rm IJ}  \equiv Tr[{\bm {\rm L}}_{\rm I} {\bm {\rm R}}_{\rm J}] = & 4 \, (\, n_c \,+\ n_t \,)~ \delta_{{\rm IJ}} \\
   \label{e:2ndChromo}
 \varphi^{(2)}_{\rm IJKL} \equiv   Tr[{\bm {\rm L}}_{\rm I} {\bm {\rm R}}_{\rm J} {\bm {\rm L}}_{\rm K} {\bm {\rm R}}_{\rm L}] = &\, 4 \, (\, n_c \,+\ n_t \,) \, (\delta_{\rm IJ}\delta_{\rm KL} - \delta_{\rm IK}\delta_{\rm JL} + \delta_{\rm IL}\delta_{\rm JK} ) \cr 
&+  4\, (\, n_c \, -\, n_t \,) \,  \e_{\rm IJKL}~~~.
\end{align}
\end{subequations}
Equations~\ref{eq:TraceConjecture} are also satisfied for larger $N=4$ representations with $d>4$ and we use them to categorize the SUSY isomer numbers for these larger representations. 
Using properties of traces and the orthogonality relationship Eq.~(\ref{e:Rdef}), one can clearly see from Eqs.~(\ref{eq:TraceConjecture}) that $\varphi^{(1)}_{\rm IJ}$ and $\varphi^{(2)}_{\rm IJKL}$ are invariant with respect to adinkra transformations such as Eqs.~(\ref{e:XLY}). Indeed, a straightforward calculation of $\varphi^{(1)}_{\rm IJ}$ and $\varphi^{(2)}_{\rm IJKL}$ for any of the adinkra representations of the chiral, vector, or tensor multiplets produces the results:
\begin{align}
	\varphi^{(1)(\text{CM})}_{\rm IJ} =& \varphi^{(1)(\text{VM})}_{\rm IJ} = \varphi^{(1)(\text{TM})}_{\rm IJ} =  4 \delta_{\rm IJ} \\
	\varphi^{(2)(\text{CM})}_{\rm IJKL} =&\, 4  \, (\delta_{\rm IJ}\delta_{\rm KL} - \delta_{\rm IK}\delta_{\rm JL} + \delta_{\rm IL}\delta_{\rm JK} ) +  4 \,  \e_{\rm IJKL}~~~, \\
	\varphi^{(2)(\text{VM})}_{\rm IJKL} =\varphi^{(2)(\text{TM})}_{\rm IJKL} =&\, 4  \, (\delta_{\rm IJ}\delta_{\rm KL} - \delta_{\rm IK}\delta_{\rm JL} + \delta_{\rm IL}\delta_{\rm JK} ) -  4 \,  \e_{\rm IJKL}~~~.
\end{align}

Comparing with Eqs.~(\ref{eq:TraceConjecture}) clearly the chiral multiplet is in the cis-adinkra representation ($n_c = 1,~n_t=0$) and both the vector and tensor multiplets are in the trans-adinkra representation ($n_c = 0,~n_t=1$). In this paper, we find that it is easiest to find the SUSY isomer numbers $n_c$ and $n_t$ for the larger multiplets of supergravity via a graphical adinkra approach. Even so, we will continue to quote the formulas for the chromocharacters for we later find in Sec.~\ref{s:synth} that the chromocharacters are useful to synthesize our results into a group theoretic framework. 

\subsection{The Real Scalar Superfield}
$~~~~$ The 4D, $\cal N$ = 1 real scalar superfield RSS is a multiplet that is very well known.  It consists of a scalar $K$,
a Majorana fermion $\zeta$, a scalar field $M$, a pseudoscalar field $N$, and axial vector field  $U$, a
Majorana fermion field $\Lambda$, and another scalar field ${\rm d}$.  All together, there are $d=8$ bosonic and $d=8$ fermionic degrees of freedom so this multiplet can be represented by $k = n_c + n_t = 2$ isomer adinkras. This multiplet was the first multiplet we studied whose adinkraic decomposition was non-trivial and required field redefinitions~\cite{Gates:2011aa}. The 4D and 0-brane Lagrangians are equivalent for this multiplet as there is no dynamics:
\begin{equation}
   {\mathcal L}_\text{RSS} = {\mathcal L}_\text{RSS}^{(0)} =-\frac{1}{2} M^2 - \frac{1}{2}N^2 + \frac{1}{2} U_\mu U^\mu - \frac{1}{2} K {\rm d} + i \frac{1}{2}\zeta_a C^{ab} \Lambda_b~~~.
\end{equation}
The 4D transformation laws and their 0-brane reductions are shown in Ref.~\cite{Gates:2011aa}. The valise adinkra for the real scalar superfield is shown in Fig.~\ref{f:RSS} with the nodal field content given by Eq.~\ref{e:RSS} where we separate the field content of the two parts with a horizontal line and number the components top to bottom, one through eight.\footnote{Here we correct for an overall minus sign error in the fermionic nodes of Fig.~4 of Ref.~\cite{Gates:2011aa} and drop an overall factor of $\sqrt{2}$ in Eq.~(\ref{e:RSS}).}  Notice that the node definitions in Eq.~(\ref{e:RSS}) contain linear combinations of fields with different engineering dimensions so that we must use integrals and derivatives in these definitions. This mixing of fields of different engineering dimensions in the adinkra nodes will be necessary for the more complicated multiplets we investigate in this paper. Though the linear combinations of fields in Eq.~(\ref{e:RSS}) are not terribly complicated, we will see that they will become more and more complicated for larger and larger multiplets. 

\begin{figure}[!ht]
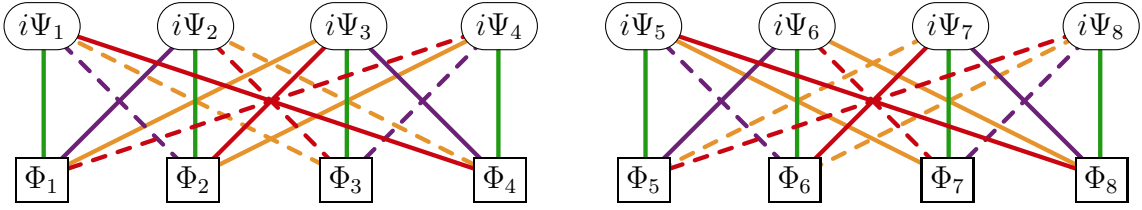
\setlength{\unitlength}{.8 mm}
\begin{center}
   \begin{picture}(200,40)(0,0)
\put(0,0){\includegraphics[width = 90\unitlength]{CisValise}}
\put(2.5,7){\fcolorbox{black}{white}{$\Phi_1$}}
\put(27.6,7){\fcolorbox{black}{white}{$\Phi_2$}}
\put(53,7){\fcolorbox{black}{white}{$\Phi_3$}}
\put(78.6,7){\fcolorbox{black}{white}{$\Phi_4$}}
\put(0.5,30){\begin{tikzpicture}
 \node[rounded rectangle,draw,fill=white!30]{$i\Psi_1$};
 \end{tikzpicture}}
\put(26,30){\begin{tikzpicture}
 \node[rounded rectangle,draw,fill=white!30]{$i\Psi_2$};
 \end{tikzpicture}}
\put(51.5,30){\begin{tikzpicture}
 \node[rounded rectangle,draw,fill=white!30]{$i\Psi_3$};
 \end{tikzpicture}}
\put(76.5,30){\begin{tikzpicture}
 \node[rounded rectangle,draw,fill=white!30]{$i\Psi_4$};
 \end{tikzpicture}}
\quad
\put(100,0){\includegraphics[width = 90\unitlength]{TransValise}}
\put(102.5,7){\fcolorbox{black}{white}{$\Phi_5$}}
\put(127.6,7){\fcolorbox{black}{white}{$\Phi_6$}}
\put(153,7){\fcolorbox{black}{white}{$\Phi_7$}}
\put(178.6,7){\fcolorbox{black}{white}{$\Phi_8$}}
\put(100.5,30){\begin{tikzpicture}
 \node[rounded rectangle,draw,fill=white!30]{$i\Psi_5$};
 \end{tikzpicture}}
\put(126,30){\begin{tikzpicture}
 \node[rounded rectangle,draw,fill=white!30]{$i\Psi_6$};
 \end{tikzpicture}}
\put(151.5,30){\begin{tikzpicture}
 \node[rounded rectangle,draw,fill=white!30]{$i\Psi_7$};
 \end{tikzpicture}}
\put(176.5,30){\begin{tikzpicture}
 \node[rounded rectangle,draw,fill=white!30]{$i\Psi_8$};
 \end{tikzpicture}}
   \end{picture}
   \end{center}
   \vspace{-20 pt}
   \caption{The valise adinkra for the real scalar superfield with the nodal field content given by Eq.~(\ref{e:RSS}). This shows explicitly how the adinkra splits into its cis- (left) and trans- (right) adinkra pieces when compared with Fig.~\ref{f:cvtvintro}.}
\label{f:RSS}
\end{figure} 
\be\label{e:RSS}
   \Phi = 
   \left(   
   \begin{array}{c}
       \int dt ~{\rm d} - \dot{K} \\
       2 M \\
       2 N \\
       2 U_0 \\
       \hline
       -2 U_2 \\
       \dot{K} + \int dt~{\rm d}\\
       2 U_1 \\
       -2 U_3                           
   \end{array}
   \right),~~~
   i\Psi = 
   \left( 
   \begin{array}{c}
        -\dot{\z}_1 -\L_2 \\
        -\dot{\z}_2 + \L_1 \\
        -\dot{\z}_3 + \L_4 \\
        \dot{\z}_4 + \L_3 \\
        \hline
        -\dot{\z}_2 - \L_1 \\
        \dot{\z}_1 - \L_2 \\
        \dot{\z}_4 - \L_3 \\
        -\dot{\z}_3 - \L_4
   \end{array}
   \right)       ~~~.           
\ee

Comparing Fig.~\ref{f:RSS} with Fig.~\ref{f:cvtvintro}, we easily see that the RSS is composed of $n_c=1$ cis-adinkra and $n_t=1$ trans-adinkra. The matrix representation of Fig.~\ref{f:RSS} is block diagonal
\be\label{e:RSSLs}
\begin{array}{llll}
\textcolor{AdinkraGreen}{{\bm {\rm L}}_1 =  {\bm {\rm  I}}_2  \otimes {\bm {\rm I}}_4} & ,& \textcolor{AdinkraViolet}{{\bm {\rm L}}_2 = i {\bm {\rm I}}_2 \otimes {\bm {\rm \beta}}_3} &, \\
\textcolor{AdinkraOrange}{{\bm {\rm L}}_3 = i \left(\begin{array}{cc}
                        1 & 0 \\
                        0 & -1 
                   \end{array}\right) \otimes  {\bm {\rm \beta}}_2}&, & \textcolor{AdinkraRed}{{\bm {\rm L}}_4 = - i {\bm {\rm I}}_2 \otimes {\bm {\rm \beta}}_1} & ,
\end{array}
\ee 
with the $\brR_{\rm I}$ matrices again defined as the transposes of the $\brL_{\rm I}$ matrices, Eq.~\ref{e:Rdef}. Together, the $\brL_{\rm I}$ and $\brR_{\rm I}$ matrices for the RSS satisfy the $\mathcal{GR}(8,4)$ algebra~(\ref{e:GRdN}) with $d=8$. 
The $\rm {D}$-operator acting on the $d=8$-plets defined in Eq.~(\ref{e:RSS}) satisfies the usual relation, Eq.~(\ref{e:LRdef}), to the $\brL_{\rm I}$ and $\brR_{\rm I}$ matrices.
Comparing with Eq.~(\ref{e:ctval}), the block diagonal matrices in Eq.~(\ref{e:RSSLs}) for the RSS are in the cis-adinkra representation in the upper left block and the trans-adinkra representation in the lower right block. A straightforward calculation of the chromocharacters,  Eqs.~(\ref{eq:TraceConjecture}), is indeed consistent with the real scalar superfield's SUSY isomer numbers $n_c = n_t = 1$:
\begin{align}
	\varphi^{(1)}_{\rm IJ} =& 8 ~\delta_{\rm IJ} \\
	\varphi^{(2)}_{\rm IJKL} =&\, 8  \, (\delta_{\rm IJ}\delta_{\rm KL} - \delta_{\rm IK}\delta_{\rm JL} + \delta_{\rm IL}\delta_{\rm JK} ) ~~~.
\end{align}
For the larger representations in this paper, we will always report the block diagonal form of the valise adinkra matrices as we have here.

\subsection{The Complex Linear Superfield}
$~~~~$ The 4D, $\cal N$ = 1 complex linear superfield multiplet (CLS) consists of a scalar $K$, a pseudoscalar $L$, a Majorana fermion $\zeta$, a  Majorana fermion auxiliary field $\rho$, a scalar auxiliary field $M$, a pseudoscalar auxiliary field $N$,  a vector auxiliary field $V$, an axial vector auxiliary field $U$, and a Majorana fermion auxiliary field $\beta$. We see from its 4D Lagrangian
\begin{align}\label{eq:LagrangianCLM}
\mathcal{L}_{\text CLS} =& -\frac{1}{2}\partial_\mu K \partial^\mu K - \frac{1}{2} \partial_\mu L \partial^\mu L - \frac{1}{2} M^2 - \frac{1}{2} N^2 +\frac{1}{4} U_\mu U^\mu  +\frac{1}{4} V_\mu V^\mu \cr
& +\frac{1}{2}i (\gamma^\mu)^{ab} \zeta_a \partial_\mu \zeta_b  + i \rho_a C^{ab}\beta_b 
\end{align}
that the CLS has the same dynamical field content as the off-shell chiral multiplet: two scalars and a Majorana fermion.  The 0-brane reduced Lagrangian for the CLS is 
\begin{align}\label{eq:LagrangianCLM0}
\mathcal{L}^{(0)}_{\text{CLS}} = &  \frac{1}{2}\dot{K}^2 + \frac{1}{2} \dot{L}^2  - \frac{1}{2}M^2 - \frac{1}{2}N^2 + \frac{1}{4} U_\mu U^{\mu} + \frac{1}{4} V_\mu V^\mu \cr
&+ i\frac{1}{2}(\z_1 \dot{\z}_1 + \z_2 \dot{\z}_2 + \z_3 \dot{\z}_3 + \z_4 \dot{\z}_4) + i (\r_2 \b_1 - \r_1 \b_2 + \r_3 \b_4 - \r_4 \b_3)~~.
\end{align}
The transformation laws that are a symmetry of the 4D Lagrangian and their 0-brane reductions are given in Ref.~\cite{Gates:2011aa}. The zero brane reduced transformation laws are encoded in the valise adinkra for the CLS in Fig.~\ref{f:CLS} with the nodes defined in Eq.~(\ref{e:CLS}) separated again by horizontal lines for the different adinkra pieces. The numbering of the components in Eq.~(\ref{e:CLS}) is top to bottom, one through 12.
\begin{figure}[!ht]
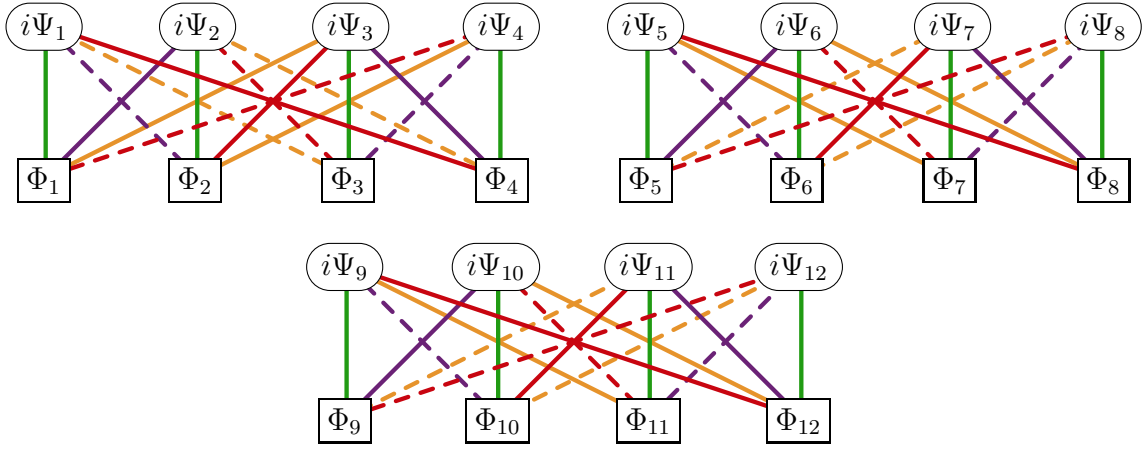
\setlength{\unitlength}{.8 mm}
\begin{center}
   \begin{picture}(200,80)(0,0)
\put(0,40){\includegraphics[width = 90\unitlength]{CisValise}}
\put(2.5,47){\fcolorbox{black}{white}{$\Phi_1$}}
\put(27.6,47){\fcolorbox{black}{white}{$\Phi_2$}}
\put(53,47){\fcolorbox{black}{white}{$\Phi_3$}}
\put(78.6,47){\fcolorbox{black}{white}{$\Phi_4$}}
\put(0.5,70){\begin{tikzpicture}
 \node[rounded rectangle,draw,fill=white!30]{$i\Psi_1$};
 \end{tikzpicture}}
\put(26,70){\begin{tikzpicture}
 \node[rounded rectangle,draw,fill=white!30]{$i\Psi_2$};
 \end{tikzpicture}}
\put(51.5,70){\begin{tikzpicture}
 \node[rounded rectangle,draw,fill=white!30]{$i\Psi_3$};
 \end{tikzpicture}}
\put(76.5,70){\begin{tikzpicture}
 \node[rounded rectangle,draw,fill=white!30]{$i\Psi_4$};
 \end{tikzpicture}}
\quad
\put(100,40){\includegraphics[width = 90\unitlength]{TransValise}}
\put(102.5,47){\fcolorbox{black}{white}{$\Phi_5$}}
\put(127.6,47){\fcolorbox{black}{white}{$\Phi_6$}}
\put(153,47){\fcolorbox{black}{white}{$\Phi_7$}}
\put(178.6,47){\fcolorbox{black}{white}{$\Phi_8$}}
\put(100.5,70){\begin{tikzpicture}
 \node[rounded rectangle,draw,fill=white!30]{$i\Psi_5$};
 \end{tikzpicture}}
\put(126,70){\begin{tikzpicture}
 \node[rounded rectangle,draw,fill=white!30]{$i\Psi_6$};
 \end{tikzpicture}}
\put(151.5,70){\begin{tikzpicture}
 \node[rounded rectangle,draw,fill=white!30]{$i\Psi_7$};
 \end{tikzpicture}}
\put(176.5,70){\begin{tikzpicture}
 \node[rounded rectangle,draw,fill=white!30]{$i\Psi_8$};
 \end{tikzpicture}}
 \put(50,0){\includegraphics[width = 90\unitlength]{TransValise}}
\put(52.5,7){\fcolorbox{black}{white}{$\Phi_9$}}
\put(77,7){\fcolorbox{black}{white}{$\Phi_{10}$}}
\put(102,7){\fcolorbox{black}{white}{$\Phi_{11}$}}
\put(127,7){\fcolorbox{black}{white}{$\Phi_{12}$}}
\put(50.5,30){\begin{tikzpicture}
 \node[rounded rectangle,draw,fill=white!30]{$i\Psi_9$};
 \end{tikzpicture}}
\put(74.7,30){\begin{tikzpicture}
 \node[rounded rectangle,draw,fill=white!30]{$i\Psi_{10}$};
 \end{tikzpicture}}
\put(100,30){\begin{tikzpicture}
 \node[rounded rectangle,draw,fill=white!30]{$i\Psi_{11}$};
 \end{tikzpicture}}
\put(125,30){\begin{tikzpicture}
 \node[rounded rectangle,draw,fill=white!30]{$i\Psi_{12}$};
 \end{tikzpicture}}
   \end{picture}
   \end{center}
   \vspace{-20 pt}
\caption{The valise adinkra for the complex linear superfield with node definitions in Eq.~(\ref{e:CLS}). All bosons have the same engineering dimension and all fermions have the same engineering dimension. The adinkra clearly splits into one cis- (upper left) and two trans- (upper right and bottom) adinkras}
\label{f:CLS}
\end{figure} 

\be\label{e:CLS}
  \Phi =
\left(
\begin{array}{c}
 -M \\
 \dot{K}-V_0 \\
 -\dot{L}-U_0 \\
 N \\
 \hline
 U_2 \\
 V_0-2 \dot{K} \\
 -U_1 \\
 U_3 \\
 \hline
 -V_3 \\
 V_1 \\
 -2 \dot{L}-U_0 \\
 V_2
\end{array}
\right)
,~~~i \Psi = \left(
\begin{array}{c}
 \frac{\dot{\r}_2}{2}-\beta _1 \\
 -\beta _2-\frac{\dot{\r}_1}{2} \\
 -\beta _3-\frac{\dot{\r}_4}{2} \\
 \beta _4-\frac{\dot{\r}_3}{2} \\
 \hline
 \beta _1-\dot{\z}_2+\frac{\dot{\r}_2}{2} \\
 \beta _2+\dot{\z}_1-\frac{\dot{\r}_1}{2} \\
 \beta _3+\dot{\z}_4-\frac{\dot{\r}_4}{2} \\
 \beta _4-\dot{\z}_3+\frac{\dot{\r}_3}{2} \\
 \hline
 \beta _1+\dot{\z}_2+\frac{\dot{\r}_2}{2} \\
 -\beta _2+\dot{\z}_1+\frac{\dot{\r}_1}{2} \\
 -\beta _3+\dot{\z}_4+\frac{\dot{\r}_4}{2} \\
 \beta _4+\dot{\z}_3+\frac{\dot{\r}_3}{2}
\end{array}
\right)
\ee

 In the valise adinkra in Fig.~\ref{f:CLS}, we have put the third piece below the first two even though all bosons have the same dimension and all the fermions have the same dimension. From here on, we will generally split up these larger valise adinkras into more than one row with the understanding that all fermions have the same dimension and all bosons have the same dimension. Comparing Fig.~\ref{f:CLS} with Fig.~\ref{f:cvtvintro}, we clearly see that the valise adinkra for the CLS splits into $n_c=1$ cis-adinkra and $n_t=2$  trans-adinkras. 
 
 The splitting is easily seen in the block diagonal matrix representation of Eq.~(\ref{e:CLSLs}) 
\be\label{e:CLSLs}
\begin{array}{llll}
\textcolor{AdinkraGreen}{{\bm {\rm L}}_1 =  {\bm {\rm  I}}_3  \otimes {\bm {\rm I}}_4} & ,& \textcolor{AdinkraViolet}{{\bm {\rm L}}_2 = i {\bm {\rm I}}_3 \otimes {\bm {\rm \beta}}_3} &, \\
\textcolor{AdinkraOrange}{{\bm {\rm L}}_3 = i \left(\begin{array}{ccc}
                        1 & 0 & 0\\
                        0 & -1 & 0 \\
                        0 & 0 & -1
                   \end{array}\right) \otimes  {\bm {\rm \beta}}_2}&, & \textcolor{AdinkraRed}{{\bm {\rm L}}_4 = - i {\bm {\rm I}}_3 \otimes {\bm {\rm \beta}}_1} & ,
\end{array}
\ee 
and again we define the $\brR_{\rm I}$ matrices as the transposes of the $\brL_{\rm I}$ matrices, Eq.~\ref{e:Rdef}, that together satisfy the $\mathcal{GR}(12,4)$ algebra, Eq.~(\ref{e:GRdN}) with $d=12$. The $\rm{D}$-operator is related to the node definitions~(\ref{e:CLS}) and the $\brL_{\rm I}$  and $\brR_{\rm I}$ matrices for the CLS in the usual way, Eq.~(\ref{e:LRdef}), with $d=12$. The block diagonal form of the matrices in Eq.~(\ref{e:CLSLs}) once again helps to easily calculate the chromocharacters, Eqs.~(\ref{eq:TraceConjecture}), for the complex linear superfield 
\begin{align}
	\varphi^{(1)}_{\rm IJ} =& 12 \delta_{\rm IJ} \\
	\varphi^{(2)}_{\rm IJKL} =&\, 12  \, (\delta_{\rm IJ}\delta_{\rm KL} -  \delta_{\rm IK}\delta_{\rm JL} + \delta_{\rm IL}\delta_{\rm JK} ) - 4 ~\epsilon_{\rm IJKL} ~~~.
\end{align}
A Comparison with the general formula in Eqs.~(\ref{eq:TraceConjecture}) shows once again that the complex linear superfield has SUSY isomer numbers $n_c=1,~n_t = 2$:

We see in Eq.~(\ref{e:CLS}) that the node field definitions are more complicated than that of the RSS, Eq.~(\ref{e:RSS}). Later, in our supergravity analysis, we will see these node definitions become even more complicated for \nmSG. In fact, we will see that the CLS adinkra is embedded within the full \nmSG ~adinkra which is consistent with the fact that the CLS is the compensator part of \nmSG ~as alluded to in Sec.~\ref{s:intro}.

\section{\texorpdfstring{4D}{4D}, \texorpdfstring{$\mathcal{N} = 1$}{N=} Minimal SG (\texorpdfstring{$n= -1/3$}{n=-1/3})}\label{s:MSG}
$~~~~$ Here we begin the presentation of the main new results of the paper: the decomposition of various supergravity representations in terms of adinkras using a new adinkraic procedure. We start with the linearized theory of 4D, $\mathcal{N} =1 $ Poincar\'e old-minimal supergravity (mSG), which contains the real component fields of a scalar auxiliary field $S$, pseudoscalar auxiliary field $P$, axial vector auxiliary field $A_\mu$, Majorana gravitino $\psi_{\mu a}$, and graviton $h_{\mu\nu}$~\cite{Siegel:1978mj,VanNieuwenhuizen:1981ae}. In this section, we will unveil our new adinkranization procedure that will always find the valise adinkra.

\subsection{Transformation Laws}

$~~~~$ We use the transformation laws
\begin{subequations}\label{eq:Dfinalnumeric}
\begin{align}
   {\rm D}_a S =&  -\frac{1}{2} ([\g^\m , \g^\n ])_{a}^{~b}\partial_\mu \psi_{\nu b} \\
   {\rm D}_a P =& \frac{1}{2}(\gamma^5[ \g^\m , \g^\n])_{a}^{~b}\partial_\mu \psi_{\nu b} \\
   {\rm D}_a A_{\mu} =& i (\gamma^5 \gamma^\nu)_a^{~b} \partial_{[\nu} \psi_{\mu] b} -\frac{1}{2} \epsilon_{\mu}^{~\nu\alpha\beta}(\gamma_\nu)_{a}^{~b} \partial_\alpha \psi_{\beta b}  \\
   {\rm D}_a h_{\mu\nu} = & \frac{1}{2} (\gamma_{(\mu})_{a}^{~b}\psi_{\nu) b} \\
   {\rm D}_a \psi_{\mu b} =&  - \frac{i}{3} (\gamma_\mu)_{ab} S -\frac{1}{3} (\gamma^5\gamma_\mu)_{ab} P + \frac{2}{3} (\gamma^5)_{ab} A_\mu +\frac{1}{6} (\gamma^5[\g_{\mu}, \g^{\nu}])_{ab}A_\nu  + \cr
   & - \frac{i}{2}([\g^{\alpha} , \g^{\beta}])_{ab}\partial_\alpha h_{\beta\mu} 
\end{align}
\end{subequations}
which are a symmetry of the Lagrangian
\begin{align}\label{eq:LParChoice}
     \mathcal{L}_{mSG} = & -\frac{1}{2}\partial_\alpha h_{\mu\nu} \partial^\alpha h^{\mu\nu} + \frac{1}{2}\partial^\alpha h \partial_\alpha h - \partial^\alpha h \partial^\beta h_{\alpha\beta} + \partial^\mu h_{\mu\nu} \partial_\alpha h^{\alpha\nu}  \cr
     &- \frac{1}{3}S^2 -\frac{1}{3}P^2 + \frac{1}{3} A_\mu A^\mu -\frac{1}{2} \psi_{\mu a} \epsilon^{\mu\nu\alpha\beta}(\gamma^5\gamma_\nu)^{ab} \partial_\alpha \psi_{\beta b}
\end{align}
up to total derivatives
\begin{align}
  {\rm D}_a \mathcal{L}_{mSG} = 0 + \mbox{total derivatives.}
\end{align}
In the above and hereafter, $h$ is the trace of the graviton
\begin{align}
  h \equiv \eta^{\mu\nu}h_{\mu\nu}~~~.
\end{align}
\noindent A direct calculation reveals the following algebra
\begin{subequations}\label{e:Algebra}
\begin{align}
   \{ {\rm D}_a, {\rm D}_b \} S = 2 i (\gamma^\mu)_{ab} \partial_\mu S~~~&,~~~  \{ {\rm D}_a, {\rm D}_b \} P = 2 i (\gamma^\mu)_{ab} \partial_\mu P~~~,\\
   \{ {\rm D}_a, {\rm D}_b \} A_\nu = & 2 i (\gamma^\mu)_{ab} \partial_\mu A_\nu~~~, \\
 \{ {\rm D}_a , {\rm D}_b \} h_{\mu\nu} = & 2 i (\gamma^\alpha)_{ab} \partial_\alpha h_{\mu\nu} - i (\gamma^\alpha)_{ab}\partial_{(\mu} h_{\nu)\alpha}~~~, \\
 \label{e:mSGDpsi}
  \{  {\rm D}_a, {\rm D}_b \} \psi_{\mu c} =& 2 i (\gamma^\alpha)_{ab} \partial_\alpha \psi_{\mu c}- i \partial_\mu \varphi_{abc}~~~.
\end{align}
\end{subequations}
The extra term on the right hand side of Eq.~(\ref{e:mSGDpsi}) is given by
\begin{align}
  \varphi_{abc} = &\frac{5}{4}(\gamma^\alpha)_{ab}\partial_\mu \psi_{\alpha c}+ \frac{1}{8} \left( (\gamma^5[\gamma^\beta, \gamma^\alpha])_{ab} (\gamma^5\gamma_\beta)_{c}^{~d} +  ([\gamma^\beta, \gamma^\alpha])_{ab} (\gamma_\beta)_{c}^{~d}  \right) \partial_\mu \psi_{\alpha d} \cr
     & + \frac{1}{8}(\gamma_\beta)_{ab}[\gamma^\beta ,\gamma^\alpha]_{c}^{~d}\partial_\mu \psi_{\alpha d} ~~~.
     \label{e:varphi}
\end{align}
Comparing the RHS's of Eqs.~(\ref{e:Algebra}) for the graviton ($ h_{\mu\nu}$) and gravitino ($\psi_{\mu a}$) to those for
the other fields, there are extra terms that
are consequences of the well known gauge symmetries of the Lagrangian
\begin{align}\label{e:hsymmetry}
   h_{\mu\nu} \to & h_{\mu\nu} + \partial_\mu \Lambda_\nu + \partial_\nu \Lambda_\mu \\
   \label{e:psisymmetry}
   \psi_{\mu a} \to & \psi_{\mu a} + \partial_\mu \epsilon_a
\end{align}
for arbitrary infinitesimal vectors $\Lambda_\mu$ and spinors $\epsilon_a$.

\subsection{One Dimensional Reduction}\label{s:0brane}
$~~~~$ In the temporal gauge
\begin{align}\label{e:TG}
   h_{0\mu} =& \, \, \psi_{0a} = 0
\end{align}
the mSG Lagrangian reduced to the 0-brane becomes
\begin{align}\label{e:L0}
  \mathcal{L}^{(0)}_{mSG} =& ~ \dot{h}_{12}^2+\dot{h}_{13}^2+\dot{h}_{23}^2- \dot{h}_{11}\dot{h}_{22} - \dot{h}_{11}\dot{h}_{33} - \dot{h}_{22}\dot{h}_{33} {~~~~}
  {~~~~} {~~~~} {~~~~} {~~~~} {~~~~} {~~~~} \cr
&-\frac{1}{3}S^2-\frac{1}{3}P^2 - \frac{1}{3} A_0^2 + \frac{1}{3} A_1^2 + \frac{1}{3}
A_2^2 + \frac{1}{3} A_3^2  \cr
&+i\left( -\psi_{31}\dot{\psi}_{12} + \psi_{32}\dot{\psi}_{11} - \psi_{33}\dot{\psi}_{14} +\psi_{34}\dot{\psi}_{13}-\psi_{11}\dot{\psi}_{23}+\psi_{12}\dot{\psi}_{24}   \right. \cr
 &\left. ~~~~~~+ \psi_{13}\dot{\psi}_{21} - \psi_{14}\dot{\psi}_{22}-\psi_{21}\dot{\psi}_{34} - \psi_{22}\dot{\psi}_{33} + \psi_{23}\dot{\psi}_{32}+\psi_{24}\dot{\psi}_{31}      \right) 
 ~~~.
\end{align}
The 0-brane reduced transformation laws are displayed in Tabs.~\ref{t:0braneb1} and~\ref{t:0branef1}. We notice linear combinations of up to three fields in these transformations. It will be necessary to define adinkra nodes with linear combinations as was done for the RSS and CLS in Sec.~\ref{s:arev}. In building the adinkra for mSG, we now start our new adinkranization procedure that will give evidence for the selection rule described in Sec.~\ref{s:intro}. We start with the cis- and trans-adinkra shown here again in Fig.~\ref{f:cvtv}. The corresponding transformation laws for both the cis- and the trans-adinkra are simultaneously displayed in Tab.~\ref{t:cistrans} using the shorthand parameter $\chi_0$ which was defined in Eq.~(\ref{e:chi0}). Our goal is to find the linear combinations of fields that collapse the 0-brane reduced transformation laws in Tabs.~\ref{t:0braneb1} and~\ref{t:0branef1} into either the cis- or the trans-adinkra, the transformations in the smaller Tab.~\ref{t:cistrans}, three different ways to compose the full $k=3$ valise adinkra for mSG. In terms of matrices, this will mean the transformation laws between these linear combinations will be describable in terms of block diagonal adinkra matrices similar to Eqs.~(\ref{e:RSSLs}) and~(\ref{e:CLSLs}).

\begin{table}[!ht]
\centering
  \caption{Transformation laws for mSG bosons in temporal gauge,~Eq.(\ref{e:TG}), and reduced to the 0-brane. }
  \renewcommand{\arraystretch}{1.3}
    {\tiny \begin{tabular}{c|cccc|}
    & $\gD$ & $\vD$ & $\oD$ & $\rD$ \\
    \hline   
   $S$ & $-\dot{\psi }_{11}-\dot{\psi }_{23}+\dot{\psi }_{32}$&$\dot{\psi
   }_{12}-\dot{\psi }_{24}+\dot{\psi }_{31}$&$\dot{\psi
   }_{13}-\dot{\psi }_{21}-\dot{\psi }_{34}$&$-\dot{\psi
   }_{14}-\dot{\psi }_{22}-\dot{\psi }_{33} $
   \\
   $A_2$ & $\frac{1}{2}\dot{\psi }_{11}-\dot{\psi }_{23}-\frac{1}{2}\dot{\psi}_{32}$&$-\frac{1}{2}\dot{\psi }_{12}-\dot{\psi}_{24}-\frac{1}{2}\dot{\psi }_{31}$&$\frac{1}{2}\dot{\psi}_{13}+\dot{\psi }_{21}-\frac{1}{2}\dot{\psi}_{34}$&$-\frac{1}{2}\dot{\psi }_{14}+\dot{\psi}_{22}-\frac{1}{2}\dot{\psi }_{33} $
   \\
   $h_{13}$ & $ \frac{1}{2} \psi _{11}+\frac{1}{2}\psi _{32}$&$-\frac{1}{2}\psi _{12}+\frac{1}{2}
   \psi _{31}$&$\frac{1}{2} \psi_{13}+\frac{1}{2}\psi _{34}$&$-\frac{1}{2}\psi_{14}+\frac{1}{2} \psi _{33}$ 
   \\
   \hline
    $h_{11}$ & $\psi _{12}$&$\psi _{11}$&$\psi _{14}$&$\psi _{13}$ 
    \\
   $h_{22}$ & $-\psi _{24}$&$\psi _{23}$&$\psi _{22}$&$-\psi _{21}$ 
   \\
   $h_{33}$ & $\psi _{31}$&$-\psi _{32}$&$\psi _{33}$&$-\psi _{34}$
   \\
   \hline  
   $A_0$ & $\dot{\psi }_{13}-\dot{\psi }_{21}-\dot{\psi }_{34}$&$-\dot{\psi
   }_{14}-\dot{\psi }_{22}-\dot{\psi }_{33}$&$\dot{\psi
   }_{11}+\dot{\psi }_{23}-\dot{\psi }_{32}$&$-\dot{\psi
   }_{12}+\dot{\psi }_{24}-\dot{\psi }_{31} $
   \\
   $A_1$ & $-\dot{\psi }_{13}-\frac{1}{2}\dot{\psi }_{21}-\frac{1}{2}\dot{\psi}_{34}$&$-\dot{\psi }_{14}+\frac{1}{2}\dot{\psi}_{22}+\frac{1}{2}\dot{\psi }_{33}$&$\dot{\psi}_{11}-\frac{1}{2}\dot{\psi }_{23}+\frac{1}{2}\dot{\psi}_{32}$&$\dot{\psi }_{12}+\frac{1}{2}\dot{\psi}_{24}-\frac{1}{2}\dot{\psi }_{31} $ 
   \\
   $h_{23}$ & $\frac{1}{2}\psi _{21}-\frac{1}{2}\psi _{34}$&$-\frac{1}{2}\psi _{22}+\frac{1}{2}\psi_{33}$&$\frac{1}{2}\psi _{23}+\frac{1}{2}\psi_{32}$&$-\frac{1}{2}\psi _{24}-\frac{1}{2}\psi _{31}
   $
   \\
  \hline 
   $P$ & $-\dot{\psi }_{14}-\dot{\psi }_{22}-\dot{\psi
   }_{33}$&$-\dot{\psi }_{13}+\dot{\psi }_{21}+\dot{\psi
   }_{34}$&$\dot{\psi }_{12}-\dot{\psi }_{24}+\dot{\psi
   }_{31}$&$\dot{\psi }_{11}+\dot{\psi }_{23}-\dot{\psi }_{32} $
   \\
  $A_3$ & $\frac{1}{2}\dot{\psi }_{14}+\frac{1}{2}\dot{\psi}_{22}-\dot{\psi }_{33}$&$-\frac{1}{2}\dot{\psi}_{13}+\frac{1}{2}\dot{\psi }_{21}-\dot{\psi}_{34}$&$-\frac{1}{2}\dot{\psi }_{12}+\frac{1}{2}\dot{\psi}_{24}+\dot{\psi }_{31}$&$\frac{1}{2}\dot{\psi}_{11}+\frac{1}{2}\dot{\psi }_{23}+\dot{\psi }_{32} $
   \\
    $h_{12}$ & $-\frac{1}{2}\psi _{14}+\frac{1}{2}\psi _{22}$&$\frac{1}{2}\psi
   _{13}+\frac{1}{2}\psi _{21}$&$\frac{1}{2}\psi _{12}+\frac{1}{2}\psi
   _{24}$&$-\frac{1}{2}\psi _{11}+\frac{1}{2}\psi _{23}$
   \\
   \hline
    \end{tabular}  }
    \label{t:0braneb1}
\end{table} 
\vskip 12 pt
\begin{table}[!ht]
\centering
\caption{Transformation laws for mSG fermions in temporal gauge,~Eq.(\ref{e:TG}), and reduced to the 0-brane. }
  \renewcommand{\arraystretch}{1.3}
    \label{t:0branef1}
    {\tiny \begin{tabular}{c|cccc|}
    & $\gD$ & $\vD$ & $\oD$ & $\rD$ \\
   \hline
   $\psi_{13}$ & $i\frac{1}{3}A_0-\frac{2}{3} i A_1$&$-\frac{1}{3}i A_3-\frac{1}{3}i P+i
   \dot{h}_{12}$&$\frac{1}{3}i A_2+\frac{1}{3}iS+i \dot{h}_{13}$&$i \dot{h}_{11}$ 
   \\
   $\psi_{21}$ & $-\frac{1}{3}i A_0-\frac{1}{3}i A_1+i \dot{h}_{23}$&$\frac{1}{3}i
   A_3+\frac{1}{3}i P+i \dot{h}_{12}$&$\frac{2 }{3}i A_2-\frac{1}{3}iS$&$-i \dot{h}_{22}$
   \\
   $\psi_{34}$ & $ -\frac{1}{3}i A_0-\frac{1}{3}i A_1-i \dot{h}_{23}$&$-\frac{2 }{3}i A_3+\frac{1}{3}i
   P$&$-\frac{1}{3}i A_2-\frac{1}{3}i S+i \dot{h}_{13}$&$-i \dot{h}_{33}$
   \\
   \hline
   $\psi_{14}$ & $\frac{1}{3}i A_3-\frac{1}{3}i P-i \dot{h}_{12}$&$-\frac{1}{3}i
   A_0-\frac{2 }{3}i A_1$&$i \dot{h}_{11}$&$-\frac{1}{3}i A_2-\frac{1}{3}i S-i
   \dot{h}_{13}$ 
   \\
   $\psi_{22}$ & $ \frac{1}{3}i A_3-\frac{1}{3}i P+i \dot{h}_{12}$&$-\frac{1}{3}i A_0+\frac{1}{3}i A_1-i \dot{h}_{23}$&$i \dot{h}_{22}$&$\frac{2}{3} i A_2-\frac{1}{3}i S$ 
   \\
   $\psi_{33}$ & $-\frac{2 }{3}i A_3-\frac{1}{3}i P$&$-\frac{1}{3}i A_0+\frac{1}{3}iA_1+i \dot{h}_{23}$&$i \dot{h}_{33}$&$-\frac{1}{3}i A_2-\frac{1}{3}i S+i
   \dot{h}_{13} $
   \\
    \hline
   $\psi_{11}$ & $i\frac{1}{3}A_2-i\frac{1}{3}S+i \dot{h}_{13}$&$i
   \dot{h}_{11}$&$i\frac{1}{3} A_0+i\frac{2}{3}A_1$&$i\frac{1 }{3}A_3+i\frac{1}{3}P-i
   \dot{h}_{12}$ 
   \\
   $\psi_{23}$ & $ -\frac{2}{3}i A_2-\frac{1}{3}i S$&$i \dot{h}_{22}$&$\frac{1}{3}i A_0-\frac{1}{3}i A_1+i \dot{h}_{23}$&$\frac{1}{3}i A_3+\frac{1}{3}i P+i
   \dot{h}_{12}$
   \\
   $\psi_{32}$ & $-\frac{1}{3}i A_2+\frac{1}{3}i S+i \dot{h}_{13}$&$-i
   \dot{h}_{33}$&$-\frac{1}{3}i A_0+\frac{1}{3}i A_1+i
   \dot{h}_{23}$&$\frac{2 }{3}i A_3-\frac{1}{3}i P $
   \\
   \hline
   $\psi_{12}$ & $i \dot{h}_{11}$&$-i\frac{1}{3}A_2+i\frac{1}{3}S-i \dot{h}_{13}$&$-i\frac{1}{3}A_3+i\frac{1}{3}P+i \dot{h}_{12}$&$-i\frac{1}{3}A_0+i\frac{2}{3}A_1$ 
   \\
   $\psi_{24}$ & $-i \dot{h}_{22}$&$-\frac{2}{3}i A_2-\frac{1}{3}i S$&$\frac{1}{3}i A_3-\frac{1}{3}i P+i \dot{h}_{12}$&$\frac{1}{3}i A_0+\frac{1}{3}i
   A_1-i \dot{h}_{23} $
   \\
   $\psi_{31}$ & $i \dot{h}_{33}$&$-\frac{1}{3}i A_2+\frac{1}{3}i S+i \dot{h}_{13}$&$\frac{2}{3}i A_3+\frac{1}{3}i P$&$-\frac{1}{3}i A_0-\frac{1}{3}i A_1-i \dot{h}_{23} $
\\
   \hline
   \end{tabular}  }
\end{table}

\begin{table}[!ht]\centering
  \caption{Transformation laws for an arbitrary cis- ($\chi_0=1$) or trans- ($\chi_0=-1$) adinkra system. The $\Phi_i$ are bosons and the $\Psi_{\hat{i}}$ are fermions. These laws are encoded by the cis- and trans-valise adinkras in Fig.~\ref{f:cvtv}.}
  \renewcommand{\arraystretch}{1.4}
  {\tiny
  \begin{tabular}{r|rrrr|rr|rrrr|}
       & $\gD$ & $\vD$ & $\oD$~~~& $\rD$ &\hspace{20 pt} && $\gD$ & $\vD$ & $\oD$~~~& $\rD$ \\ 
  \cline{1-5}\cline{7-11}
   $\Phi_1 $&$ i \Psi _1 $&$ i \Psi _2$ &$ i \chi_0 \Psi _3 $&$ -i \Psi_4 $&
  \hspace{20 pt} &$\Psi_1 $&$ \dot{\Phi}_1 $&$ -\dot{\Phi}_2 $& $-\chi_0 \dot{\Phi}_3 $&$ \dot{\Phi}_4 $\\$\Phi _2 $&$ i \Psi_2 $&$ -i \Psi_1 $&$ i\chi_0\Psi_4 $&$ i \Psi_3 $&
  \hspace{20 pt} &$\Psi_2 $&$ \dot{\Phi}_2 $&$ \dot{\Phi}_1 $&$-\chi_0 \dot{\Phi}_4 $&$ -\dot{\Phi}_3 $
  \\
 $\Phi _3 $&$ i \Psi_3 $&$ -i \Psi_4 $&$ -i\chi_0 \Psi_1 $&$ -i \Psi_2 $&
 \hspace{20 pt} &$\Psi_3 $&$ \dot{\Phi}_3 $&$ \dot{\Phi}_4 $&$ \chi_0\dot{\Phi}_1 $&$ \dot{\Phi}_2 $\\
 $\Phi_4 $&$ i \Psi_4 $&$ i \Psi_3 $&$ -i\chi_0 \Psi_2 $&$ i \Psi_1 $&
 \hspace{20 pt} &$\Psi_4 $&$ \dot{\Phi}_4 $&$ -\dot{\Phi}_3 $&$ \chi_0\dot{\Phi}_2 $&$ -\dot{\Phi}_1 $\\
 \cline{1-5}\cline{7-11}
  \end{tabular}}
  \label{t:cistrans}
\end{table}

\begin{figure}[!ht]
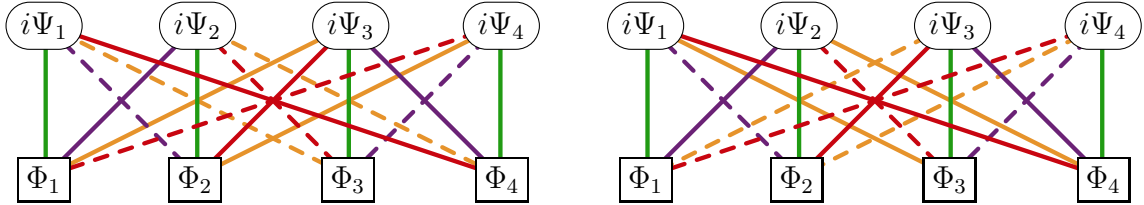
\setlength{\unitlength}{.8 mm}
\begin{center}
   \begin{picture}(200,40)(0,0)
\put(0,0){\includegraphics[width = 90\unitlength]{CisValise}}
\put(2.5,7){\fcolorbox{black}{white}{$\Phi_1$}}
\put(27.6,7){\fcolorbox{black}{white}{$\Phi_2$}}
\put(53,7){\fcolorbox{black}{white}{$\Phi_3$}}
\put(78.6,7){\fcolorbox{black}{white}{$\Phi_4$}}
\put(0.5,30){\begin{tikzpicture}
 \node[rounded rectangle,draw,fill=white!30]{$i\Psi_1$};
 \end{tikzpicture}}
\put(26,30){\begin{tikzpicture}
 \node[rounded rectangle,draw,fill=white!30]{$i\Psi_2$};
 \end{tikzpicture}}
\put(51.5,30){\begin{tikzpicture}
 \node[rounded rectangle,draw,fill=white!30]{$i\Psi_3$};
 \end{tikzpicture}}
\put(76.5,30){\begin{tikzpicture}
 \node[rounded rectangle,draw,fill=white!30]{$i\Psi_4$};
 \end{tikzpicture}}
\quad
\put(100,0){\includegraphics[width = 90\unitlength]{TransValise}}
\put(102.5,7){\fcolorbox{black}{white}{$\Phi_1$}}
\put(127.6,7){\fcolorbox{black}{white}{$\Phi_2$}}
\put(153,7){\fcolorbox{black}{white}{$\Phi_3$}}
\put(178.6,7){\fcolorbox{black}{white}{$\Phi_4$}}
\put(100.5,30){\begin{tikzpicture}
 \node[rounded rectangle,draw,fill=white!30]{$i\Psi_1$};
 \end{tikzpicture}}
\put(126,30){\begin{tikzpicture}
 \node[rounded rectangle,draw,fill=white!30]{$i\Psi_2$};
 \end{tikzpicture}}
\put(151.5,30){\begin{tikzpicture}
 \node[rounded rectangle,draw,fill=white!30]{$i\Psi_3$};
 \end{tikzpicture}}
\put(176.5,30){\begin{tikzpicture}
 \node[rounded rectangle,draw,fill=white!30]{$i\Psi_4$};
 \end{tikzpicture}}
   \end{picture}
   \end{center}
   \vspace{-20 pt}
\caption{The cis- (left) and trans- (right) valise adinkras encoding the transformation laws in Tab.~\ref{t:cistrans}. }
\label{f:cvtv}
\end{figure}

From Fig.~\ref{f:cvtv}, we can see that each D-transformation for an adinkraic representation is a bijective map between fermions and bosons. The transformations in Tab.~\ref{t:0braneb1} map twelve bosons into twelve linear combinations of fermions that are \emph{different} for each D-transformation. Similarly, Tab.~\ref{t:0branef1} maps twelve fermions into twelve linear combinations of bosons that are \emph{different}  for each D-transformation. For the  $(12|12)$ mSG system to be represented in terms of the irreducible $(4|4)$ cis- and trans-valise adinkras, there must be three distinct sets of linear combinations that define the nodes in the cis- or trans-valise adinkra and collapse Tabs.~\ref{t:0braneb1} and~\ref{t:0branef1} into the smaller Tab.~\ref{t:cistrans}.

Tables~\ref{t:0braneb1} and~\ref{t:0branef1} are organized horizontally into groups of three fields that can combine in linear combinations to form the nodes of an adinkra.  For instance, we see that various combinations of $\dot{\psi}_{13}$, $\dot{\psi}_{21}$, and $\dot{\psi}_{34}$ transform into various combinations of only $A_0$, $A_1$, and $\dot{h}_{23}$ under the \gD ~transformation. Finding the linear combinations that collapse Tabs.~\ref{t:0braneb1} and~\ref{t:0branef1} to either the $(4|4)$ cis-adinkra or  the $(4|4)$ trans-adinkra as depicted in the smaller Tab.~\ref{t:cistrans} and Fig.~\ref{f:cvtv} will automatically give us the SUSY isomer numbers $n_c$ and $n_t$ for mSG. Our strategy is to find the number $n_c$ of ways Tabs.~\ref{t:0braneb1} and~\ref{t:0branef1} collapse to Tab.~\ref{t:cistrans} under the cis-adinkra choice ($\chi_0 = 1$) and the number $n_t$ of ways Tabs.~\ref{t:0braneb1} and~\ref{t:0branef1} collapse to Tab.~\ref{t:cistrans} under the trans-adinkra choice ($\chi_0 = -1$). 

 Starting with a seed linear combination:
\begin{align}
   i \Psi_1 = & - u_1 \dot{\psi}_{13} + u_2 \dot{\psi}_{21} + u_3 \dot{\psi}_{34}
   \label{e:Psi1}
\end{align}
where $u_1$, $u_2$, and $u_3$ are arbitrary constants, we can calculate the D-operator on the $\Psi_1$ node through both Tabs.~\ref{t:0branef1} and~\ref{t:cistrans} to define another node. We continue the process until we have defined all the nodes. We will see that after four iterations, each with a different color, choosing the cis- or trans-adinkra representation, $\chi_0= 1$ or $\chi_0 = -1$ respectively, will lead to constraints on $u_1$, $u_2$, and  $u_3$. A simple analysis of these constraints will tell us:
\begin{enumerate}
  \item The nodal field content of the cis- and trans-valise adinkras in the representation.
  \item The number of cis- ($n_c$) and trans-valise ($n_c$) adinkras that compose the representation.
\end{enumerate} 
We proceed by calculating $\gD\Psi_1$ from Tab.~\ref{t:cistrans} and equating this to $\gD\Psi_1$ calculated from Eq.~(\ref{e:Psi1}) and Tab.~\ref{t:0branef1}. This defines $\Phi_1$:
\begin{align}
    \gD \Psi_1 = & \dot{\Phi}_1  \nonumber\\*
   =\gD \Psi_1 = & \frac{1}{3}(-u_1 - u_2 - u_3)\dot{A}_0 + \frac{1}{3}(2 u_1 - u_2 - u_3) \dot{A}_1 + (u_2 - u_3) \ddot{h}_{23}~~~ \cr
   \Rightarrow \Phi_1 = & \frac{1}{3}(-u_1 - u_2 - u_3)A_0 + \frac{1}{3}(2 u_1 - u_2 - u_3) A_1 + (u_2 - u_3) \dot{h}_{23}~~~.
               \label{e:it1}
\end{align}  
Next, we calculate $\vD \Phi_1$ through both Tabs.~\ref{t:0braneb1} and~\ref{t:cistrans}, defining $\Psi_2$:
\begin{align}
     \vD \Phi_1 = & i \Psi_2 \nonumber\\*
   = \vD \Phi_1 = &  \frac{1}{3}(- u_1 + 2 u_2 + 2 u_3) \dot{\psi}_{14} + \frac{1}{3}(2 u_1 - u_2 + 2 u_3) \dot{\psi}_{22} \nonumber\\*
                &+ \frac{1}{3}(2 u_1 + 2 u_2 - u_3 ) \dot{\psi}_{33}~~~ \cr
\Rightarrow i \Psi_2 =  & \frac{1}{3} (- u_1 + 2 u_2 +  2 u_3) \dot{\psi}_{14} + \frac{1}{3}(2 u_1 - u_2 + 2 u_3) \dot{\psi}_{22}  \cr
				&+ \frac{1}{3}(2 u_1 + 2 u_2 - u_3 ) \dot{\psi}_{33} ~~~.
                \label{e:it2}
\end{align}
Following this with a calculation of $\oD \Psi_2$ gives us $\Phi_4$:
\begin{align}
    \oD \Psi_2  = & -\chi_0 \dot{\Phi}_4 \nonumber\\*
    = \oD \Psi_2  = & \frac{1}{3}(- u_1 + 2 u_2 + 2 u_3) \ddot{h}_{11} + \frac{1}{3}(2 u_1 - u_2 + 2 u_3) \ddot{h}_{22} \nonumber\\*
                &+ \frac{1}{3}(2 u_1 + 2 u_2 - u_3 ) \ddot{h}_{33}~~~ \cr
                \Rightarrow  \Phi_4 = &-\frac{\chi_0}{3}(- u_1 + 2 u_2 + 2 u_3) \dot{h}_{11} - \frac{\chi_0}{3}(2 u_1 - u_2 + 2 u_3) \dot{h}_{22} \nonumber\\*
                &- \frac{\chi_0}{3}(2 u_1 + 2 u_2 - u_3 ) \dot{h}_{33}~~~
                \label{e:it3}
\end{align}
where we have used $\chi_0=\pm 1$ to move $\chi_0$ to the RHS of the last line. 
Our fourth iteration with a different color, $\rD \Phi_4$, gives us back $\Psi_1$:
\begin{align}
    \rD \Phi_4 = & i \Psi_1   \cr
    = \rD \Phi_4 = & -\frac{\chi_0}{3}(- u_1 + 2 u_2 + 2 u_3) \dot{\psi}_{13} + \frac{\chi_0}{3}(2 u_1 - u_2 + 2 u_3) \dot{\psi}_{21}  \cr
    &+ \frac{\chi_0}{3}(2 u_1 + 2 u_2 - u_3 ) \dot{\psi}_{34} \cr
    \Rightarrow i \Psi_1 = &-\frac{\chi_0}{3}(- u_1 + 2 u_2 + 2 u_3) \dot{\psi}_{13} + \frac{\chi_0}{3}(2 u_1 - u_2 + 2 u_3) \dot{\psi}_{21}  \cr
    &+ \frac{\chi_0}{3}(2 u_1 + 2 u_2 - u_3 ) \dot{\psi}_{34}
\end{align}
and when compared with the seed relation, Eq.~(\ref{e:Psi1}), forces a constraint as promised:
\begin{align}
      u_1 \dot{\psi}_{13} - u_2 \dot{\psi}_{21} - u_3 \dot{\psi}_{34} 
              = &  \frac{\chi_0}{3}(- u_1 + 2 u_2 +  2 u_3) \dot{\psi}_{13} - \frac{\chi_0}{3}(2 u_1 - u_2 + 2 u_3) \dot{\psi}_{21} \cr 
     & - \frac{\chi_0}{3}(2 u_1 + 2 u_2 - u_3 ) \dot{\psi}_{34}~~~.\label{e:Psi1Constraint}
\end{align}
This constraint simplifies to the following under the choice of cis or trans:
\begin{align}\label{e:cischoice}
   \text{cis:}& ~~~\chi_0 = 1 & \hspace{-60 pt}\Rightarrow & ~~~u_1 = u_2 = u_3~~~,\\*
    \label{e:transchoice}
   \text{trans:}&~~~\chi_0 = -1  &  \hspace{-60 pt} \Rightarrow & ~~~u_1 + u_2 + u_3 = 0~~~.
\end{align}
The four iterations, Eqs.~(\ref{e:Psi1}) through~(\ref{e:Psi1Constraint}), that lead us to the constraint Eqs.~(\ref{e:cischoice}) and~(\ref{e:transchoice}) can be succinctly depicted with the adinkras in Fig.~\ref{f:cistransit}. 
\begin{figure}[!ht]\setlength{\unitlength}{.8 mm}
\begin{center}
   \begin{picture}(200,50)(0,0)
\put(0,0){\includegraphics[width = 92\unitlength]{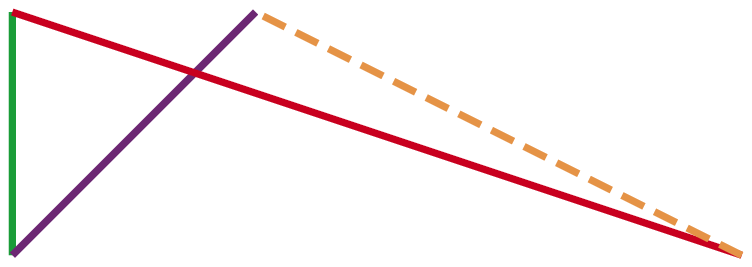}}
\put(-5,43){(\ref{e:Psi1}),(\ref{e:Psi1Constraint})}
\put(5,31){\textcolor{AdinkraGreen}{\vector(0,-1){17}}}
\put(0,-1){(\ref{e:it1})}
\put(13,12){\textcolor{AdinkraViolet}{\vector(1,1){19}}}
\put(27.5,43){(\ref{e:it2})}
\put(41,33){\textcolor{AdinkraOrange}{\vector(2,-1){38}}}
\put(77.5,-1){(\ref{e:it3})}
\put(79,9){\textcolor{AdinkraRed}{\vector(-3,1){67}}}
\put(2.5,7){\fcolorbox{black}{white}{$\Phi_1$}}
\put(79.6,7){\fcolorbox{black}{white}{$\Phi_4$}}
\put(1,32){\begin{tikzpicture}
 \node[rounded rectangle,draw,fill=white!30]{$i\Psi_1$};
 \end{tikzpicture}}
\put(27,32){\begin{tikzpicture}
 \node[rounded rectangle,draw,fill=white!30]{$i\Psi_2$};
 \end{tikzpicture}}
\put(100,0){\includegraphics[width = 92\unitlength]{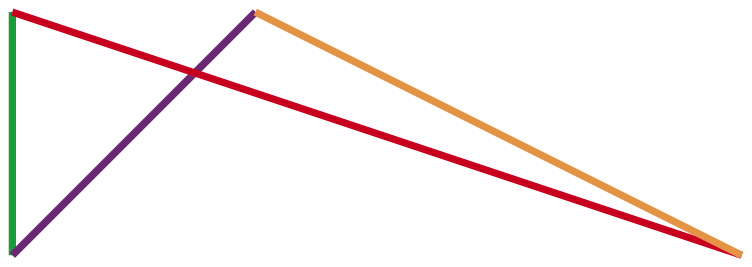}}
\put(95,43){(\ref{e:Psi1}),(\ref{e:Psi1Constraint})}
\put(105,31){\textcolor{AdinkraGreen}{\vector(0,-1){17}}}
\put(100,-1){(\ref{e:it1})}
\put(113,12){\textcolor{AdinkraViolet}{\vector(1,1){19}}}
\put(127.5,43){(\ref{e:it2})}
\put(141,33){\textcolor{AdinkraOrange}{\vector(2,-1){38}}}
\put(177.5,-1){(\ref{e:it3})}
\put(179.5,9){\textcolor{AdinkraRed}{\vector(-3,1){67}}}
\put(102.5,7){\fcolorbox{black}{white}{$\Phi_1$}}
\put(179.6,7){\fcolorbox{black}{white}{$\Phi_4$}}
\put(101,32){\begin{tikzpicture}
 \node[rounded rectangle,draw,fill=white!30]{$i\Psi_1$};
 \end{tikzpicture}}
\put(127,32){\begin{tikzpicture}
 \node[rounded rectangle,draw,fill=white!30]{$i\Psi_2$};
 \end{tikzpicture}}
   \end{picture}
   \end{center}
\caption{On the left (right), the iterative procedure in Eqs.~(\ref{e:Psi1}) through~(\ref{e:Psi1Constraint}) that leads to the cis (trans) constraint in Eq.~(\ref{e:cischoice}). This is the cis-valise (trans-valise) adinkra in Fig.~\ref{f:cvtv} with the nodes and links that do not enter the iteration removed. }
\label{f:cistransit}
\end{figure}

Further iterations define the other four nodes of each adinkraic representation but lead to no further constraints on $u_1$, $u_2$, or $u_3$. These resulting nodal definitions for the cis and trans choices are
\begin{align}
      \Phi =   \left(
\begin{array}{c}
 -u_1 A_0  \\
 -u_1 P \\
 u_1 S \\
 -u_1 \dot{h}
\end{array}
\right)~~~,&~~~i \Psi = \left(
\begin{array}{c}
 -u_1(\dot{\psi} _{13}-\dot{\psi} _{21}-\dot{\psi} _{34})  \\
 -u_1( -\dot{\psi} _{14}-\dot{\psi} _{22}-\dot{\psi} _{33}) \\
 -u_1 (\dot{\psi} _{11}+\dot{\psi} _{23}-\dot{\psi} _{32}) \\
 -u_1(\dot{\psi} _{12}-\dot{\psi} _{24}+\dot{\psi} _{31})
\end{array}
\right) ~~~,
\nonumber\\*
\nonumber\\*
\mbox{cis:}~\chi_0 = 1~~~,&~~~u_1~\mbox{unconstrained,}
\label{e:mSGcis}
\end{align} 
and 
\begin{align}
      \Phi =    \left(
\begin{array}{c}
u_1 A_1  +(u_2-u_3) \dot{h}_{23}  \\
u_3 A_3 + (u_1-u_2)\dot{h}_{12}  \\
 u_2 A_2  + (u_3 - u_1) \dot{h}_{31}  \\
 -u_1\dot{h}_{11} -u_2\dot{h}_{22} -u_3\dot{h}_{33}  
\end{array}
\right)~~~,&~~~i \Psi =  \left(
\begin{array}{c}
 -u_1 \dot{\psi} _{13}+u_2 \dot{\psi} _{21}+u_3 \dot{\psi} _{34} \\
 -u_1 \dot{\psi} _{14}-u_2 \dot{\psi} _{22}-u_3\dot{\psi} _{33} \\
 -u_1 \dot{\psi} _{11}-u_2 \dot{\psi} _{23}+u_3 \dot{\psi} _{32} \\
 -u_1 \dot{\psi} _{12}+u_2 \dot{\psi} _{24}-u_3 \dot{\psi} _{31} \\
\end{array}
\right) ~~~,
\nonumber\\*
\nonumber\\*
\mbox{trans:}~\chi_0 = -1~~~,&~~~u_1 + u_2 + u_3 = 0~~~.
\label{e:mSGtrans}
\end{align}
\noindent The cis choice, Eq.~\ref{e:mSGcis}, has only one free parameter, $u_1$, which encodes an overall scale freedom in the definition of the nodes.   We can, with no loss of generality, set this parameters to $u_1=-1$. The cis-adinkra submultiplet is therefore unique and so we have $n_c = 1$ for mSG. We see in Eq.~(\ref{e:mSGcis}) that the cis definition of $\Phi$ contains the trace of the graviton. This is related to the rotational symmetry fixed by the solution, Eq.~(\ref{e:cischoice}). Notice that the two solutions~(\ref{e:cischoice}) and~(\ref{e:transchoice}) form a line and a plane through the origin in $\mathbb{R}^3$, respectively, that are perpendicular to each other.  In contrast to there being one unique cis choice, the trans choice Eq.~(\ref{e:mSGtrans}) leads to two independent parameters that describe two linearly independent nodal field definitions.  We see then that mSG has the SUSY isomer numbers
\begin{align}
   n_c = 1~~~,~~~n_t = 2~~~,
\end{align} 
which are the same as those for the complex linear supermultiplet~\cite{Gates:2011aa}. 

The equations
 \begin{equation}
\Phi  =  \left(
\text{{\tiny $
\begin{array}{c}
A_0 \\
P \\
-S \\
\dot{h}\\
\hline
\vspace{-7 pt} \\
 u_1 A_1
   +(u_2-u_3 )\dot{h}_{23} \\
 u_3 A_3 +
   (u_1-u_2)\dot{h}_{12} \\
 u_2 A_2 +(u_3-u_1 )\dot{h}_{31}
   \\
-u_1\dot{h}_{11} -u_2\dot{h}_{22} -u_3\dot{h}_{33}
   \\
   \hline
\vspace{-7 pt} \\
v_1 A_1
   +(v_2-v_3 )\dot{h}_{23}  \\
 v_3 A_3 +
   (v_1-v_2)\dot{h}_{12} \\
 v_2 A_2 +(v_3-v_1 )\dot{h}_{31}
   \\
-v_1\dot{h}_{11} -v_2\dot{h}_{22} -v_3\dot{h}_{33}
\end{array}
$}}
\right)~~~,
~~~i \Psi = \left(
\text{{\tiny $
\begin{array}{c}
\dot{\psi} _{13}-\dot{\psi} _{21}-\dot{\psi} _{34} \\
-\dot{\psi} _{14}-\dot{\psi} _{22}-\dot{\psi} _{33} \\
 \dot{\psi} _{11}+\dot{\psi} _{23}-\dot{\psi} _{32} \\
 \dot{\psi} _{12}-\dot{\psi} _{24}+\dot{\psi} _{31} \\
\hline
\vspace{-7 pt} \\
 -u_1 \dot{\psi }_{13}+u_2
   \dot{\psi }_{21}+u_3 \dot{\psi }_{34} \\
-u_1 \dot{\psi }_{14}-u_2 \dot{\psi }_{22}-u_3 \dot{\psi
   }_{33}\\
-u_1 \dot{\psi }_{11}-u_2 \dot{\psi }_{23}+u_3 \dot{\psi
   }_{32}\\
 -u_1 \dot{\psi }_{12}+u_2
   \dot{\psi }_{24}-u_3 \dot{\psi }_{31}\\
 \hline
\vspace{-7 pt} \\
 -v_1 \dot{\psi }_{13}+v_2
   \dot{\psi }_{21}+v_3 \dot{\psi }_{34} \\
-v_1 \dot{\psi }_{14}-v_2 \dot{\psi }_{22}-v_3 \dot{\psi
   }_{33} \\
-v_1 \dot{\psi }_{11}-v_2 \dot{\psi }_{23}+v_3 \dot{\psi
   }_{32} \\
 -v_1 \dot{\psi }_{12}+v_2
   \dot{\psi }_{24}-v_3 \dot{\psi }_{31}
\end{array}
$}}
\right)
\label{e:mSGPhiPsi}
\end{equation}
with components numbered top to bottom, one to 12 define the nodal field content for the complete adinkraic decomposition for mSG where $v_1$, $v_2$, and $v_3$ parameterize the second trans-adinkra submultiplet and satisfy the same constraint as $u_1$,  $u_2$, and $u_3$:
\begin{align}\label{e:uconstraint}
   u_1 + u_2 + u_3 = v_1 + v_2 + v_3 = 0
\end{align} 
Furthermore, considering our free parameters to be $\vec{u} = (u_1,u_2)$ and $\vec{v} = (v_1,v_2)$, we must also enforce linear independence: $\vec{v} \not\propto \vec{u}$, otherwise they would be describing the exact same field content up to an overall rescaling.

\begin{table}[!htbp]
\centering
  \caption{Zero-brane reduced mSG transformation rules in the adinkraic representation in Eqs.~(\ref{e:mSGPhiPsi}).}
  \renewcommand{\arraystretch}{1.3}
  {\tiny
 \begin{tabular}{r|rrrr|rr|rrrr|}
   & $\gD$ & $\vD$ & $\oD$ & $\rD$ &\hspace{50 pt}& & $\gD$ & $\vD$ & $\oD$ & $\rD$ \\
   \cline{1-5}\cline{7-11}
 $ \Phi _1 $&$ i \Psi _1 $&$ i \Psi _2 $&$ i \Psi _3 $&$ -i \Psi _4 $&\hspace{50 pt}& $\Psi _1 $&$ \dot{\Phi }_1 $&$ -\dot{\Phi }_2 $&$ -\dot{\Phi }_3 $&$ \dot{\Phi }_4 $\\$
 \Phi _2 $&$ i \Psi _2 $&$ -i \Psi _1 $&$ i \Psi _4 $&$ i \Psi _3 $&\hspace{50 pt}& $ \Psi _2 $&$ \dot{\Phi }_2 $&$ \dot{\Phi }_1 $&$ -\dot{\Phi }_4 $&$ -\dot{\Phi }_3 $\\$
 \Phi _3 $&$ i \Psi _3 $&$ -i \Psi _4 $&$ -i \Psi _1 $&$ -i \Psi _2 $&\hspace{50 pt}& $ \Psi _3 $&$ \dot{\Phi }_3 $&$ \dot{\Phi }_4 $&$ \dot{\Phi }_1 $&$ \dot{\Phi }_2 $\\$
 \Phi _4 $&$ i \Psi _4 $&$ i \Psi _3 $&$ -i \Psi _2 $&$ i \Psi _1 $&\hspace{50 pt}& $ \Psi _4 $&$ \dot{\Phi }_4 $&$ -\dot{\Phi }_3 $&$ \dot{\Phi }_2 $&$ -\dot{\Phi }_1 $\\
  \cline{1-5}\cline{7-11}$
 \Phi _5 $&$ i \Psi _5 $&$ i \Psi _6 $&$ -i \Psi _7 $&$ -i \Psi _8 $&\hspace{50 pt}& $ \Psi _5 $&$ \dot{\Phi }_5 $&$ -\dot{\Phi }_6 $&$ \dot{\Phi }_7 $&$ \dot{\Phi }_8 $\\$
 \Phi _6 $&$ i \Psi _6 $&$ -i \Psi _5 $&$ -i \Psi _8 $&$ i \Psi _7 $&\hspace{50 pt}& $ \Psi _6 $&$ \dot{\Phi }_6 $&$ \dot{\Phi }_5 $&$ \dot{\Phi }_8 $&$ -\dot{\Phi }_7 $\\$
 \Phi _7 $&$ i \Psi _7 $&$ -i \Psi _8 $&$ i \Psi _5 $&$ -i \Psi _6 $&\hspace{50 pt}& $ \Psi _7 $&$ \dot{\Phi }_7 $&$ \dot{\Phi }_8 $&$ -\dot{\Phi }_5 $&$ \dot{\Phi }_6 $\\$
 \Phi _8 $&$ i \Psi _8 $&$ i \Psi _7 $&$ i \Psi _6 $&$ i \Psi _5 $&\hspace{50 pt}& $ \Psi _8 $&$ \dot{\Phi }_8 $&$ -\dot{\Phi }_7 $&$ -\dot{\Phi }_6 $&$ -\dot{\Phi }_5 $\\
  \cline{1-5}\cline{7-11}$
 \Phi _9 $&$ i \Psi _9 $&$ i \Psi _{10} $&$ -i \Psi _{11} $&$ -i \Psi _{12} $&\hspace{50 pt}& $ \Psi _9 $&$ \dot{\Phi }_9 $&$ -\dot{\Phi }_{10} $&$ \dot{\Phi }_{11} $&$ \dot{\Phi }_{12} $\\$
 \Phi _{10} $&$ i \Psi _{10} $&$ -i \Psi _9 $&$ -i \Psi _{12} $&$ i \Psi _{11} $&\hspace{50 pt}& $ \Psi _{10} $&$ \dot{\Phi }_{10} $&$ \dot{\Phi }_9 $&$ \dot{\Phi }_{12} $&$ -\dot{\Phi }_{11} $\\$
 \Phi _{11} $&$ i \Psi _{11} $&$ -i \Psi _{12} $&$ i \Psi _9 $&$ -i \Psi _{10} $&\hspace{50 pt}& $ \Psi _{11} $&$ \dot{\Phi }_{11} $&$ \dot{\Phi }_{12} $&$ -\dot{\Phi }_9 $&$ \dot{\Phi }_{10} $\\$
 \Phi _{12} $&$ i \Psi _{12} $&$ i \Psi _{11} $&$ i \Psi _{10} $&$ i \Psi _9 $&\hspace{50 pt}& $ \Psi _{12} $&$ \dot{\Phi }_{12} $&$ -\dot{\Phi }_{11} $&$ -\dot{\Phi }_{10} $&$ -\dot{\Phi }_9$\\
   \cline{1-5}\cline{7-11}
  \end{tabular}}
  \label{t:0brane2}
\end{table}

The node definitions~(\ref{e:mSGPhiPsi}) collapse Tabs.~\ref{t:0braneb1} and~\ref{t:0branef1} into the smaller Tab.~\ref{t:0brane2} three different ways, which is encoded in the valise-adinkra for the full mSG multiplet in Fig.~\ref{f:SGAdinkra}. Comparing Fig.~\ref{f:SGAdinkra} with Fig.~\ref{f:cvtvintro}, we clearly see that the mSG valise adinkra is composed of $n_c=1$ cis-adinkra and $n_t=2$ trans-adinkras.
\begin{figure}[!ht]
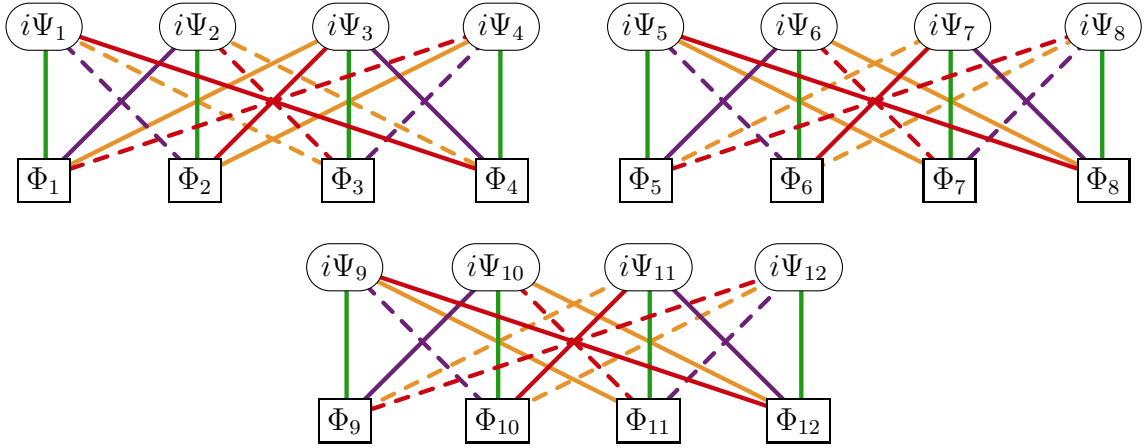
\setlength{\unitlength}{.8 mm}
\begin{center}
   \begin{picture}(200,80)(0,0)
\put(0,40){\includegraphics[width = 90\unitlength]{CisValise}}
\put(2.5,47){\fcolorbox{black}{white}{$\Phi_1$}}
\put(27.6,47){\fcolorbox{black}{white}{$\Phi_2$}}
\put(53,47){\fcolorbox{black}{white}{$\Phi_3$}}
\put(78.6,47){\fcolorbox{black}{white}{$\Phi_4$}}
\put(0.5,70){\begin{tikzpicture}
 \node[rounded rectangle,draw,fill=white!30]{$i\Psi_1$};
 \end{tikzpicture}}
\put(26,70){\begin{tikzpicture}
 \node[rounded rectangle,draw,fill=white!30]{$i\Psi_2$};
 \end{tikzpicture}}
\put(51.5,70){\begin{tikzpicture}
 \node[rounded rectangle,draw,fill=white!30]{$i\Psi_3$};
 \end{tikzpicture}}
\put(76.5,70){\begin{tikzpicture}
 \node[rounded rectangle,draw,fill=white!30]{$i\Psi_4$};
 \end{tikzpicture}}
\quad
\put(100,40){\includegraphics[width = 90\unitlength]{TransValise}}
\put(102.5,47){\fcolorbox{black}{white}{$\Phi_5$}}
\put(127.6,47){\fcolorbox{black}{white}{$\Phi_6$}}
\put(153,47){\fcolorbox{black}{white}{$\Phi_7$}}
\put(178.6,47){\fcolorbox{black}{white}{$\Phi_8$}}
\put(100.5,70){\begin{tikzpicture}
 \node[rounded rectangle,draw,fill=white!30]{$i\Psi_5$};
 \end{tikzpicture}}
\put(126,70){\begin{tikzpicture}
 \node[rounded rectangle,draw,fill=white!30]{$i\Psi_6$};
 \end{tikzpicture}}
\put(151.5,70){\begin{tikzpicture}
 \node[rounded rectangle,draw,fill=white!30]{$i\Psi_7$};
 \end{tikzpicture}}
\put(176.5,70){\begin{tikzpicture}
 \node[rounded rectangle,draw,fill=white!30]{$i\Psi_8$};
 \end{tikzpicture}}
 \put(50,0){\includegraphics[width = 90\unitlength]{TransValise}}
\put(52.5,7){\fcolorbox{black}{white}{$\Phi_9$}}
\put(77,7){\fcolorbox{black}{white}{$\Phi_{10}$}}
\put(102,7){\fcolorbox{black}{white}{$\Phi_{11}$}}
\put(127,7){\fcolorbox{black}{white}{$\Phi_{12}$}}
\put(50.5,30){\begin{tikzpicture}
 \node[rounded rectangle,draw,fill=white!30]{$i\Psi_9$};
 \end{tikzpicture}}
\put(74.7,30){\begin{tikzpicture}
 \node[rounded rectangle,draw,fill=white!30]{$i\Psi_{10}$};
 \end{tikzpicture}}
\put(100,30){\begin{tikzpicture}
 \node[rounded rectangle,draw,fill=white!30]{$i\Psi_{11}$};
 \end{tikzpicture}}
\put(125,30){\begin{tikzpicture}
 \node[rounded rectangle,draw,fill=white!30]{$i\Psi_{12}$};
 \end{tikzpicture}}
   \end{picture}
   \end{center}
   \vspace{-20 pt}
\caption{The mSG valise adinkra is composed of one cis-adinkra (upper left) and two trans-adinkras (upper right and bottom).  The mSG multiplet therefore has SUSY isomer numbers $n_c = 1$, $n_t = 2$. This is the exact same valise adinkra as that of the complex linear supermultiplet shown in Fig.~\ref{f:CLS}. The engineering dimensions of all bosons are the same and the engineering dimensions of all fermions are the same.}
\label{f:SGAdinkra}
\end{figure} 
\noindent In Sec.~\ref{s:CSG}, we will see that the cSG adinkra is composed precisely of the two trans-adinkra submultiplets of mSG. We identify the cis-adinkra submultiplet of mSG as the chiral compensator, $\sigma$, as explained in Sec.~\ref{s:intro}.  This is consistent with prior results~\cite{Gates:2009me, Gates:2011aa} that found the chiral multiplet to be the cis-adinkra in valise form, with SUSY isomer numbers $n_c = 1$, $n_t = 0$. 

We can lower all fermionic nodes and the node that contains the trace of the graviton $h = h_{11} + h_{22} + h_{33}$, and swap a sign on $S$ and $i\Psi_2 = -\dot{\psi} _{14}-\dot{\psi} _{22}-\dot{\psi} _{33}$ to arrive at the extended adinkra in Fig.~\ref{f:LAMin}.
\begin{figure}[!htb]
\centering
   \begin{picture}(300,130)(0,0)
    \put(0,0){\includegraphics[width = 300\unitlength]{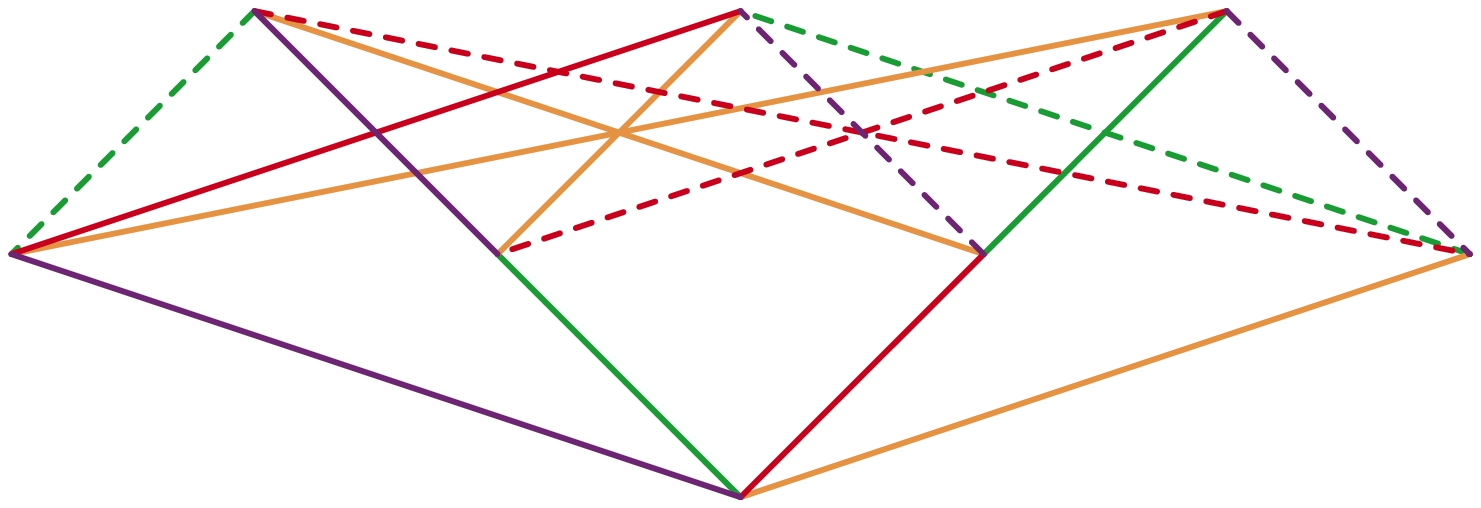}}
    \put(144,14){\fcolorbox{black}{white}{$h$}}
    \put(52,104){\fcolorbox{black}{white}{$S$}}
    \put(143,104){\fcolorbox{black}{white}{$P$}}
    \put(234,104){\fcolorbox{black}{white}{$A_0$}}
\put(-25,51){\begin{tikzpicture}
 \node[rounded rectangle,draw,fill=white!30]{ $\psi_{11} + \psi_{23} - \psi_{32}$};
 \end{tikzpicture}}
 \put(65,51){\begin{tikzpicture}
 \node[rounded rectangle,draw,fill=white!30]{ $\psi_{12} - \psi_{24} + \psi_{31}$};
 \end{tikzpicture}}
 \put(155,51){\begin{tikzpicture}
 \node[rounded rectangle,draw,fill=white!30]{ $\psi_{13} - \psi_{21} - \psi_{34}$};
 \end{tikzpicture}}
 \put(245,51){\begin{tikzpicture}
 \node[rounded rectangle,draw,fill=white!30]{ $\psi_{14} + \psi_{22} + \psi_{33}$};
 \end{tikzpicture}}
   \end{picture}
   \vspace{-10 pt}
   \caption{The chiral compensator submultiplet of mSG. It is the cis-valise adinkra in Fig.~\ref{f:SGAdinkra} with all fermion nodes lowered and the graviton trace node lowered. Also, the $S$ node and $\psi_{14} + \psi_{22} + \psi_{33}$ nodes have the opposite sign as in the cis-adinkra.}
   \label{f:LAMin}
\end{figure}
\noindent  Notice the nodal field content is now without derivatives. The nodal field definitions for the trans-valise adinkras on the the other hand  \emph{can not} be chosen to be linear combinations of fields of the same engineering dimensions.

Table~\ref{t:0brane2} and the adinkra in Fig.~\ref{f:SGAdinkra} can be succinctly written as Eq.~(\ref{e:LRdef}) with adinkra matrices
\be\label{eq:SUGRAValise}
\begin{array}{llll}
\textcolor{AdinkraGreen}{{\bm {\rm L}}_1 =  {\bm {\rm  I}}_3  \otimes {\bm {\rm I}}_4} & ,& \textcolor{AdinkraViolet}{{\bm {\rm L}}_2 = i {\bm {\rm I}}_3 \otimes {\bm {\rm \beta}}_3} &, \\
\textcolor{AdinkraOrange}{{\bm {\rm L}}_3 = i \left(\begin{array}{ccc}
                        1 & 0 & 0 \\
                        0 & -1 & 0 \\
                        0 & 0 & -1
                   \end{array}\right) \otimes  {\bm {\rm \beta}}_2}&, & \textcolor{AdinkraRed}{{\bm {\rm L}}_4 = - i {\bm {\rm I}}_3 \otimes {\bm {\rm \beta}}_1} & .
\end{array}
\ee
We have written the adinkra matrices~(\ref{eq:SUGRAValise}) in their block diagonal form in terms of the $SO(4)$ generators, Eq.~(\ref{eq:SO4Generators}.
The adinkra matrices~(\ref{eq:SUGRAValise}) satisfy the orthogonal relationship, Eq.~(\ref{e:Rdef}), and the $\mathcal{GR}(12,4)$ algebra, Eq.~(\ref{e:GRdN}) with $d=12$. We note that the adinkra matrices for mSG, Eq.~(\ref{eq:SUGRAValise}), are identical to those for the complex linear supermultiplet, Eq.~(\ref{e:CLSLs}).

The chromocharacters, Eqs.~(\ref{eq:TraceConjecture}), for mSG are once again easily calculated via their block diagonal form as
\begin{align}\label{eq:TraceMSG}
    \varphi^{(1)}_{\rm IJ} = & 12~ \delta_{{\rm IJ}} \\
   \varphi^{(2)}_{\rm IJKL} = &\, 12 (\delta_{\rm IJ}\delta_{\rm KL} - \delta_{\rm IK}\delta_{\rm JL} + \delta_{\rm IL}\delta_{\rm JK} ) -  4\,  \,  \e_{\rm IJKL}~~~,
\end{align} the same as for the complex linear supermultiplet, with SUSY isomer numbers $n_c = 1, n_t = 2$. The top left adinkra in the full valise of Fig.~\ref{f:SGAdinkra} is the cis-adinkra, and the top right and bottom adinkras are the two trans-adinkras. As explained in Sec.~\ref{s:intro}, at the level of representations, mSG is a chiral compensator superfield added to cSG. Since we have already identified the cis-adinkra in Fig.~\ref{f:LAMin} as the $(4|4)$ chiral compensator, we would be led to believe that the two trans-adinkras compose the $(8|8)$ cSG. We will indeed find this to be true in Sec.~\ref{s:CSG}.

\section{\texorpdfstring{4D}{4D}, \texorpdfstring{${\mathcal N}=1$}{N=1} Non-Minimal SG (\texorpdfstring{$n\ne -1/3,0$}{n=-1/3})}\label{s:NMSG}
$~~~~$ The linearized theory of 4D, $\mathcal{N} =1 $ Poincar\'e non-minimal supergravity (\nmSG) contains the real component fields of a scalar auxiliary field $\tilde{S}$, pseudoscalar auxiliary field $\tilde{P}$, axial vector auxiliary field $\tilde{A}_\mu$, Majorana gravitino $\psi_{\mu a}$, graviton $h_{\mu\nu}$, vector auxiliary field~$\tilde{V}_\mu$, axial vector auxiliary field~$\tilde{W}_\mu$, and Majorana auxiliary fermions $\tilde{\lambda}_a$ and $\tilde{\beta}_a$~\cite{Siegel:1978mj}. 
\subsection{Transformation Laws}
$~~~~$ To proceed with adinkra analysis, we must write the transformation laws for \nmSG~\cite{Siegel:1978mj} in Majorana notation.  These transformation laws depend on the parameter $n \ne -1/3, 0$ and are given by
\begin{subequations}
\begin{eqnarray}
D_a \tilde{S} &=& \frac{1}{4 N}( [ \gamma^\mu , \gamma^\nu ] )_a{}^d \partial_{\mu} \psi_{\nu d}-\frac{nN}{4}\tilde{\beta}_a + (\gamma^\nu)_a{}^d \partial_\nu \tilde{\lambda}_d\\*
D_a \tilde{P} &=& i \frac{1}{4 N} (\gamma^5 [ \gamma^\mu , \gamma^\nu ] )_a{}^d \partial_{\mu} \psi_{\nu d} -i\frac{nN}{4} (\gamma^5)_a{}^d \tilde{\beta}_d + i (\gamma^5 \gamma^\nu)_a{}^d \partial_\nu \tilde{\lambda}_d\\
D_a \tilde{A}_\mu &=& -i\frac{1}{2} (\gamma^5 \gamma^\nu)_a{}^d \partial_{[\nu} \psi_{\mu]d} + \frac{1}{4} \epsilon_\mu{}^{\nu\alpha\beta} (\gamma_\nu)_a{}^d \partial_{\alpha} \psi_{\beta d} + i 2 N (\gamma^5)_a{}^d \partial_\mu \tilde{\lambda}_d \\
D_a h_{\mu\nu} &=& \frac 12 \,  (\gamma_{(\mu})_{a}{}^d \psi_{\nu)d}   \\
D_a \tilde{W}_\mu &=& i \frac{nN}{4} (\gamma^5 \gamma_\mu)_a{}^d \tilde{\beta}_d -i\frac{2 N}{3 } (\gamma^5)_a{}^d \partial_\mu \tilde{\lambda}_d-i\frac{1}{6} (\gamma^5 \gamma^\nu)_a{}^d \partial_{[\mu} \psi_{\nu] d} +\cr
&& + i  (\gamma^5 \gamma^\nu \gamma_\mu)_a{}^d \partial_\nu \tilde{\lambda}_d + \frac{1}{6}  \epsilon_{\mu}^{~\nu\alpha\beta} (\gamma_\beta )_a{}^d  \partial_{\nu} \psi_{\alpha d}   \\
D_a \tilde{V}_\mu &=& -\frac{nN}{4} (\gamma_\mu)_a{}^d \tilde{\beta}_d  + (\gamma^\nu \gamma_\mu)_a{}^d \partial_\nu \tilde{\lambda}_d 
  \\
D_a \psi_{\mu c} &=& -i 2 nN (\gamma_\mu)_{ac} \tilde{S} + 2 nN (\gamma^5 \gamma_\mu)_{ac} \tilde{P} -2 (\gamma^5)_{ac} \tilde{A}_\mu + \frac{2}{3} (\gamma^5 \gamma^\nu \gamma_\mu )_{ac} \tilde{A}_\nu \nonumber\\*
 &&- i \frac{1}{2} ( [ \gamma^\alpha , \gamma^\beta] )_{ac} \partial_\alpha h_{\beta\mu}   + 2 nN (\gamma^5 \gamma^\nu \gamma_\mu)_{ac} \tilde{W}_\nu  \\
D_a \tilde{\lambda}_c &=& -i \frac{nN}{2} C_{ac} \tilde{S} + \frac{nN}{2}  (\gamma^5)_{ac} \tilde{P} + \frac{3 n}{2}(\gamma^5 \gamma^\mu)_{ac} \tilde{W}_\mu + i \frac{1}{2} (\gamma^\mu)_{ac} \tilde{V}_\mu  \\
D_a \tilde{\beta}_c &=& -i 2 (\gamma^\mu)_{ac} \partial_\mu \tilde{S} + 2 (\gamma^5 \gamma^\mu)_{ac} \partial_\mu \tilde{P} -\frac{3 }{N} (\gamma^5  [\gamma^{\mu} , \gamma^{\nu} ])_{ac} \partial_{\mu} \tilde{W}_{\nu}  + i \frac{2}{nN} C_{ac} \partial^\mu \tilde{V}_\mu  \cr
&&   + \frac{4}{3 nN} (\gamma^5)_{ac} \partial_\mu \tilde{A}^\mu + 2\frac{nN +1}{nN} (\gamma^5)_{ac} \partial^\mu \tilde{W}_\mu  + i \frac{1}{nN}([\gamma^{\mu} , \gamma^{\nu}])_{ac} \partial_{\mu} \tilde{V}_{\nu}  
\end{eqnarray}
\end{subequations}
where
\begin{align}
  N = \frac{3n+1}{n}~~~.
\end{align}

These are an invariance of the same Lagrangian as was presented in Ref.~\cite{Siegel:1978mj} in the linearized limit
\begin{align}
\mathcal{L}_{\text{\nmSG}} =& -\, \frac 12 \,  \partial_\alpha h_{\mu\nu} \partial^\alpha h^{\mu\nu} + \frac 12 \partial_\alpha h \partial^\alpha h -  \partial^\alpha h \partial^\beta h_{\alpha\beta} +  \partial^\mu h_{\mu\nu} \partial_\alpha h^{\alpha\nu}+ \frac{4}{3} \tilde{A}_\mu \tilde{A}^\mu  \cr
& + \frac{4 (3 n+1)^2}{n}( \tilde{S}^2 +  \tilde{P}^2) - \frac{4 (3 n+1)}{n}  \tilde{V}_\mu \tilde{V}^\mu - 12 (3 n+1)  \tilde{W}_\mu \tilde{W}^\mu  \cr
&- \frac 12 \,  \psi_{\mu a} \epsilon^{\mu\nu\alpha\beta} (\gamma^5 \gamma_\nu)^{ab} \partial_\alpha \psi_{\beta b}-  i \frac{4 (3 n+1)^2}{n}  C^{a \, b} \tilde{\lambda}_a \tilde{\beta}_b ~~~.
\end{align}
The algebra closes on the auxiliary fields $\tilde{X} = (\tilde{S},\tilde{P},\tilde{A}_\mu, \tilde{V}_\mu, \tilde{W}_\mu, \tilde{\lambda}_a, \tilde{\beta}_a)$ as
\begin{align}
    \{ {\rm D}_a , {\rm D}_b \} \tilde{X} = 2 i (\gamma^\mu)_{ab} \partial_\mu \tilde{X}
\end{align}
and on the physical fields $\psi_{\mu a}$ and $h_{\mu\nu}$ as
\begin{align}
    \{ {\rm D}_a , {\rm D}_b \} h_{\mu\nu} = & 2 i (\gamma^\alpha)_{ab} \partial_\alpha h_{\mu\nu} - i (\gamma^\alpha)_{ab}\partial_{(\mu} h_{\nu)\alpha} \cr
     \{ {\rm D}_a , {\rm D}_b \} \psi_{\mu c} = & 2 i (\gamma^\alpha)_{ab} \partial_\alpha  \psi_{\mu c} - i\partial_\mu \varphi_{abc} - i\partial_\mu \chi_{abc} ~~
\end{align}
where $\varphi_{abc}$ is the same as for the minimal case, Eq.~(\ref{e:varphi}). The new term on the right hand side of the gravitino algebra
\begin{align}
   \chi_{abc} = & \frac{2 (3n+1)}{8n} \left( 8 (\gamma^\alpha)_{ab}(\gamma_\alpha)_{c}^{~d} + [\gamma^\alpha,\gamma^\beta]_{ab} [\gamma_\alpha , \gamma_\beta]_{c}^{~d} \right)\tilde{\lambda}_{d }
   \label{e:chi}
\end{align}
is in terms of the new auxiliary fermion $\lambda_a$ and is also a consequence of the gauge symmetry Eq.~(\ref{e:psisymmetry}) of the Lagrangian. 

To facilitate finding the {\nmSG} adinkra, it will be advantageous to move to a basis in the 4D,
$\mathcal{N}=1$ theory where the transformation laws take a simpler form. Performing the following field redefinitions 
\begin{subequations}
\begin{align}
    \beta_a &= \frac{n}{2} [ \g^\m,\g^\n]_a^{~d} \partial_\m \psi_{\n d} - \frac{(3n +1)^2}{2} \tilde{\beta}_a +  2\partial_\n \tilde{\l}_a ~~~, \\*
        \lambda_a &= 4 \frac{3n+1}{n} \tilde{\lambda}_a ~~~,\\     
    S &= 2 (3n+1) \tilde{S}~~~,~~~ P = 2 (3n +1) \tilde{P}~~~, \\
    W_\mu &= 2 (3n+1) \tilde{W}_\mu + \frac{2}{3}\tilde{A}_\mu ~~~,\\
    V_\mu &= 2\frac{3n+1}{n} \tilde{V}_\mu ~~~,~~~ A_\mu = -2 \tilde{A}_\mu ~~~,
\end{align}
\end{subequations}
the transformation laws become
\begin{subequations}\label{e:DNMEasy}
\begin{eqnarray}
D_a S &=& \beta_a\\*
D_a P &=&  i  (\gamma^5)_a{}^d \beta_d \\
D_a A_\mu &=& i (\gamma^5 \gamma^\nu)_a{}^d \partial_{[\nu} \psi_{\mu]d} - \frac{1}{2} \epsilon_\mu{}^{\nu\alpha\beta} (\gamma_\nu)_a{}^d \partial_{\alpha} \psi_{\beta d} - i  (\gamma^5)_a{}^d \partial_\mu \lambda_d \\
D_a h_{\mu\nu} &=& \frac 12 \,  (\gamma_{(\mu})_{a}{}^d \psi_{\nu)d}  \\
D_a W_\mu &=& -i  (\gamma^5 \gamma_\mu)_a{}^d \beta_d  
+\frac{1}{2}  \epsilon_{\mu}^{~\nu\alpha\beta} (\gamma_\beta )_a{}^d  \partial_{\nu} \psi_{\alpha d}   \\
D_a V_\mu &=& n^{-1} (\gamma_\mu)_a{}^d \beta_d - \partial_\mu \lambda_a - (\gamma^\nu)_a{}^d \partial_{[\mu} \psi_{\nu]d} +  (\gamma^\nu \gamma_\mu)_a{}^d \partial_\nu \lambda_d \cr
&&- i  \epsilon_{\mu}^{~\nu\alpha\beta} (\gamma^5 \gamma_\beta )_a{}^d  \partial_{\nu} \psi_{\alpha d}
  \\
D_a \psi_{\mu c} &=& -i (\gamma_\mu)_{ac} S +  (\gamma^5 \gamma_\mu)_{ac} P + (\gamma^5)_{ac} A_\mu   \cr
&&- i \frac{1}{2} ( [ \gamma^\alpha , \gamma^\beta] )_{ac} \partial_\alpha h_{\beta\mu}  +  (\gamma^5 \gamma^\nu \gamma_\mu)_{ac} W_\nu \\
D_a \lambda_c &=& -i N C_{ac} S + N (\gamma^5)_{ac} P + 3 (\gamma^5 \gamma^\mu)_{ac} W_\mu + \cr
&&+ i  (\gamma^\mu)_{ac} V_\mu + (\gamma^5 \gamma^\mu)_{ac} A_\mu \\
D_a \beta_c &=& i  (\gamma^\mu)_{ac} \partial_\mu S - (\gamma^5 \gamma^\mu)_{ac} \partial_\mu P - (\gamma^5)_{ac} \partial^\mu W_\mu \cr &&
- i n C_{ac} \partial^\mu V_\mu + i C_{ac} \partial^\mu \partial_{[\nu} h_{\mu]}^{~~\nu} ~~~.
\end{eqnarray}
\end{subequations}
The Lagrangian that is invariant with respect to these simplified transformation laws is:
\begin{align}\label{e:LNMSimple}
\mathcal{L}_{\text{\nmSG}} =& -\, \frac 12 \,  \partial_\alpha h_{\mu\nu} \partial^\alpha h^{\mu\nu} + \frac 12 \partial_\alpha h \partial^\alpha h -  \partial^\alpha h \partial^\beta h_{\alpha\beta} +  \partial^\mu h_{\mu\nu} \partial_\alpha h^{\alpha\nu} + \frac{1}{n} S^2  \cr
&+   \frac{1}{n} P^2 -\frac{2}{3 n+1} W_\m A^\m + \frac{ n}{3 n+1}  A_\mu A^\mu - \frac{ n}{3 n+1}  V_\mu V^\mu  \cr
&  - \frac{3}{3 n+1}  W_\mu W^\mu - \frac 12 \,  \psi_{\mu a} \epsilon^{\mu\nu\alpha\beta} (\gamma^5 
\gamma_\nu)^{ab} \partial_\alpha \psi_{\beta b} + i\frac{2}{3 n+1}   \lambda_a \beta^a   \cr
&- i \frac{n}{3 n+1} \l_b (\g^\m\g^\n)^{bc}\partial_{[\m}\psi_{\n] c} -i\frac{n}{3 n+1}\l_b (\g^\n)^{bc}
\partial_\nu \l_c ~~~.
\end{align}

\subsection{One-Dimensional Reduction}\label{s:NM1D}
$~~~~$ Once more we work in the temporal gauge (\ref{e:TG}) for the graviton and
gravitino so that the
Lagrangian~(\ref{e:LNMSimple}) reduced to the 0-brane becomes 
\pagebreak
\begin{align}\label{e:L0NM}
  \mathcal{L}^{(0)}_{\text{\nmSG}} =& \dot{h}_{12}^2+\dot{h}_{13}^2+\dot{h}_{23}^2- \dot{h}_{11}\dot{h}_{22} - \dot{h}_{11}\dot{h}_{33} - \dot{h}_{22}\dot{h}_{33} + \frac{1}{n} S^2 + \frac{1}{n} P^2  \cr
  &+\frac{1}{3 n+1} \bigg(2 \left(-W_1 A_1 -  W_2 A_2 - W_3 A_3+W_0 A_0\right)   \cr
 &~~~~~~~~~~~~~~~+ n  (A_1 A_1 + A_2 A_2 + A_3 A_3-A_0 A_0) \cr
 &~~~~~~~~~~~~~~~- n  (V_1 V_1 + V_2 V_2 + V_3 V_3-V_0 V_0) \cr
  &~~~~~~~~~~~~~~~- 3 (W_1 W_1 + W_2 W_2 + W_3 W_3-W_0 W_0 ) \bigg)  \cr
 &+i\left( -\psi_{31}\dot{\psi}_{12} + \psi_{32}\dot{\psi}_{11} - \psi_{33}\dot{\psi}_{14} +\psi_{34}\dot{\psi}_{13}-\psi_{11}\dot{\psi}_{23}+\psi_{12}\dot{\psi}_{24}   \right. \cr
  &\left. + \psi_{13}\dot{\psi}_{21} - \psi_{14}\dot{\psi}_{22}-\psi_{21}\dot{\psi}_{34} - \psi_{22}\dot{\psi}_{33} + \psi_{23}\dot{\psi}_{32}+\psi_{24}\dot{\psi}_{31}      \right)  \cr
 &+ i\frac{1}{3 n+1}  (2 (\l_2 \b_1-\l_1 \b_2 - \l_4 \b_3 + \l_3 \b_4)-n (\l_1 \dot{l}_1 + \l_2 \dot{l}_2+\l_3 \dot{l}_3+\l_4 \dot{l}_4 ) ) \cr
 & - i \frac{2n}{3 n+1}\left(\lambda _2 \dot{\psi }_{11}+\lambda _1 \dot{\psi }_{12}+\lambda _4 \dot{\psi }_{13}+\lambda _3 \dot{\psi }_{14}-\lambda _4 \dot{\psi }_{21}+\lambda _3 \dot{\psi
   }_{22}\right. \cr
   &\left. +\lambda _2 \dot{\psi }_{23}-\lambda _1 \dot{\psi }_{24}+\lambda _1 \dot{\psi }_{31}-\lambda _2 \dot{\psi }_{32}+\lambda _3 \dot{\psi }_{33}-\lambda _4
   \dot{\psi }_{34}\right) ~~~.
\end{align}
The 0-brane reduced transformation laws can be succinctly displayed as in Tabs.~\ref{t:NM0braneB1}~and~\ref{t:NM0braneF1}. 
\begin{table}[!b]
    \caption{{\nmSG} bosonic transformation laws in temporal gauge,~Eq.(\ref{e:TG}), and reduced to the 0-brane. }
 \hspace{-7 pt}
 \renewcommand{\arraystretch}{1.3}
  {\tiny \begin{tabular}{c|cccc|}
    & $\gD$ & $\vD$ & $\oD$ & $\rD$ \\
    \hline
    $h_{11}$ & $\psi _{12}$&$\psi _{11}$&$\psi _{14}$&$\psi _{13}$ \\
   $h_{22}$ & $-\psi _{24}$&$\psi _{23}$&$\psi _{22}$&$-\psi _{21}$ \\
   $h_{33}$ & $\psi _{31}$&$-\psi _{32}$&$\psi _{33}$&$-\psi _{34}
   $\\
    $V_0$ & $\dot{\psi
   }_{24}-\dot{\psi }_{31}-\dot{\psi }_{12}-\frac{\beta _2}{n}$&$-\dot{\psi
   }_{11}-\dot{\psi }_{23}+\dot{\psi }_{32}+\frac{\beta _1}{n}$&$-\dot{\psi }_{14}-\dot{\psi }_{22}-\dot{\psi
   }_{33}+\frac{\beta
   _4}{n}$&$-\dot{\psi }_{13}+\dot{\psi
   }_{21}+\dot{\psi }_{34}-\frac{\beta _3}{n}$\\
    $V_1$ & $\dot{\psi
   }_{12}-\dot{\psi }_{24}+\dot{\psi }_{31}+\frac{\beta _2}{n}+\dot{\lambda }_1$&$\dot{\psi }_{32}-\dot{\psi }_{11}-\dot{\psi
   }_{23}+\frac{\beta
   _1}{n}-\dot{\lambda }_2$&$\frac{\beta _4}{n}-\dot{\lambda
   }_3-\dot{\psi }_{14}-\dot{\psi }_{22}-\dot{\psi
   }_{33}$&$\dot{\psi
   }_{13}-\dot{\psi }_{21}-\dot{\psi }_{34}+\frac{\beta _3}{n}+\dot{\lambda }_4$\\
   \hline
   $h_{13}$ & $ \frac{1}{2} \psi _{11}+\frac{1}{2}\psi _{32}$&$-\frac{1}{2}\psi _{12}+\frac{1}{2}
   \psi _{31}$&$\frac{1}{2} \psi_{13}+\frac{1}{2}\psi _{34}$&$-\frac{1}{2}\psi_{14}+\frac{1}{2} \psi _{33}$ \\
   $S$ & $\beta _1$&$\beta _2$&$\beta_3$&$\beta _4$ \\
    $W_2$ & $\frac{1}{2} \left(\dot{\psi }_{32}-\dot{\psi
   }_{11}\right)-\beta _1$&$\frac{1}{2} \left(\dot{\psi
   }_{12}+\dot{\psi }_{31}\right)-\beta _2$&$\frac{1}{2}
   \left(\dot{\psi }_{34}-\dot{\psi }_{13}\right)+\beta _3$&$\frac{1}{2} \left(\dot{\psi }_{14}+\dot{\psi
   }_{33}\right)+\beta
   _4$\\
    $V_3$ & $\frac{\beta _1}{n}-\dot{\lambda }_2-\dot{\psi
   }_{11}-\dot{\psi }_{23}+\dot{\psi }_{32}$&$\dot{\psi
   }_{24}-\dot{\psi }_{12}-\dot{\psi }_{31}-\frac{\beta
   _2}{n}-\dot{\lambda }_1$&$\frac{\beta _3}{n}+\dot{\lambda
   }_4+\dot{\psi }_{13}-\dot{\psi }_{21}-\dot{\psi
   }_{34}$&$\dot{\psi
   }_{14}+\dot{\psi }_{22}+\dot{\psi }_{33}-\frac{\beta _4}{n}+\dot{\lambda }_3$\\
    $A_2$ & $-\frac{1}{2} \left(-\dot{\psi }_{11}+2 \dot{\psi
   }_{23}+\dot{\psi }_{32}\right)$&$-\frac{1}{2} \left(\dot{\psi
   }_{12}+2 \dot{\psi }_{24}+\dot{\psi }_{31}\right)$&$-\frac{1}{2}
   \left(-\dot{\psi }_{13}-2 \dot{\psi }_{21}+\dot{\psi
   }_{34}\right)$&$-\frac{1}{2} \left(\dot{\psi }_{14}-2 \dot{\psi
   }_{22}+\dot{\psi }_{33}\right)$\\
   \hline
    $h_{12}$ & $-\frac{1}{2}\psi _{14}+\frac{1}{2}\psi _{22}$&$\frac{1}{2}\psi
   _{13}+\frac{1}{2}\psi _{21}$&$\frac{1}{2}\psi _{12}+\frac{1}{2}\psi
   _{24}$&$-\frac{1}{2}\psi _{11}+\frac{1}{2}\psi _{23}$\\
   $P$ & $-\beta _4$&$\beta_3$&$-\beta _2$&$\beta_1$\\
    $W_3$ & $\frac{1}{2} \left(-\dot{\psi }_{14}-\dot{\psi
   }_{22}\right)-\beta _4$&$\frac{1}{2} \left(\dot{\psi
   }_{13}-\dot{\psi }_{21}\right)-\beta _3$&$\frac{1}{2}
   \left(\dot{\psi }_{12}-\dot{\psi }_{24}\right)-\beta
   _2$&$\frac{1}{2} \left(-\dot{\psi }_{11}-\dot{\psi
   }_{23}\right)-\beta _1$\\
    $V_2$ & $\dot{\psi
   }_{14}+\dot{\psi }_{22}+\dot{\psi }_{33}-\frac{\beta _4}{n}+\dot{\lambda }_3$&$\dot{\psi }_{13}-\dot{\psi
   }_{21}-\dot{\psi }_{34}+\frac{\beta
   _3}{n}+\dot{\lambda }_4$&$\dot{\psi }_{12}-\dot{\psi }_{24}+\dot{\psi}_{31}+\frac{\beta _2}{n}+\dot{\lambda
   }_1$&$\dot{\psi
   }_{11}+\dot{\psi }_{23}-\dot{\psi }_{32}-\frac{\beta _1}{n}+\dot{\lambda }_2$\\
    $A_3$ & $-\frac{1}{2} \left(-\dot{\psi }_{14}-\dot{\psi }_{22}+2
   \dot{\psi }_{33}\right)$&$-\frac{1}{2} \left(\dot{\psi
   }_{13}-\dot{\psi }_{21}+2 \dot{\psi }_{34}\right)$&$-\frac{1}{2}
   \left(\dot{\psi }_{12}-\dot{\psi }_{24}-2 \dot{\psi
   }_{31}\right)$&$-\frac{1}{2} \left(-\dot{\psi }_{11}-\dot{\psi
   }_{23}-2 \dot{\psi }_{32}\right)$\\
   \hline
   $h_{23}$ & $\frac{1}{2}\psi _{21}-\frac{1}{2}\psi _{34}$&$-\frac{1}{2}\psi _{22}+\frac{1}{2}\psi_{33}$&$\frac{1}{2}\psi _{23}+\frac{1}{2}\psi_{32}$&$-\frac{1}{2}\psi _{24}-\frac{1}{2}\psi _{31}
   $ \\
   $W_0$ & $-\beta _3$&$-\beta _4$&$\beta _1$&$\beta _2$\\
   $W_1$ & $\frac{1}{2} \left(\dot{\psi }_{21}+\dot{\psi
   }_{34}\right)+\beta _3$&$\frac{1}{2} \left(-\dot{\psi }_{22}-\dot{\psi
   }_{33}\right)-\beta _4$&$\frac{1}{2} \left(\dot{\psi
   }_{23}-\dot{\psi }_{32}\right)+\beta _1$&$\frac{1}{2} \left(\dot{\psi
   }_{31}-\dot{\psi }_{24}\right)-\beta _2$
     \\
    $A_0$ & $\dot{\psi }_{13}-\dot{\psi }_{21}-\dot{\psi
   }_{34}+\dot{\lambda }_4$&$-\dot{\psi }_{14}-\dot{\psi
   }_{22}-\dot{\psi }_{33}-\dot{\lambda }_3$&$\dot{\psi
   }_{11}+\dot{\psi }_{23}-\dot{\psi }_{32}+\dot{\lambda }_2$&$-\dot{\psi }_{12}+\dot{\psi }_{24}-\dot{\psi }_{31}-\dot{\lambda}_1$\\
    $A_1$ & $-\frac{1}{2} \left(2 \dot{\psi }_{13}+\dot{\psi
   }_{21}+\dot{\psi }_{34}\right)$&$-\frac{1}{2} \left(2 \dot{\psi
   }_{14}-\dot{\psi }_{22}-\dot{\psi }_{33}\right)$&$-\frac{1}{2}
   \left(-2 \dot{\psi }_{11}+\dot{\psi }_{23}-\dot{\psi
   }_{32}\right)$&$-\frac{1}{2} \left(-2 \dot{\psi }_{12}-\dot{\psi
   }_{24}+\dot{\psi }_{31}\right)$ \\
   \hline
    \end{tabular}  }
    \label{t:NM0braneB1}
\end{table} 
\begin{table}[!h]
\centering
    \caption{{\nmSG} fermionic transformation laws in temporal gauge,~Eq.(\ref{e:TG}), and reduced to the 0-brane. }
  \renewcommand{\arraystretch}{1.3}
   {\tiny \begin{tabular}{c|cccc|}
    & $\gD$ & $\vD$ & $\oD$ & $\rD$ \\
    \hline
   $\psi_{11}$ & $i \dot{h}_{13}-i S-i W_2$&$i
   \dot{h}_{11}$&$i A_1-i W_0+i
   W_1$&$-i \dot{h}_{12}-i P-i
   W_3$ \\
    $\psi_{23}$ & $-i A_2-i S-i W_2$&$i
   \dot{h}_{22}$&$i
   \dot{h}_{23}-i W_0+i W_1$&$i
   \dot{h}_{12}-i P-i
   W_3 $ \\
   $\psi_{32}$ & $i \dot{h}_{13}+i S+i
   W_2$&$-i \dot{h}_{33}$&$i
   \dot{h}_{23}+i W_0-i W_1$&$i
   A_3+i P+i W_3$ \\
    $\lambda_2$ & $i A_2+iN
   S-i V_3+3 i W_2$&$i
   \left(V_0-V_1\right)$&$i
   A_0-i A_1+3 i W_0-3 i
   W_1$&$i A_3+iN
   P+i V_2+3 i W_3$ \\
    $\beta_1$ & $i \dot{S}$&$i n \left(\ddot{h}+ \dot{V}_0\right)$&$i \dot{W}_0$&$i \dot{P}$ \\
      \hline
   $\psi_{12}$ & $i \dot{h}_{11}$&$-i
   \dot{h}_{13}+i S+i W_2$&$i
   \dot{h}_{12}-i P+i W_3$&$i A_1+i
   W_0+i W_1$ \\
    $\psi_{24}$ & $-i \dot{h}_{22}$&$-iA_2-i
   S-i W_2$&$i \dot{h}_{12}+i
   P-i W_3$&$-i \dot{h}_{23}-i
   W_0-i W_1$ \\
    $\psi_{31}$ & $i \dot{h}_{33}$&$i
   \dot{h}_{13}+i S+i W_2$&$i
   A_3-i P+i W_3$&$-i
   \dot{h}_{23}+i W_0+i
   W_1$ \\
    $\lambda_1$ & $i
   \left(V_0+V_1\right)$&$-i
   A_2- i N S-i
   V_3-3 i W_2$&$-i A_3+i
   NP+i V_2-3 i
   W_3$&$-i A_0-i A_1-3 i W_0-3 i
   W_1 $ \\
   $\beta_2$ & $-i n
   \left(\ddot{h}+ \dot{V}_0\right)$&$i \dot{S}$&$-i \dot{P}$&$i\dot{W}_0 $ 
   \\
   \hline
    $\psi_{13}$ & $-i A_1-i W_0-i W_1$&$i
   \dot{h}_{12}+i P+i W_3$&$i
   \dot{h}_{13}+i S-i W_2$&$i
   \dot{h}_{11}$
    \\
     $\psi_{21}$ & $i \dot{h}_{23}+i W_0+i
   W_1$&$i \dot{h}_{12}-i P-i
   W_3$&$i A_2-i S+i W_2$&$-i
   \dot{h}_{22}$ \\
    $\psi_{34}$ & $-i \dot{h}_{23}+i W_0+i
   W_1$&$-i A_3-i P-i W_3$&$i
   \dot{h}_{13}-i S+i W_2$&$-i
   \dot{h}_{33}$ \\
    $\lambda_4$ & $i A_0+i A_1+3 i W_0+3
   i W_1$&$-i A_3-i NP+i V_2-3 i W_3$&$i
   A_2-i NS+i
   V_3+3 i W_2$&$i
   \left(V_0+V_1\right)$ \\
 $\beta_3$ & $-i \dot{W}_0$&$i \dot{P}$&$i \dot{S}$&$-i n \left(\ddot{h}+ \dot{V}_0\right)$ \\
      \hline
    $\psi_{14}$ & $-i \dot{h}_{12}+i P-i
   W_3$&$-i A_1+i W_0-i W_1$&$i
   \dot{h}_{11}$&$-i
   \dot{h}_{13}-i S+i
   W_2$\\
    $\psi_{22}$ & $i \dot{h}_{12}+i P-i
   W_3$&$-i \dot{h}_{23}+i W_0-i
   W_1$&$i \dot{h}_{22}$&$i A_2-i
   S+i W_2$ \\
    $\psi_{33}$ & $-i A_3+i P-i W_3$&$i
   \dot{h}_{23}+i W_0-i W_1$&$i
   \dot{h}_{33}$&$i
   \dot{h}_{13}-i S+i
   W_2$ \\
    $\lambda_3$ & $i A_3-i N
   P+i V_2+3 i W_3$&$-i A_0+i
   A_1-3 i W_0+3 i W_1$&$i
   \left(V_0-V_1\right)$&$-i
   A_2+i NS+i
   V_3-3 i W_2$ \\
   $\beta_4$ & $-i \dot{P}$&$-i \dot{W}_0$&$i n  \left(\ddot{h}+ \dot{V}_0\right)$&$i \dot{S}$ \\
   \hline
    \end{tabular}  }
    \label{t:NM0braneF1}
\end{table} 

\pagebreak
\noindent Following the same iterative procedure depicted in Fig.~\ref{f:cistransit} leads us again to one unique cis-adinkra submultiplet, and a set of choices for trans-adinkra submultiplets, the nodal content of which are
\begin{align}
      \Phi =   \left(
\begin{array}{c}
 W_0 \\
 P \\
 S \\
 n
   \left(\dot{h}+V_0\right)
\end{array}
\right)~~~,&~~~i \Psi = \left(
\begin{array}{c}
 -\beta _3 \\
 -\beta _4 \\
 \beta _1 \\
 -\beta _2
\end{array}
\right) ~~~,~~~
\mbox{cis:}~\chi_0 = 1~~~,
\label{e:nmSGcis}
\end{align} 
\begin{align}
      \Phi =  &  \left(
\begin{array}{c}
u_1 A_1
   +(u_2-u_3 )\dot{h}_{23} + \frac{u_4}{N} \left(A_0+A_1-\frac{W_0}{n}-\frac{W_1}{n}\right)-u_5 W_1 \\
 u_3 A_3 +
   (u_1-u_2)\dot{h}_{12}+\frac{u_4}{N} \left(A_3-\frac{W_3}{n}-V_2\right)-u_5 W_3 \\
 u_2 A_2 +(u_3-u_1 )\dot{h}_{31}
   +\frac{u_4}{N} \left(A_2-\frac{W_2}{n}+V_3\right)-u_5 W_2 \\
-u_1\dot{h}_{11} -u_2\dot{h}_{22} -u_3\dot{h}_{33}
   +\frac{u_4}{N} (V_0+V_1)+ u_5 n (\dot{h}+
   V_0)
\end{array}
\right)~~~,
\nonumber\\ i \Psi = & \left(
\begin{array}{c}
 -u_1 \dot{\psi }_{13}+u_2
   \dot{\psi }_{21}+u_3 \dot{\psi }_{34}+\frac{u_4}{N} \dot{\lambda }_4-u_5 \beta _3 \\
-u_1 \dot{\psi }_{14}-u_2 \dot{\psi }_{22}-u_3 \dot{\psi
   }_{33} +2 \frac{u_4}{N}\left(\frac{\beta _4}{n}-\frac{\dot{\lambda }_3}{2}-\dot{\psi
   }_{14}-\dot{\psi }_{22}-\dot{\psi }_{33}\right)+u_5 \beta
   _4\\
-u_1 \dot{\psi }_{11}-u_2 \dot{\psi }_{23}+u_3 \dot{\psi
   }_{32}+ 2 \frac{u_4}{N} \left(\frac{ \beta _1}{n}-\frac{\dot{\lambda }_2}{2}-\dot{\psi
   }_{11}- \dot{\psi }_{23}+ \dot{\psi }_{32}\right)+u_5 \beta
   _1 \\
 -u_1 \dot{\psi }_{12}+u_2
   \dot{\psi }_{24}-u_3 \dot{\psi }_{31}+\frac{u_4}{N} \dot{\lambda }_1-u_5 \beta _2
\end{array}
\right)~~~,
\nonumber\\*
\nonumber\\*
& \mbox{trans:}~\chi_0 = -1~~~,~~~u_1 + u_2 + u_3 + u_4 + u_5 = 0~~.
\label{e:nmSGtrans}
\end{align}  
Since there is one unique cis choice, we have $n_c = 1$ for \nmSG ~as was the case for mSG. The trans choice, Eq.~(\ref{e:nmSGtrans}), has four independent parameters that lead to four linearly independent nodal field definitions. We therefore have $n_t = 4$ for \nmSG. In summary, the SUSY isomer numbers for \nmSG ~are
\begin{align}
  n_c = 1~~~,~~~n_t = 4~~~.
\end{align}

To fill out the nodal field content of the $(20|20)$ \nmSG ~multiplet, we must make three more copies of the trans-submultiplet, Eq.~(\ref{e:nmSGtrans}), by creating three more sets of parameters, $v_i$, $q_i$, and $p_i$, that are constrained exactly as the $u_i$ are:
\begin{subequations}\label{e:uvpqconstraints}
\begin{eqnarray}
   u_1 + u_2 + u_3 + u_4 + u_5 &=& 0 \\
    v_1 + v_2 + v_3 + v_4 + v_5 &=& 0 \\
    p_1 + p_2 + p_3 + p_4 + p_5 &=& 0 \\
    q_1 + q_2 + q_3 + q_4 + q_5 &=& 0~~~. 
    \end{eqnarray}
\end{subequations}
 The resulting field content for all nodes of the \nmSG~adinkra with components numbered top to bottom, one through 20 is
\begin{align}
\Phi = & \left(
\text{{\tiny 
$\begin{array}{c}
 W_0 \\
 P \\
 S \\
 n
   \left(\dot{h}+V_0\right)\\
   \hline
\vspace{-7 pt} \\
u_1 A_1
   +(u_2-u_3 )\dot{h}_{23} + \frac{u_4}{N} \left(A_0+A_1-\frac{W_0}{n}-\frac{W_1}{n}\right)-u_5 W_1 \\
 u_3 A_3 +
   (u_1-u_2)\dot{h}_{12}+\frac{u_4}{N} \left(A_3-\frac{W_3}{n}-V_2\right)-u_5 W_3 \\
 u_2 A_2 +(u_3-u_1 )\dot{h}_{31}
   +\frac{u_4}{N} \left(A_2-\frac{W_2}{n}+V_3\right)-u_5 W_2 \\
-u_1\dot{h}_{11} -u_2\dot{h}_{22} -u_3\dot{h}_{33}
   +\frac{u_4}{N} (V_0+V_1)+ u_5 n (\dot{h}+
   V_0) \\
   \hline
\vspace{-7 pt} \\
v_1 A_1
   +(v_2-v_3 )\dot{h}_{23} + \frac{v_4}{N} \left(A_0+A_1-\frac{W_0}{n}-\frac{W_1}{n}\right)-v_5 W_1 \\
 v_3 A_3 +
   (v_1-v_2)\dot{h}_{12}+\frac{v_4}{N} \left(A_3-\frac{W_3}{n}-V_2\right)-v_5 W_3 \\
 v_2 A_2 +(v_3-v_1 )\dot{h}_{31}
   +\frac{v_4}{N} \left(A_2-\frac{W_2}{n}+V_3\right)-v_5 W_2 \\
-v_1\dot{h}_{11} -v_2\dot{h}_{22} -v_3\dot{h}_{33}
   +\frac{v_4}{N} (V_0+V_1)+ v_5 n (\dot{h}+
   V_0)\\
   \hline
\vspace{-7 pt} \\
q_1 A_1
   +(q_2-q_3 )\dot{h}_{23} + \frac{q_4}{N} \left(A_0+A_1-\frac{W_0}{n}-\frac{W_1}{n}\right)-q_5 W_1 \\
 q_3 A_3 +
   (q_1-q_2)\dot{h}_{12}+\frac{q_4}{N} \left(A_3-\frac{W_3}{n}-V_2\right)-q_5 W_3 \\
 q_2 A_2 +(q_3-q_1 )\dot{h}_{31}
   +\frac{q_4}{N} \left(A_2-\frac{W_2}{n}+V_3\right)-q_5 W_2 \\
-q_1\dot{h}_{11} -q_2\dot{h}_{22} -q_3\dot{h}_{33}
   +\frac{q_4}{N} (V_0+V_1)+ q_5 n (\dot{h}+
   V_0)  \\
   \hline
\vspace{-7 pt} \\
p_1 A_1
   +(p_2-p_3 )\dot{h}_{23} + \frac{p_4}{N} \left(A_0+A_1-\frac{W_0}{n}-\frac{W_1}{n}\right)-p_5 W_1 \\
 p_3 A_3 +
   (p_1-p_2)\dot{h}_{12}+\frac{p_4}{N} \left(A_3-\frac{W_3}{n}-V_2\right)-p_5 W_3 \\
 p_2 A_2 +(p_3-p_1 )\dot{h}_{31}
   +\frac{p_4}{N} \left(A_2-\frac{W_2}{n}+V_3\right)-p_5 W_2 \\
-p_1\dot{h}_{11} -p_2\dot{h}_{22} -p_3\dot{h}_{33}
   +\frac{p_4}{N} (V_0+V_1)+ p_5 n (\dot{h}+
   V_0) 
\end{array}
$}}
\right)~~~,
    \label{e:nmSGPhi}
\end{align}

\begin{equation}
    i\Psi =   \left(
\text{{\tiny $
    \begin{array}{c}
 -\beta _3 \\
 -\beta _4 \\
 \beta _1 \\
 -\beta _2 \\
 \hline
\vspace{-6 pt} \\
 -u_1 \dot{\psi }_{13}+u_2
   \dot{\psi }_{21}+u_3 \dot{\psi }_{34}+\frac{u_4}{N} \dot{\lambda }_4-u_5 \beta _3 \\
-u_1 \dot{\psi }_{14}-u_2 \dot{\psi }_{22}-u_3 \dot{\psi
   }_{33} +2 \frac{u_4}{N} \left(\frac{\beta _4}{n}-\frac{\dot{\lambda }_3}{2}-\dot{\psi
   }_{14}-\dot{\psi }_{22}-\dot{\psi }_{33}\right)+u_5 \beta
   _4\\
-u_1 \dot{\psi }_{11}-u_2 \dot{\psi }_{23}+u_3 \dot{\psi
   }_{32}+ 2 \frac{u_4}{N} \left(\frac{ \beta _1}{n}-\frac{\dot{\lambda }_2}{2}-\dot{\psi
   }_{11}- \dot{\psi }_{23}+ \dot{\psi }_{32}\right)+u_5 \beta
   _1 \\
 -u_1 \dot{\psi }_{12}+u_2
   \dot{\psi }_{24}-u_3 \dot{\psi }_{31}+\frac{u_4}{N} \dot{\lambda }_1-u_5 \beta _2 \\
 \hline
\vspace{-6 pt} \\
 -v_1 \dot{\psi }_{13}+v_2
   \dot{\psi }_{21}+v_3 \dot{\psi }_{34}+\frac{v_4}{N} \dot{\lambda }_4-v_5 \beta _3 \\
-v_1 \dot{\psi }_{14}-v_2 \dot{\psi }_{22}-v_3 \dot{\psi
   }_{33} +2 \frac{v_4}{N} \left(\frac{\beta _4}{n}-\frac{\dot{\lambda }_3}{2}-\dot{\psi
   }_{14}-\dot{\psi }_{22}-\dot{\psi }_{33}\right)+v_5 \beta
   _4\\
-v_1 \dot{\psi }_{11}-v_2 \dot{\psi }_{23}+v_3 \dot{\psi
   }_{32}+ 2 \frac{v_4}{N} \left(\frac{ \beta _1}{n}-\frac{\dot{\lambda }_2}{2}-\dot{\psi
   }_{11}- \dot{\psi }_{23}+ \dot{\psi }_{32}\right)+v_5 \beta
   _1 \\
 -v_1 \dot{\psi }_{12}+v_2
   \dot{\psi }_{24}-v_3 \dot{\psi }_{31}+\frac{v_4}{N} \dot{\lambda }_1-v_5 \beta _2 \\
 \hline
\vspace{-6 pt} \\
 -q_1 \dot{\psi }_{13}+q_2
   \dot{\psi }_{21}+q_3 \dot{\psi }_{34}+\frac{q_4}{N} \dot{\lambda }_4-q_5 \beta _3 \\
-q_1 \dot{\psi }_{14}-q_2 \dot{\psi }_{22}-q_3 \dot{\psi
   }_{33} +2 \frac{q_4}{N} \left(\frac{\beta _4}{n}-\frac{\dot{\lambda }_3}{2}-\dot{\psi
   }_{14}-\dot{\psi }_{22}-\dot{\psi }_{33}\right)+q_5 \beta
   _4\\
-q_1 \dot{\psi }_{11}-q_2 \dot{\psi }_{23}+q_3 \dot{\psi
   }_{32}+ 2 \frac{q_4}{N} \left(\frac{ \beta _1}{n}-\frac{\dot{\lambda }_2}{2}-\dot{\psi
   }_{11}- \dot{\psi }_{23}+ \dot{\psi }_{32}\right)+q_5 \beta
   _1 \\
 -q_1 \dot{\psi }_{12}+q_2
   \dot{\psi }_{24}-q_3 \dot{\psi }_{31}+\frac{q_4}{N} \dot{\lambda }_1-q_5 \beta _2  \\
 \hline
\vspace{-6 pt} \\
 -p_1 \dot{\psi }_{13}+p_2
   \dot{\psi }_{21}+p_3 \dot{\psi }_{34}+\frac{p_4}{N} \dot{\lambda }_4-p_5 \beta _3 \\
-p_1 \dot{\psi }_{14}-p_2 \dot{\psi }_{22}-p_3 \dot{\psi
   }_{33} +2 \frac{p_4}{N} \left(\frac{\beta _4}{n}-\frac{\dot{\lambda }_3}{2}-\dot{\psi
   }_{14}-\dot{\psi }_{22}-\dot{\psi }_{33}\right)+p_5 \beta
   _4\\
-p_1 \dot{\psi }_{11}-p_2 \dot{\psi }_{23}+p_3 \dot{\psi
   }_{32}+ 2 \frac{p_4}{N} \left(\frac{ \beta _1}{n}-\frac{\dot{\lambda }_2}{2}-\dot{\psi
   }_{11}- \dot{\psi }_{23}+ \dot{\psi }_{32}\right)+p_5 \beta
   _1 \\
 -p_1 \dot{\psi }_{12}+p_2
   \dot{\psi }_{24}-p_3 \dot{\psi }_{31}+\frac{p_4}{N} \dot{\lambda }_1-p_5 \beta _2
\end{array}
$ }}
    \right)    ~~~,
    \label{e:nmSGPsi}
\end{equation}
where horizontal lines separate the five submultiplets. The node definitions in Eqs.~(\ref{e:nmSGPhi}) and~(\ref{e:nmSGPsi}) collapse Tabs.~\ref{t:NM0braneB1} and~\ref{t:NM0braneF1} into the smaller Tab.~\ref{t:0braneNM2} which is succinctly displayed as the valise adinkra for \nmSG ~in Fig.~\ref{f:NMAdinkra}.  Comparing Fig.~\ref{f:NMAdinkra} to Fig.~\ref{f:cvtvintro}, we clearly see that the valise adinkra for {\nmSG} is composed of $n_c=1$ cis-adinkra and $n_t=4$ trans-adinkras.

\begin{table}[!hb]
\centering
  \caption{Zero-brane reduced {\nmSG} transformation rules in the adinkraic representation defined in Eqs.~(\ref{e:nmSGPhi}) and~(\ref{e:nmSGPsi}).}
  {\tiny
  \begin{tabular}{r|rrrr|rr|rrrr|}
   & $\gD$ & $\vD$ & $\oD$ & $\rD$ &\hspace{50 pt}& & $\gD$ & $\vD$ & $\oD$ & $\rD$ \\
   \cline{1-5}\cline{7-11}
 $ \Phi _1 $&$ i \Psi _1 $&$ i \Psi _2 $&$ i \Psi _3 $&$ -i \Psi _4 $&\hspace{50 pt}&$\Psi _1 $&$ \dot{\Phi }_1 $&$ -\dot{\Phi }_2 $&$ -\dot{\Phi }_3 $&$ \dot{\Phi }_4 $\\$
 \Phi _2 $&$ i \Psi _2 $&$ -i \Psi _1 $&$ i \Psi _4 $&$ i \Psi _3 $&\hspace{50 pt}&$ \Psi _2 $&$ \dot{\Phi }_2 $&$ \dot{\Phi }_1 $&$ -\dot{\Phi }_4 $&$ -\dot{\Phi }_3 $\\$
 \Phi _3 $&$ i \Psi _3 $&$ -i \Psi _4 $&$ -i \Psi _1 $&$ -i \Psi _2 $&\hspace{50 pt}&$ \Psi _3 $&$ \dot{\Phi }_3 $&$ \dot{\Phi }_4 $&$ \dot{\Phi }_1 $&$ \dot{\Phi }_2 $\\$
 \Phi _4 $&$ i \Psi _4 $&$ i \Psi _3 $&$ -i \Psi _2 $&$ i \Psi _1 $&\hspace{50 pt}&$ \Psi _4 $&$ \dot{\Phi }_4 $&$ -\dot{\Phi }_3 $&$ \dot{\Phi }_2 $&$ -\dot{\Phi }_1 $\\
  \cline{1-5}\cline{7-11}$
 \Phi _5 $&$ i \Psi _5 $&$ i \Psi _6 $&$ -i \Psi _7 $&$ -i \Psi _8 $&\hspace{50 pt}&$ \Psi _5 $&$ \dot{\Phi }_5 $&$ -\dot{\Phi }_6 $&$ \dot{\Phi }_7 $&$ \dot{\Phi }_8 $\\$
 \Phi _6 $&$ i \Psi _6 $&$ -i \Psi _5 $&$ -i \Psi _8 $&$ i \Psi _7 $&\hspace{50 pt}&$ \Psi _6 $&$ \dot{\Phi }_6 $&$ \dot{\Phi }_5 $&$ \dot{\Phi }_8 $&$ -\dot{\Phi }_7 $\\$
 \Phi _7 $&$ i \Psi _7 $&$ -i \Psi _8 $&$ i \Psi _5 $&$ -i \Psi _6 $&\hspace{50 pt}&$ \Psi _7 $&$ \dot{\Phi }_7 $&$ \dot{\Phi }_8 $&$ -\dot{\Phi }_5 $&$ \dot{\Phi }_6 $\\$
 \Phi _8 $&$ i \Psi _8 $&$ i \Psi _7 $&$ i \Psi _6 $&$ i \Psi _5 $&\hspace{50 pt}&$ \Psi _8 $&$ \dot{\Phi }_8 $&$ -\dot{\Phi }_7 $&$ -\dot{\Phi }_6 $&$ -\dot{\Phi }_5 $\\
  \cline{1-5}\cline{7-11}
  $\Phi _9 $&$ i \Psi _9 $&$ i \Psi _{10} $&$ -i \Psi _{11} $&$ -i \Psi _{12} $&\hspace{50 pt}&$ \Psi _9 $&$ \dot{\Phi }_9 $&$ -\dot{\Phi }_{10} $&$ \dot{\Phi }_{11} $&$ \dot{\Phi }_{12} $\\$
 \Phi _{10} $&$ i \Psi _{10} $&$ -i \Psi _9 $&$ -i \Psi _{12} $&$ i \Psi _{11} $&\hspace{50 pt}&$ \Psi _{10} $&$ \dot{\Phi }_{10} $&$ \dot{\Phi }_9 $&$ \dot{\Phi }_{12} $&$ -\dot{\Phi }_{11} $\\$
 \Phi _{11} $&$ i \Psi _{11} $&$ -i \Psi _{12} $&$ i \Psi _9 $&$ -i \Psi _{10} $&\hspace{50 pt}&$ \Psi _{11} $&$ \dot{\Phi }_{11} $&$ \dot{\Phi }_{12} $&$ -\dot{\Phi }_9 $&$ \dot{\Phi }_{10} $\\$
 \Phi _{12} $&$ i \Psi _{12} $&$ i \Psi _{11} $&$ i \Psi _{10} $&$ i \Psi _9 $&\hspace{50 pt}&$ \Psi _{12} $&$ \dot{\Phi }_{12} $&$ -\dot{\Phi }_{11} $&$ -\dot{\Phi }_{10} $&$ -\dot{\Phi }_9$\\
  \cline{1-5}\cline{7-11}$
 \Phi _{13} $&$ i \Psi _{13} $&$ i \Psi _{14} $&$ -i \Psi _{15} $&$ -i \Psi _{16} $&\hspace{50 pt}&$ \Psi _{13} $&$ \dot{\Phi }_{13} $&$ -\dot{\Phi }_{14} $&$ \dot{\Phi }_{15} $&$ \dot{\Phi }_{16} $\\$
 \Phi _{14} $&$ i \Psi _{14} $&$ -i \Psi _{13} $&$ -i \Psi _{16} $&$ i \Psi _{15} $&\hspace{50 pt}&$ \Psi _{14} $&$ \dot{\Phi }_{14} $&$ \dot{\Phi }_{13} $&$ \dot{\Phi }_{16} $&$ -\dot{\Phi }_{15} $\\$
 \Phi _{15} $&$ i \Psi _{15} $&$ -i \Psi _{16} $&$ i \Psi _{13} $&$ -i \Psi _{14} $&\hspace{50 pt}&$ \Psi _{15} $&$ \dot{\Phi }_{15} $&$ \dot{\Phi }_{16} $&$ -\dot{\Phi }_{13} $&$ \dot{\Phi }_{14} $\\$
 \Phi _{16} $&$ i \Psi _{16} $&$ i \Psi _{15} $&$ i \Psi _{14} $&$ i \Psi _{13} $&\hspace{50 pt}&$ \Psi _{16} $&$ \dot{\Phi }_{16} $&$ -\dot{\Phi }_{15} $&$ -\dot{\Phi }_{14} $&$ -\dot{\Phi }_{13}$\\
  \cline{1-5}\cline{7-11}$
 \Phi _{17} $&$ i \Psi _{17} $&$ i \Psi _{18} $&$ -i \Psi _{19} $&$ -i \Psi _{20} $&\hspace{50 pt}&$ \Psi _{17} $&$ \dot{\Phi }_{17} $&$ -\dot{\Phi }_{18} $&$ \dot{\Phi }_{19} $&$ \dot{\Phi }_{20} $\\$
 \Phi _{18} $&$ i \Psi _{18} $&$ -i \Psi _{17} $&$ -i \Psi _{20} $&$ i \Psi _{19} $&\hspace{50 pt}&$ \Psi _{18} $&$ \dot{\Phi }_{18} $&$ \dot{\Phi }_{17} $&$ \dot{\Phi }_{20} $&$ -\dot{\Phi }_{19} $\\$
 \Phi _{19} $&$ i \Psi _{19} $&$ -i \Psi _{20} $&$ i \Psi _{17} $&$ -i \Psi _{18} $&\hspace{50 pt}&$ \Psi _{19} $&$ \dot{\Phi }_{19} $&$ \dot{\Phi }_{20} $&$ -\dot{\Phi }_{17} $&$ \dot{\Phi }_{18} $\\$
 \Phi _{20} $&$ i \Psi _{20} $&$ i \Psi _{19} $&$ i \Psi _{18} $&$ i \Psi _{17} $&\hspace{50 pt}&$ \Psi _{20} $&$ \dot{\Phi }_{20} $&$ -\dot{\Phi }_{19} $&$ -\dot{\Phi }_{18} $&$ -\dot{\Phi }_{17}$ \\
 \cline{1-5}\cline{7-11}
  \end{tabular}}
  \label{t:0braneNM2}
\end{table}

\begin{figure}[!ht]
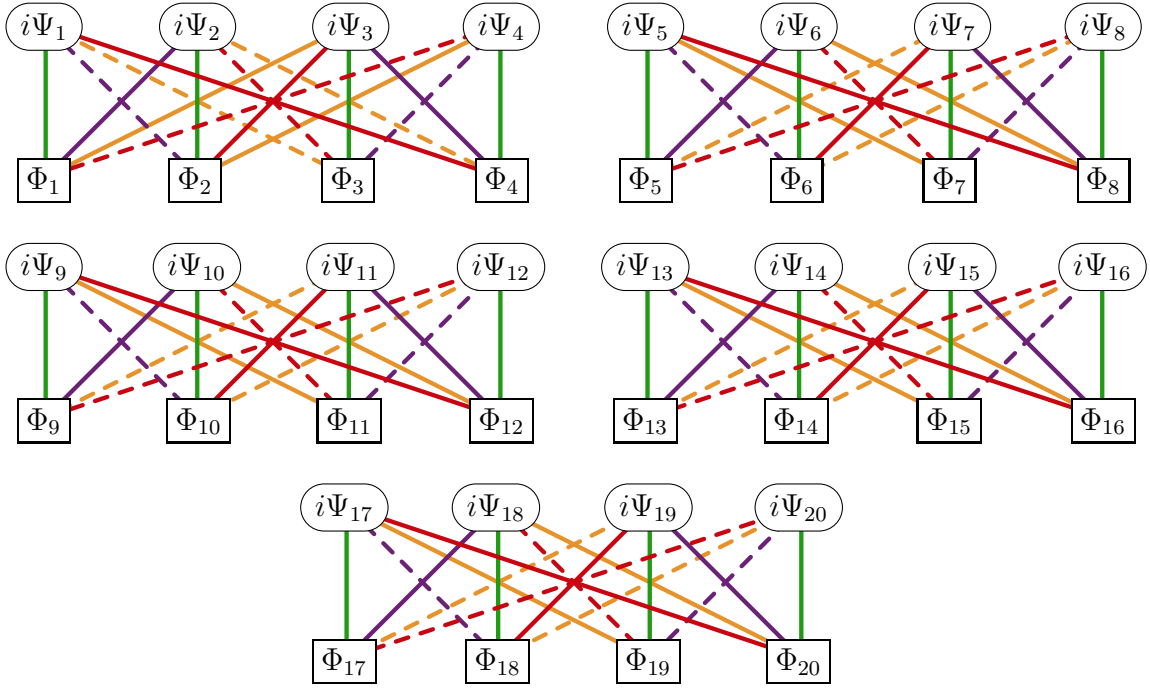
\setlength{\unitlength}{.8 mm}
\begin{center}
   \begin{picture}(200,120)(0,0)
\put(0,80){\includegraphics[width = 90\unitlength]{CisValise}}
\put(2.5,87){\fcolorbox{black}{white}{$\Phi_1$}}
\put(27.6,87){\fcolorbox{black}{white}{$\Phi_2$}}
\put(53,87){\fcolorbox{black}{white}{$\Phi_3$}}
\put(78.6,87){\fcolorbox{black}{white}{$\Phi_4$}}
\put(0.5,110){\begin{tikzpicture}
 \node[rounded rectangle,draw,fill=white!30]{$i\Psi_1$};
 \end{tikzpicture}}
\put(26,110){\begin{tikzpicture}
 \node[rounded rectangle,draw,fill=white!30]{$i\Psi_2$};
 \end{tikzpicture}}
\put(51.5,110){\begin{tikzpicture}
 \node[rounded rectangle,draw,fill=white!30]{$i\Psi_3$};
 \end{tikzpicture}}
\put(76.5,110){\begin{tikzpicture}
 \node[rounded rectangle,draw,fill=white!30]{$i\Psi_4$};
 \end{tikzpicture}}
\quad
\put(100,80){\includegraphics[width = 90\unitlength]{TransValise}}
\put(102.5,87){\fcolorbox{black}{white}{$\Phi_5$}}
\put(127.6,87){\fcolorbox{black}{white}{$\Phi_6$}}
\put(153,87){\fcolorbox{black}{white}{$\Phi_7$}}
\put(178.6,87){\fcolorbox{black}{white}{$\Phi_8$}}
\put(100.5,110){\begin{tikzpicture}
 \node[rounded rectangle,draw,fill=white!30]{$i\Psi_5$};
 \end{tikzpicture}}
\put(126,110){\begin{tikzpicture}
 \node[rounded rectangle,draw,fill=white!30]{$i\Psi_6$};
 \end{tikzpicture}}
\put(151.5,110){\begin{tikzpicture}
 \node[rounded rectangle,draw,fill=white!30]{$i\Psi_7$};
 \end{tikzpicture}}
\put(176.5,110){\begin{tikzpicture}
 \node[rounded rectangle,draw,fill=white!30]{$i\Psi_8$};
 \end{tikzpicture}}
\put(0,40){\includegraphics[width = 90\unitlength]{TransValise}}
\put(2.5,47){\fcolorbox{black}{white}{$\Phi_9$}}
\put(27.3,47){\fcolorbox{black}{white}{$\Phi_{10}$}}
\put(52.3,47){\fcolorbox{black}{white}{$\Phi_{11}$}}
\put(77.6,47){\fcolorbox{black}{white}{$\Phi_{12}$}}
\put(0.5,70){\begin{tikzpicture}
 \node[rounded rectangle,draw,fill=white!30]{$i\Psi_{9}$};
 \end{tikzpicture}}
\put(25,70){\begin{tikzpicture}
 \node[rounded rectangle,draw,fill=white!30]{$i\Psi_{10}$};
 \end{tikzpicture}}
\put(50.5,70){\begin{tikzpicture}
 \node[rounded rectangle,draw,fill=white!30]{$i\Psi_{11}$};
 \end{tikzpicture}}
\put(75.5,70){\begin{tikzpicture}
 \node[rounded rectangle,draw,fill=white!30]{$i\Psi_{12}$};
 \end{tikzpicture}}
\quad
\put(100,40){\includegraphics[width = 90\unitlength]{TransValise}}
\put(101.5,47){\fcolorbox{black}{white}{$\Phi_{13}$}}
\put(126.6,47){\fcolorbox{black}{white}{$\Phi_{14}$}}
\put(152,47){\fcolorbox{black}{white}{$\Phi_{15}$}}
\put(177.6,47){\fcolorbox{black}{white}{$\Phi_{16}$}}
\put(99.5,70){\begin{tikzpicture}
 \node[rounded rectangle,draw,fill=white!30]{$i\Psi_{13}$};
 \end{tikzpicture}}
\put(125,70){\begin{tikzpicture}
 \node[rounded rectangle,draw,fill=white!30]{$i\Psi_{14}$};
 \end{tikzpicture}}
\put(150.5,70){\begin{tikzpicture}
 \node[rounded rectangle,draw,fill=white!30]{$i\Psi_{15}$};
 \end{tikzpicture}}
\put(175.5,70){\begin{tikzpicture}
 \node[rounded rectangle,draw,fill=white!30]{$i\Psi_{16}$};
 \end{tikzpicture}}
 \put(50,0){\includegraphics[width = 90\unitlength]{TransValise}}
\put(51.5,7){\fcolorbox{black}{white}{$\Phi_{17}$}}
\put(77,7){\fcolorbox{black}{white}{$\Phi_{18}$}}
\put(102,7){\fcolorbox{black}{white}{$\Phi_{19}$}}
\put(127,7){\fcolorbox{black}{white}{$\Phi_{20}$}}
\put(49.5,30){\begin{tikzpicture}
 \node[rounded rectangle,draw,fill=white!30]{$i\Psi_{17}$};
 \end{tikzpicture}}
\put(74.7,30){\begin{tikzpicture}
 \node[rounded rectangle,draw,fill=white!30]{$i\Psi_{18}$};
 \end{tikzpicture}}
\put(100,30){\begin{tikzpicture}
 \node[rounded rectangle,draw,fill=white!30]{$i\Psi_{19}$};
 \end{tikzpicture}}
\put(125,30){\begin{tikzpicture}
 \node[rounded rectangle,draw,fill=white!30]{$i\Psi_{20}$};
 \end{tikzpicture}}
   \end{picture}
   \end{center}
   \vspace{-20 pt}
\caption{The {\nmSG} valise adinkra. It is composed of one cis-adinkra and four trans-adinkras so therefore has the SUSY isomer numbers $n_c=  1,~n_t = 4$. All bosons have the same engineering dimension and all fermions have the same engineering dimension.}
\label{f:NMAdinkra}
\end{figure} 
\newpage
\noindent Eliminating the fifth parameter from each of the four submultiplets via the constraint Eq.~(\ref{e:uvpqconstraints}), the other four parameters form vectors in a parameters space
\begin{align}\label{e:nmvecs}
\vec{u} = (u_1, u_2, u_3, u_4)~~~,&~~~ \vec{v} = (v_1, v_2, v_3, v_4)~~~, \nonumber\\*
\vec{p} = (p_1, p_2, p_3, p_4)~~~,&~~~ \vec{q} = (q_1, q_2, q_3, q_4)~~~,
\end{align}
where we must choose $\vec{u}$,  $\vec{v}$, $\vec{p}$, and $\vec{q}$ to be linearly independent. This is, as before, so that each trans-submultiplet has unique nodal field content.

As discussed in Sec.~\ref{s:intro}, \nmSG ~is composed of cSG plus a complex linear compensating superfield, $\Sigma$. We will now show how a certain solution in our parameter space exposes the cSG valise adinkra as a submultiplet of the {\nmSG} valise adinkra; a feature that was automatically realized for mSG in Sec.~\ref{s:MSG}.

Consider the choice: 
\begin{align}\label{e:nmParChoice}
   u_4 = u_5 = v_4 = v_5 = 0 &~~~\Rightarrow~~~ u_1 + u_2 + u_3 = v_1 + v_2 + v_3 = 1~, \\*
   q_1 = q_2 = q_3 = - \frac{q_4}{N} = 1 &~~~\Rightarrow~~~ q_5 = n^{-1} ~,\\*
   p_1 = p_2 = p_3 = 1~,~p_4 = 0 &~~~\Rightarrow~~~ p_5 = -3 ~~~,
\end{align}
that reduces the nodal field content from Eqs.~(\ref{e:nmSGPhi}) and~(\ref{e:nmSGPsi}) to
\begin{equation}
    \Phi =\left(
   \mbox{{\tiny $
   \begin{array}{c}
 W_0 \\
 P \\
 S \\
 n
   \left(\dot{h}+V_0\right)\\
   \hline
u_1 A_1
   +(u_2-u_3 )\dot{h}_{23} \\
 u_3 A_3 +
   (u_1-u_2)\dot{h}_{12} \\
 u_2 A_2 +(u_3-u_1 )\dot{h}_{31}
   \\
-u_1\dot{h}_{11} -u_2\dot{h}_{22} -u_3\dot{h}_{33}
   \\
   \hline
v_1 A_1
   +(v_2-v_3 )\dot{h}_{23}  \\
 v_3 A_3 +
   (v_1-v_2)\dot{h}_{12} \\
 v_2 A_2 +(v_3-v_1 )\dot{h}_{31}
   \\
-v_1\dot{h}_{11} -v_2\dot{h}_{22} -v_3\dot{h}_{33}
    \\
   \hline
\frac{W_0}{n}-A_0 \\
 V_2 \\
 -V_3 \\
 -V_1 \\
   \hline
 A_1+3 W_1 \\
 A_3+3 W_3 \\
 A_2+3 W_2 \\
-(3 n+1) \dot{h} -3 n V_0
\end{array}
$}}\right)
,~~~i \Psi =  \left( 
\mbox{{\tiny $
    \begin{array}{c}
 -\beta _3 \\
 -\beta _4 \\
 \beta _1 \\
 -\beta _2 \\
 \hline
 -u_1 \dot{\psi }_{13}+u_2
   \dot{\psi }_{21}+u_3 \dot{\psi }_{34} \\
-u_1 \dot{\psi }_{14}-u_2 \dot{\psi }_{22}-u_3 \dot{\psi
   }_{33}\\
-u_1 \dot{\psi }_{11}-u_2 \dot{\psi }_{23}+u_3 \dot{\psi
   }_{32}\\
 -u_1 \dot{\psi }_{12}+u_2
   \dot{\psi }_{24}-u_3 \dot{\psi }_{31}\\
 \hline
 -v_1 \dot{\psi }_{13}+v_2
   \dot{\psi }_{21}+v_3 \dot{\psi }_{34} \\
-v_1 \dot{\psi }_{14}-v_2 \dot{\psi }_{22}-v_3 \dot{\psi
   }_{33} \\
-v_1 \dot{\psi }_{11}-v_2 \dot{\psi }_{23}+v_3 \dot{\psi
   }_{32} \\
 -v_1 \dot{\psi }_{12}+v_2
   \dot{\psi }_{24}-v_3 \dot{\psi }_{31}\\
 \hline
 -\frac{\beta _3}{n}-\dot{\lambda }_4-\dot{\psi }_{13}+\dot{\psi
   }_{21}+\dot{\psi }_{34} \\
 -\frac{\beta _4}{n}+\dot{\lambda }_3+\dot{\psi }_{14}+\dot{\psi
   }_{22}+\dot{\psi }_{33} \\
 -\frac{\beta _1}{n}+\dot{\lambda }_2+\dot{\psi }_{11}+\dot{\psi
   }_{23}-\dot{\psi }_{32} \\
 -\frac{\beta _2}{n}-\dot{\lambda }_1-\dot{\psi }_{12}+\dot{\psi
   }_{24}-\dot{\psi }_{31} \\
 \hline
 3 \beta _3-\dot{\psi }_{13}+\dot{\psi }_{21}+\dot{\psi }_{34}
   \\
 -3 \beta _4-\dot{\psi }_{14}-\dot{\psi }_{22}-\dot{\psi }_{33}
   \\
 -3 \beta _1-\dot{\psi }_{11}-\dot{\psi }_{23}+\dot{\psi }_{32}
   \\
 3 \beta _2-\dot{\psi }_{12}+\dot{\psi }_{24}-\dot{\psi }_{31}
\end{array}
$}}
    \right)~~~.
\label{e:NMPhiPsi}
\end{equation}
Under the parameter choice, Eq.~(\ref{e:nmParChoice}), the submultiplet 
\begin{equation}
(\Phi_{5},\Phi_{6},\Phi_{7},\Phi_{8},\Phi_{9},\Phi_{10},\Phi_{11},\Phi_{12}| \Psi_{5},\Psi_{6},\Psi_{7},\Psi_{8},\Psi_{9},\Psi_{10},\Psi_{11},\Psi_{12})
\label{e:CSGsub}
\end{equation}
is \emph{identical} for \nmSG ~and mSG.  We shall identify this submultiplet with that of cSG in Sec.~\ref{s:CSG}. The ($12|12$) submultiplet parameterized by the other nodes of \nmSG ~compose an $n_c=1$, $n_t=2$ system, and we identify this as the complex linear compensator, per our discussion in Sec.~\ref{s:intro}. This is evidenced in that this submultiplet is spanned by one cis- and two trans-valise adinkras, the upper left and bottom rightmost two adinkras, respectively, in Fig.~\ref{f:NMAdinkra}. Indeed, the ($12|12$) complex linear superfield was found in Ref.~\cite{Gates:2011aa} to have the SUSY isomer numbers $n_c=1$, $n_t=2$. 

Table~\ref{t:0braneNM2} and the adinkra shown in Fig.~\ref{f:NMAdinkra} can be written as Eq.~(\ref{e:LRdef}) with $20 \times 20$ adinkra matrices, written in terms of the $SO(4)$ generators in Eq.~(\ref{eq:SO4Generators}), given by
\be\label{eq:NMSGValise}
\begin{array}{llll}
\textcolor{AdinkraGreen}{{\bm {\rm L}}_1 =  {\bm {\rm  I}}_5  \otimes {\bm {\rm I}}_4} & ,& \textcolor{AdinkraViolet}{{\bm {\rm L}}_2 = i {\bm {\rm I}}_5 \otimes {\bm {\rm \beta}}_3} &, \\
\textcolor{AdinkraOrange}{{\bm {\rm L}}_3 = i \left(\begin{array}{ccccc}
                        1 & 0 & 0 &0&0\\
                        0 & -1 & 0 &0&0\\
                        0 & 0 & -1 &0 &0\\
                        0 & 0 & 0 & -1 & 0\\
                        0 & 0 & 0 & 0 & -1
                   \end{array}\right) \otimes  {\bm {\rm \beta}}_2}&, &\textcolor{AdinkraRed}{{\bm {\rm L}}_4 = - i {\bm {\rm I}}_5 \otimes {\bm {\rm \beta}}_1} & .
\end{array}
\ee
The {\nmSG} adinkra matrices satisfy the orthogonal relationship, Eq.~(\ref{e:Rdef}) and the ${\cal GR}(20,4)$ garden algebra, Eq.~(\ref{e:GRdN}) with $d=20$.
The {\nmSG} chromocharacters, Eqs.~(\ref{eq:TraceConjecture}), are 
\be\label{eq:TraceNMSG}
\eqalign{
  \varphi^{(1)}_{\rm IJ} = & 20~ \delta_{{\rm IJ}} \cr
   \varphi^{(2)}_{\rm IJKL} = &\, 20 (\delta_{\rm IJ}\delta_{\rm KL} - \delta_{\rm IK}\delta_{\rm JL} + \delta_{\rm IL}\delta_{\rm JK} ) -  12\,  \,  \e_{\rm IJKL}
}\ee
and so comparing with Eqs.~(\ref{eq:TraceConjecture}) we see once again that {\nmSG} has the SUSY isomer numbers $n_c = 1$, $n_t = 4$.

\section{\texorpdfstring{4D}{4D}, \texorpdfstring{${\mathcal N} = 1$}{N=1} Conformal Supergravity}\label{s:CSG}
$~~~~$ The linearized theory of 4D, $\mathcal{N} =1 $ conformal supergravity (cSG) contains the real component fields of an axial vector \emph{gauge} field $A_\mu$, Majorana gravitino $\psi_{\mu a}$, and graviton $h_{\mu\nu}$. We will see in this Section that the adinkras for cSG are indeed submultiplets of both the mSG and {\nmSG} adinkras, with isomer numbers $n_c = 0$, $n_t = 2$.

\subsection{Transformation Laws}
$~~~~$ The transformation Laws for cSG are easily found by removing the fields  $S$ and $P$ from the mSG laws, ~Eq.~(\ref{e:DNMEasy}).
\begin{subequations}\label{eq:DConformal}
\begin{align}
   {\rm D}_a A_{\mu} =& i (\gamma^5 \gamma^\nu)_a^{~b} \partial_{[\nu} \psi_{\mu] b} -\frac{1}{2} \epsilon_{\mu}^{~\nu\alpha\beta}(\gamma_\nu)_{a}^{~b} \partial_\alpha \psi_{\beta b}  \\
   {\rm D}_a h_{\mu\nu} = & \frac{1}{2} (\gamma_{(\mu})_{|a|}^{~b}\psi_{\nu) b} \\
   {\rm D}_a \psi_{\mu b} =&   \frac{2}{3} (\gamma^5)_{ab} A_\mu +\frac{1}{6} (\gamma^5[\g_{\mu}, \g^{\nu}])_{ab}A_\nu  - \frac{i}{2}([\g^{\alpha} , \g^{\beta}])_{ab}\partial_\alpha h_{\beta\mu} 
   \end{align}
\end{subequations}
These are a symmetry of the cSG Lagrangian
\begin{align}
   {\mathcal L}_{cSG} = & \frac{1}{2} h^{\mu\nu} \square^2 h_{\mu\nu} - h^{\mu\nu}\square \partial_{\nu}\partial_{\alpha}h^{\alpha}_{~\mu} +\frac{1}{3} h^{\mu\nu}\partial_{\alpha}\partial_{\beta}\partial_{\mu}\partial_{\nu}h^{\alpha\beta} - \frac{1}{6} h \square^2 h +  \cr
   &+ \frac{1}{3} h \square \partial_{\mu}\partial_{\nu} h^{\mu\nu} -\frac{1}{6} F_{\mu\nu} F^{\mu\nu} -i\frac{1}{3}\psi_{\nu b} \partial_\mu\partial^\nu \partial^\beta (\gamma^\mu)^{bc} \psi_{\beta c} + \cr
   &+ i\frac{1}{3}\psi_{\nu b} (\gamma^\mu)^{bc} \square \partial_{\mu} \psi^\nu_{~c} -\frac{1}{6}\epsilon^{\nu\beta\sigma\mu}(\gamma^5 \gamma_{\sigma})^{bc} \psi_{\nu b} \square \partial_\mu\psi_{\beta c} 
   \label{e:LCSG}
\end{align}
where the canonical $U(1)$ field strength is
\begin{equation}
   F_{\mu\nu} = \partial_\mu A_\nu - \partial_\nu A_\mu ~~~.
\end{equation}
With a bit of work, it can be shown that the graviton part of the cSG Lagrangian is the square of the Weyl tensor 
\begin{align}
C_{\alpha\mu\beta\nu}C^{\alpha\mu\beta\nu} = & \left(R_{\alpha\mu\beta\n} - \frac{1}{2}(g_{\a[\b}R_{\n]\m} - \eta_{\m[\b}R_{\n]\a}) + \frac{1}{6} R g_{\a[\b}g_{\n]\m} \right) \cr
& \times \left(R^{\alpha\mu\beta\n} - \frac{1}{2}(g^{\a[\b}R^{\n]\m} - \eta^{\m[\b}R^{\n]\a}) + \frac{1}{6} R g^{\a[\b}g^{\n]\m} \right) \cr
 = & R_{\mu\nu\a\b}R^{\mu\n\a\b} - 2 R_{\mu\nu}R^{\mu\nu} + \frac{1}{3}R^2 \cr
=& \frac{1}{2} h^{\mu\nu} \square^2 h_{\mu\nu} - h^{\mu\nu}\square \partial_{\nu}\partial_{\alpha}h^{\alpha}_{~\mu} +\frac{1}{3} h^{\mu\nu}\partial_{\alpha}\partial_{\beta}\partial_{\mu}\partial_{\nu}h^{\alpha\beta} +  \cr
   & - \frac{1}{6} h \square^2 h + \frac{1}{3} h \square \partial_{\mu}\partial_{\nu} h^{\mu\nu}
\end{align}
in the linear limit $g_{\mu\nu} = \eta_{\mu\nu} + h_{\mu\nu}$ where indices are raised and lowered with the Minkowski metric $\eta_{\mu\nu}$. This is just as is expected for conformal gravity. The following linear expansions are useful in the preceding calculation
\begin{subequations}
\begin{align}
     R_{\a\m\b\n} = & \frac{1}{2} (\partial_{\m}\partial_{[\n} h_{\b]\a}- \partial_\a \partial_{[\n}  h_{\b] \m}) \\
     R_{\mu\nu} =& \frac{1}{2} \square h_{\mu\nu} + \frac{1}{2} \partial_\m \partial_\n h - \frac{1}{2} \partial_\a \partial_{(\n} h^{\a}{}_{\m)} \\
     R  = & \square h - \partial_{\mu} \partial_\n h^{\m\n} ~~~.
\end{align}
\end{subequations}

The cSG Lagrangian possesses the linear limit conformal symmetries
\begin{align}\label{e:CSymm}
   \delta h_{\mu\nu} = & B \eta_{\mu\nu} + \partial_{\mu} \Lambda_\nu + \partial_\nu \Lambda_\mu \cr
   \delta A_\mu = & \partial_\mu \rho \cr
   \delta \psi_{\mu a} = & \partial_\mu \epsilon_a + (\gamma_\mu)_{a}^{~b} \sigma_b~~~.
\end{align}
The cSG transformation laws satisfy the algebra
\begin{align}
   \{ {\rm D}_a , {\rm D}_b \} h_{\mu\nu} = & 2 i (\gamma^\alpha)_{ab} \partial_\alpha h_{\mu\nu} - i (\gamma^\alpha)_{ab}\partial_{(\mu} h_{\nu)\alpha}\cr
  \{ {\rm D}_a , {\rm D}_b \} A_{\mu} = & 2 i (\gamma^\nu)_{ab} \partial_\nu A_{\mu} \cr
   \{ {\rm D}_a , {\rm D}_b \} \psi_{\mu c} = & 2 i (\gamma^\alpha)_{ab} \partial_\alpha  \psi_{\mu c} - i \partial_\m \varphi_{abc}- i (\g_\m)_c^{~d} \s_{abd} ~~~
\end{align}
where $\varphi_{abc}$ is as before in both the minimal and non-minimal representations, Eq.~(\ref{e:varphi}), and another piece has arisen on the right hand side of the gravitino algebra
\begin{align}
     \s_{abd} = & \frac{1}{3}\left( (\g^{[\a})_{ab} (\g^{\b]})_{d}^{~e} + i \e^{\nu\r\a\b}(\g_\n)_{ab} (\g^5 \g_\r)_{d}^{~e}  \right) \partial_\a \psi_{\b e}~~~.
\end{align}
The term involving $\s_{abd}$ is proportional to $(\g_\m)_c{}^d$ and so is a consequence of the related symmetry of the cSG Lagrangian depicted in the last line of Eq.~(\ref{e:CSymm}). The auxiliary fields in both the minimal and non-minimal cases, serve to remove this term, which necessarily reduces the full conformal symmetry group to Poincar\'e.

\subsection{One-Dimensional Reduction}
$~~~~$ We use temporal gauge, Eq.~(\ref{e:TG}), as before for $h_{\mu\nu}$ and $\psi_{\mu a}$, but now also for the axial $U(1)$ vector
 \begin{align}\label{e:TGA}
   A_0 = & 0~~~.
\end{align}
 In this gauge, the Lagrangian~(\ref{e:LCSG}) reduced to the 0 brane becomes
 \begin{align}
    \mathcal{L}^{(0)}_{cSG}  =& \frac{1}{2} \sum_{i=1}^{3} \sum_{j=1}^3 \ddot{h}_{ij} \ddot{h}_{ij} - \frac{1}{6} \ddot{h}^2 - \frac{1}{3} \sum_{i=1}^3 \dot{A}_i \dot{A}_i + \frac{i}{3} \sum_{i=1}^3 \sum_{b = 1}^4 \dot{\psi}_{ib} \ddot{\psi}_{i b} + \cr
    & - \frac{i}{3} \left( \dot{\psi}_{11} \ddot{\psi}_{23} -\dot{\psi}_{13} \ddot{\psi}_{21} - \dot{\psi}_{12} \ddot{\psi}_{24} + \dot{\psi}_{14} \ddot{\psi}_{22} \right) + \cr
    & - \frac{i}{3} \left( \dot{\psi}_{31} \ddot{\psi}_{12} -\dot{\psi}_{32} \ddot{\psi}_{11} + \dot{\psi}_{33} \ddot{\psi}_{14} - \dot{\psi}_{34} \ddot{\psi}_{13} \right) +\cr
    & - \frac{i}{3} \left( \dot{\psi}_{21} \ddot{\psi}_{34} -\dot{\psi}_{24} \ddot{\psi}_{31} + \dot{\psi}_{22} \ddot{\psi}_{33} - \dot{\psi}_{23} \ddot{\psi}_{32} \right)~~~.
 \label{e:CSymm00}   
 \end{align}
Its symmetries, Eqs.~(\ref{e:CSymm}), become
 \begin{align}
   \delta h_{ij} = & B \delta_{ij}~~~,~~~\delta A_i = 0~~~,~~~ i,j = 1,2,3, \cr
   \delta \psi_{1 1} = & \sigma_2 ~~~,~~~\delta \psi_{1 2} =  \sigma_1 ~~~,~~~\delta \psi_{1 3} =  \sigma_4 ~~~,~~~\delta \psi_{1 4} =  \sigma_3 ~~~,  \cr
   \delta \psi_{2 1} = & -\sigma_4 ~~~,~~~\delta \psi_{2 2} =  \sigma_3 ~~~,~~~\delta \psi_{2 3} =  \sigma_2 ~~~,~~~\delta \psi_{2 4} =  -\sigma_1 ~~~, \cr
   \delta \psi_{3 1} = & \sigma_1 ~~~,~~~\delta \psi_{3 2} =  -\sigma_2 ~~~,~~~\delta \psi_{3 3} =  \sigma_3 ~~~,~~~\delta \psi_{3 4} =  -\sigma_4 ~~~,\label{e:CSymm0}
\end{align}
with the constraints
\begin{align}
   B =& 2 \dot{\Lambda}_0~~~,~~~\Lambda_i = \mbox{constant}~~~,~~~\rho = \mbox{constant} \cr
   \dot{\epsilon}_1 =& - \sigma_2 ~~~,~~~\dot{\epsilon}_2 =  \sigma_1 ~~~,~~~\dot{\epsilon}_3 =  \sigma_4 ~~~,~~~\dot{\epsilon}_4 =  -\sigma_3 ~~~,
\end{align}
that serve to maintain temporal gauge, Eqs.~(\ref{e:TG}) and~(\ref{e:TGA}).

The adinkranization proceeds precisely as in Sec.~\ref{s:MSG}, but now we set the compensator terms to zero, $S = P = 0$, and choose temporal gauge $A_0 = 0$. The cis-adinkra choice, Eq.~(\ref{e:mSGcis}), is then removed in reducing the mSG transformation laws to the cSG transformation laws. The node definitions for cSG are then the two trans-adinkra sets from mSG but here with components numbered top to bottom one through eight:
\begin{align}
\Phi  = & \left(
\mbox{{\tiny $
\begin{array}{c}
 u_1 A_1
   +(u_2-u_3 )\dot{h}_{23} \\
 u_3 A_3 +
   (u_1-u_2)\dot{h}_{12} \\
 u_2 A_2 +(u_3-u_1 )\dot{h}_{31}
   \\
-u_1\dot{h}_{11} -u_2\dot{h}_{22} -u_3\dot{h}_{33}
   \\
   \hline
v_1 A_1
   +(v_2-v_3 )\dot{h}_{23}  \\
 v_3 A_3 +
   (v_1-v_2)\dot{h}_{12} \\
 v_2 A_2 +(v_3-v_1 )\dot{h}_{31}
   \\
-v_1\dot{h}_{11} -v_2\dot{h}_{22} -v_3\dot{h}_{33}
\end{array}
$}}
\right)~~~,~~~i \Psi = \left(
\mbox{{\tiny $
\begin{array}{c}
 -u_1 \dot{\psi }_{13}+u_2
   \dot{\psi }_{21}+u_3 \dot{\psi }_{34} \\
-u_1 \dot{\psi }_{14}-u_2 \dot{\psi }_{22}-u_3 \dot{\psi
   }_{33}\\
-u_1 \dot{\psi }_{11}-u_2 \dot{\psi }_{23}+u_3 \dot{\psi
   }_{32}\\
 -u_1 \dot{\psi }_{12}+u_2
   \dot{\psi }_{24}-u_3 \dot{\psi }_{31}\\
 \hline
 -v_1 \dot{\psi }_{13}+v_2
   \dot{\psi }_{21}+v_3 \dot{\psi }_{34} \\
-v_1 \dot{\psi }_{14}-v_2 \dot{\psi }_{22}-v_3 \dot{\psi
   }_{33} \\
-v_1 \dot{\psi }_{11}-v_2 \dot{\psi }_{23}+v_3 \dot{\psi
   }_{32} \\
 -v_1 \dot{\psi }_{12}+v_2
   \dot{\psi }_{24}-v_3 \dot{\psi }_{31}
\end{array}
$}}
\right)~~~
\label{e:PhiPsiCS}
\end{align}
where again horizontal lines separate the nodes of the different adinkra pieces.
The node definitions~(\ref{e:PhiPsiCS}) have the 0-brane reduced cSG transformation laws shown in Tab.~\ref{t:0braneCS} and encoded succinctly in the valise adinkra for cSG in Fig.~\ref{f:CS}. 

\begin{table}[!h]
\centering
  \caption{Zero-brane reduced cSG transformation rules in the adinkraic representation in Eqs.~(\ref{e:PhiPsiCS}).}
  {\tiny 
 \begin{tabular}{r|rrrr|rr|rrrr|}
   & $\gD$ & $\vD$ & $\oD$ & $\rD$ &\hspace{50 pt}& & $\gD$ & $\vD$ & $\oD$ & $\rD$ \\
  \cline{1-5}\cline{7-11}$
 \Phi _1 $&$ i \Psi _1 $&$ i \Psi _2 $&$ -i \Psi _3 $&$ -i \Psi _4 $&\hspace{50 pt}&$ \Psi _1 $&$ \dot{\Phi }_1 $&$ -\dot{\Phi }_2 $&$ \dot{\Phi }_3 $&$ \dot{\Phi }_4 $\\$
 \Phi _2 $&$ i \Psi _2 $&$ -i \Psi _1 $&$ -i \Psi _4 $&$ i \Psi _3 $&\hspace{50 pt}&$ \Psi _2 $&$ \dot{\Phi }_2 $&$ \dot{\Phi }_1 $&$ \dot{\Phi }_4 $&$ -\dot{\Phi }_3 $\\$
 \Phi _3 $&$ i \Psi _3 $&$ -i \Psi _4 $&$ i \Psi _1 $&$ -i \Psi _2 $&\hspace{50 pt}&$ \Psi _3 $&$ \dot{\Phi }_3 $&$ \dot{\Phi }_4 $&$ -\dot{\Phi }_1 $&$ \dot{\Phi }_2 $\\$
 \Phi _4 $&$ i \Psi _4 $&$ i \Psi _3 $&$ i \Psi _2 $&$ i \Psi _1 $&\hspace{50 pt}&$ \Psi _4 $&$ \dot{\Phi }_4 $&$ -\dot{\Phi }_3 $&$ -\dot{\Phi }_2 $&$ -\dot{\Phi }_1 $\\
  \cline{1-5}\cline{7-11}$
 \Phi _5 $&$ i \Psi _5 $&$ i \Psi _{6} $&$ -i \Psi _{7} $&$ -i \Psi _{8} $&\hspace{50 pt}&$ \Psi _5 $&$ \dot{\Phi }_5 $&$ -\dot{\Phi }_{6} $&$ \dot{\Phi }_{7} $&$ \dot{\Phi }_{8} $\\$
 \Phi _{6} $&$ i \Psi _{6} $&$ -i \Psi _5 $&$ -i \Psi _{8} $&$ i \Psi _{7} $&\hspace{50 pt}&$ \Psi _{6} $&$ \dot{\Phi }_{6} $&$ \dot{\Phi }_5 $&$ \dot{\Phi }_{8} $&$ -\dot{\Phi }_{7} $\\$
 \Phi _{7} $&$ i \Psi _{7} $&$ -i \Psi _{8} $&$ i \Psi _5 $&$ -i \Psi _{6} $&\hspace{50 pt}&$ \Psi _{7} $&$ \dot{\Phi }_{7} $&$ \dot{\Phi }_{8} $&$ -\dot{\Phi }_5 $&$ \dot{\Phi }_{6} $\\$
 \Phi _{8} $&$ i \Psi _{8} $&$ i \Psi _{7} $&$ i \Psi _{6} $&$ i \Psi _5 $&\hspace{50 pt}&$ \Psi _{8} $&$ \dot{\Phi }_{8} $&$ -\dot{\Phi }_{7} $&$ -\dot{\Phi }_{6} $&$ -\dot{\Phi }_5$\\
 \cline{1-5}\cline{7-11}
  \end{tabular}}
  \label{t:0braneCS}
\end{table}

\begin{figure}[!ht]
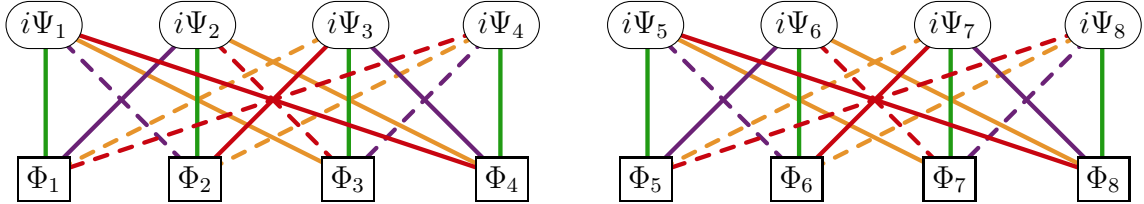
\setlength{\unitlength}{.8 mm}
\begin{center}
   \begin{picture}(200,40)(0,0)
\put(0,0){\includegraphics[width = 90\unitlength]{TransValise}}
\put(2.5,7){\fcolorbox{black}{white}{$\Phi_1$}}
\put(27.6,7){\fcolorbox{black}{white}{$\Phi_2$}}
\put(53,7){\fcolorbox{black}{white}{$\Phi_3$}}
\put(78.6,7){\fcolorbox{black}{white}{$\Phi_4$}}
\put(0.5,30){\begin{tikzpicture}
 \node[rounded rectangle,draw,fill=white!30]{$i\Psi_1$};
 \end{tikzpicture}}
\put(26,30){\begin{tikzpicture}
 \node[rounded rectangle,draw,fill=white!30]{$i\Psi_2$};
 \end{tikzpicture}}
\put(51.5,30){\begin{tikzpicture}
 \node[rounded rectangle,draw,fill=white!30]{$i\Psi_3$};
 \end{tikzpicture}}
\put(76.5,30){\begin{tikzpicture}
 \node[rounded rectangle,draw,fill=white!30]{$i\Psi_4$};
 \end{tikzpicture}}
\quad
\put(100,0){\includegraphics[width = 90\unitlength]{TransValise}}
\put(102.5,7){\fcolorbox{black}{white}{$\Phi_5$}}
\put(127.6,7){\fcolorbox{black}{white}{$\Phi_6$}}
\put(153,7){\fcolorbox{black}{white}{$\Phi_7$}}
\put(178.6,7){\fcolorbox{black}{white}{$\Phi_8$}}
\put(100.5,30){\begin{tikzpicture}
 \node[rounded rectangle,draw,fill=white!30]{$i\Psi_5$};
 \end{tikzpicture}}
\put(126,30){\begin{tikzpicture}
 \node[rounded rectangle,draw,fill=white!30]{$i\Psi_6$};
 \end{tikzpicture}}
\put(151.5,30){\begin{tikzpicture}
 \node[rounded rectangle,draw,fill=white!30]{$i\Psi_7$};
 \end{tikzpicture}}
\put(176.5,30){\begin{tikzpicture}
 \node[rounded rectangle,draw,fill=white!30]{$i\Psi_8$};
 \end{tikzpicture}}
   \end{picture}
   \end{center}
   \vspace{-20 pt}
   \caption{The cSG valise adinkra. It has the exact same nodal content as the two mSG trans-adinkra pieces in Fig.~\ref{f:SGAdinkra} and two of the four {\nmSG} trans-adinkra pieces in Fig.~\ref{f:NMAdinkra} under the parameter choice~(\ref{e:nmParChoice}). The cSG SUSY isomer numbers are therefore $n_c = 0$, $n_t = 2$. The engineering dimensions of all bosons are the same and the engineering dimensions of all fermions are the same.}
\label{f:CS}
\end{figure} 

\noindent We conclude that cSG has SUSY isomer numbers
 \begin{align}
  n_c = 0~~~,~~~n_t = 2~~~.
   \label{e:cSGenanti}
\end{align}
The parameters in Eqs.~(\ref{e:PhiPsiCS}) are constrained as in Eq.~(\ref{e:uconstraint}).
These node definitions have all the 0-brane symmetries of the cSG Lagrangian, Eq.~(\ref{e:CSymm0}). Note that the nodes in the cSG valise adinkra are precisely the submultiplet, (\ref{e:CSGsub}), that showed up in the mSG and the \nmSG ~under the parameter choice Eq.~(\ref{e:nmParChoice}). From the node definitions in Eqs.~(\ref{e:PhiPsiCS}), it would appear at first glance that there are more degrees of freedom in the original fields than the number of nodes. This is not the case precisely because the nodes are invariant with respect to the 0-brane symmetries, Eq.~(\ref{e:CSymm0}). These residual symmetries remove one more degree of freedom from the graviton, leaving it with a total of five, and four more from the gravitino, leaving it with eight. Adding the three degrees of freedom from the completely gauge fixed $A_{i}$ field brings the degrees of freedom to $(8|8)$ once all gauge degrees of freedom encoded in Eq.~(\ref{e:CSymm}) have been removed.

The cSG adinkra matrices can be read off either Tab.~\ref{t:0braneCS} or Fig.~\ref{f:CS}
\be\label{eq:CSGValise}
\begin{array}{llll}
\textcolor{AdinkraGreen}{{\bm {\rm L}}_1 =  {\bm {\rm  I}}_2  \otimes {\bm {\rm I}}_4} & ,& \textcolor{AdinkraViolet}{{\bm {\rm L}}_2 = i {\bm {\rm I}}_2 \otimes {\bm {\rm \beta}}_3} &, \\
\textcolor{AdinkraOrange}{{\bm {\rm L}}_3 = -i {\bm {\rm I}}_2 \otimes  {\bm {\rm \beta}}_2}&, &\textcolor{AdinkraRed}{{\bm {\rm L}}_4 = - i {\bm {\rm I}}_2 \otimes {\bm {\rm \beta}}_1} & .
\end{array}
\ee
These satisfy the orthogonality relationship, Eq.~(\ref{e:Rdef}), and the ${\mathcal G}{\mathcal R}(8,4)$ algebra, Eq.~(\ref{e:GRdN}) with  $d=8$.
The chromocharacters, Eqs.~(\ref{eq:TraceConjecture}), for cSG are 
\be\label{eq:TraceCSG}
\eqalign{
   \varphi^{(1)}_{\rm IJ} = & 8~ \delta_{{\rm IJ}} \cr
   \varphi^{(2)}_{\rm IJKL} = &\, 8 (\delta_{\rm IJ}\delta_{\rm KL} - \delta_{\rm IK}\delta_{\rm JL} + \delta_{\rm IL}\delta_{\rm JK} ) -  8\,  \,  \e_{\rm IJKL}~~~.
}\ee
Comparing with our chromocharacter formulas in Eqs.~(\ref{eq:TraceConjecture}), once again we see that the cSG SUSY isomer numbers are $n_c = 0$, $n_t = 2$.
\subsection{The View From The Conformal Perspective}\label{s:cSGsynth}
$~~~~$ We saw in Sec.~\ref{s:NM1D} that lines five through 12 of the bosons and fermions in both the mSG nodes (\ref{e:mSGPhiPsi}) and the \nmSG ~nodes (\ref{e:NMPhiPsi}) all correspond 
to the same structure.  Let us rename these structures according to the definitions
\begin{align}
{\cal H}_{i \, \Phi}^{(V)}  = & \left(
\mbox{{\tiny $
\begin{array}{c}
 u_1 A_1
   +(u_2-u_3 )\dot{h}_{23} \\
 u_3 A_3 +
   (u_1-u_2)\dot{h}_{12} \\
 u_2 A_2 +(u_3-u_1 )\dot{h}_{31}
   \\
-u_1\dot{h}_{11} -u_2\dot{h}_{22} -u_3\dot{h}_{33}
   \\
   \hline
v_1 A_1
   +(v_2-v_3 )\dot{h}_{23}  \\
 v_3 A_3 +
   (v_1-v_2)\dot{h}_{12} \\
 v_2 A_2 +(v_3-v_1 )\dot{h}_{31}
   \\
-v_1\dot{h}_{11} -v_2\dot{h}_{22} -v_3\dot{h}_{33}
\end{array}
$}}
\right)~~~,~~~i{\cal H}_{{\hat{j}} \, \Psi}^{(V)} = \left(
\mbox{{\tiny $
\begin{array}{c}
 -u_1 \dot{\psi }_{13}+u_2
   \dot{\psi }_{21}+u_3 \dot{\psi }_{34} \\
-u_1 \dot{\psi }_{14}-u_2 \dot{\psi }_{22}-u_3 \dot{\psi
   }_{33}\\
-u_1 \dot{\psi }_{11}-u_2 \dot{\psi }_{23}+u_3 \dot{\psi
   }_{32}\\
 -u_1 \dot{\psi }_{12}+u_2
   \dot{\psi }_{24}-u_3 \dot{\psi }_{31}\\
 \hline
 -v_1 \dot{\psi }_{13}+v_2
   \dot{\psi }_{21}+v_3 \dot{\psi }_{34} \\
-v_1 \dot{\psi }_{14}-v_2 \dot{\psi }_{22}-v_3 \dot{\psi
   }_{33} \\
-v_1 \dot{\psi }_{11}-v_2 \dot{\psi }_{23}+v_3 \dot{\psi
   }_{32} \\
 -v_1 \dot{\psi }_{12}+v_2
   \dot{\psi }_{24}-v_3 \dot{\psi }_{31}
\end{array}
$}}
\right)
\label{e:PhiPsiCSz}
\end{align}
that are still subject to the constraint Eq.~(\ref{e:uconstraint}) 
and re-express the supersymmetrical D-algebra in the form of two equations
\be\label{eq:cSGadnkVaL}
   { \rm D}_{\rm I}  {\cal H}{}_{i \, \Phi}^{(V)}= i ( {\rm L }_{\rm I})_{i\hat{j}}
  {\cal H}_{{\hat{j}} \, \Psi}^{(V)} 
 ~~,~~~{ \rm D}_{\rm I} {\cal H}{}_{{\hat{j}} \, \Psi}^{(V)}  = 
 ({\rm R }_{\rm I})_{\hat{j}i}  {{d~} \over {dt}}  {\cal H}{}_{i \, \Phi}^{(V)}
\ee
where the L-matrices and corresponding R-matrices are defined by Eq.~(\ref{eq:CSGValise}).  This particular valise describes 4D, $\cal N$ = 1 conformal
supergravity.  The fact that it universally occurs in each of the supergravity
formulation is equivalent to two well known facts in other approaches:
\begin{enumerate}\renewcommand{\theenumi}{\alph{enumi}}
  \item In the superspace approach of \cite{Siegel:1978mj} there was presented a
description of supergravity in terms of unconstrained `prepotential'
superfields.  One of the features of the construction was to
show that in fact, all off-shell versions of SG theory when
written in terms of unconstrained superfields can be split
into a `conformal prepotential superfield' and a `conformal
compensator superfield.'  Although the former is unique, the
latter was shown not to be.  This possibility of different
choices for the conformal compensator accounts for the different
auxiliary field structures that can occur in off-shell descriptions.
This work was also the first to show that even in the confines
of a Poincar\' e theory, there is an important role for the
symmetries of a conformal theory.
\item In the conformal component-level approach of \cite{Bergshoeff:1980is} which
began after the work of Ref.~\cite{Siegel:1978mj}, once more one sees that there is
a sub-multiplet in all off-shell supergravity theories that
consists solely of the fields required to describe conformal
supergravity.  In this approach the component fields that appear
over and above these arise (as they do in the superfield approach)
as the result of the breaking of conformal symmetry.
\end{enumerate}

The task of understanding how the spacetime superconformal group
is embedded with the approach of an adinkra-based formulation
is an important task to be carried out in future research along
these lines.

\section{Synthesis}\label{s:synth}
$~~~~$ In this section we synthesize the main results of the paper into a cohesive framework, including relations to the previous two Refs.~\cite{Gates:2009me,Gates:2011aa}. 
Consider the super-character \cite{Kinney:2005ej}
\begin{align}
\chi_\rho (t, a, b) = \mathrm{Tr}_\rho \left[ (-1)^F t^{2\Delta} \mathrm e^{i a M_{01} } \mathrm e^{i b M_{23} }\right] .
\end{align}
Here $F$ is the fermion number, $M_{01}$ and $M_{23}$ are Lorentz generators, and $\Delta$ computes the engineering dimension of the field in the representation $\rho$ normalized so that a physical spinor has dimension $\Delta = \frac32$. In all cases of interest to us, $\Delta$ is an integer iff $F=0$ so that we may combine the first two factors to give
\begin{align}
\chi_\rho (t, a, b) = \mathrm{Tr}_\rho \left[ (-t)^{2\Delta} \mathrm e^{i a M_{01} } \mathrm e^{i b M_{23} }\right] .
\end{align}
If we are uninterested in the dimension of the fields in the representation $\rho$ and if, furthermore, we forget about the grading by fermion number, we may set $t=-1$ and study the resulting $Spin(3,1)$ character instead. This quantity is related by Wick rotation to the chromocharacters, Eq.~(\ref{eq:TraceConjecture}).\footnote{An advantage of this compactification is that the representations are replaced with finite-dimensional analogues which possess no gauge freedom.} That is, we consider instead of $\chi$, a ``twisted'' version $\tilde \chi$ on $Spin(4)_R$
\begin{align}
    \tilde{\chi}_\rho = & \mathrm{Tr}_\rho \left[ (-t)^{2\Delta} \mathrm e^{i a M_{12}} \mathrm e^{i b M_{34} }\right]  \\
    \label{e:M}
   ( M_{{\rm I}{\rm J}})_{i}{}^{j} = & \frac{i}{4} \left [ ({\rm L}_{{\rm I}})_{i}{}^{\hat{k}} ({\rm R}_{{\rm J}})_{\hat{k}}{}^{j} - ({\rm L}_{{\rm J}})_{i}{}^{\hat{k}} ({\rm R}_{{\rm I}})_{\hat{k}}{}^{j} \right]~~~.
\end{align} 
Notice that in our definition of `twisted,' the usual Lorentz generators are replaced by the matrices defined in Eq.~(\ref{e:M}) that are not Lorentz generators, but we will use them as if they are. Then, for example, any garden algebra satisfying the orthogonality relation~(\ref{e:Rdef}) and the chromocharacter formulas~(\ref{eq:TraceConjecture}) will have
\begin{align}
\left.  {\partial \over \partial a} {\partial \over \partial b} \tilde \chi \right|_{a,b=0} &= 4(n_c - n_t) \cr
\left. \left({\partial \over \partial a}\right)^2 \tilde \chi \right|_{a,b=0} =  \left. \left( {\partial \over \partial b}\right)^2 \tilde \chi \right|_{a,b=0} &= -4(n_c + n_t) 
\end{align}
and we recover the isomer numbers. 
In this way, ``adinkranization'' may be thought of as an algorithm for computing the characters of the (Wick-rotated) super-Poincar\'e group. 

More explicitly, the calculation of the twisted character for the chiral (cis), real-linear (trans), real-unconstrained ($\mathbb{R}$), complex-unconstrained ($\mathbb{C}$),\footnote{The complex unconstrained adinkras were not explicitly reported in~\cite{Gates:2011aa}, however this is clearly just two copies of the real unconstrained superfield. The adinkra matrices are therefore two block diagonal copies and all traces would simply double.} cSG, mSG and complex linear superfield ($\Sigma$), and \nmSG ~are
\begin{subequations}\label{e:twistedchars}
\begin{align}
\label{e:twistedcharscis}
\tilde{\chi}_\mathrm{cis}(a,b)&=4 \cos\left(\frac{a}{2}\right) \cos \left(\frac{b}{2}\right) + 4 \sin 
	\left(\frac{a}{2}\right) \sin \left(\frac{b}{2}\right)\\
\tilde{\chi}_\mathrm{trans}(a,b)&=4 \cos \left(\frac{a}{2}\right) \cos \left(\frac{b}{2}\right)-4 \sin
   \left(\frac{a}{2}\right) \sin \left(\frac{b}{2}\right)\\
\tilde{\chi}_\mathbb{R}(a,b)&=8 \cos \left(\frac{a}{2}\right) \cos \left(\frac{b}{2}\right)\\
\tilde{\chi}_\mathbb{C}(a,b)&=16 \cos \left(\frac{a}{2}\right) \cos \left(\frac{b}{2}\right)\\
\tilde{\chi}_\mathrm{cSG}(a,b)&=8 \cos \left(\frac{a}{2}\right) \cos \left(\frac{b}{2}\right)-8 \sin
   \left(\frac{a}{2}\right) \sin \left(\frac{b}{2}\right)\\
\tilde{\chi}_\mathrm{mSG}(a,b) = \tilde{\chi}_\Sigma (a,b) &= 12 \cos \left(\frac{a}{2}\right) \cos \left(\frac{b}{2}\right)-4 \sin
   \left(\frac{a}{2}\right) \sin \left(\frac{b}{2}\right)\\
\tilde{\chi}_{\not\mathrm{mSG}}(a,b)&=20 \cos \left(\frac{a}{2}\right) \cos \left(\frac{b}{2}\right)-12 \sin
   \left(\frac{a}{2}\right) \sin \left(\frac{b}{2}\right).
\end{align}
\end{subequations}
In fact, these representations all fit into the formula:
\begin{align}
     \tilde{\chi}(a,b) = & 4 (n_c + n_t) \cos \left(\frac{a}{2} \right)\cos \left(\frac{b}{2} \right) + 4 (n_c - n_t) \sin \left(\frac{a}{2} \right)\sin \left(\frac{b}{2} \right) ~~~.
\end{align}
By taking partial derivatives, we clearly find the correct $(n_c,n_t)$ numbers.
By way of comparison, the characters for the 2-component (anti-)Weyl, the analogue of the (cis-) trans-representation, and 4-component Dirac spinor representations of $Spin(3,1)$ are
\begin{subequations}
\begin{align}
\chi_\mathrm{Weyl}(a,b)&= 2 \cosh\left(\frac{a}{2}\right) \cos \left(\frac{b}{2}\right)
	+2i\sinh\left(\frac{a}{2}\right) \sin \left(\frac{b}{2}\right)\\
\chi_\mathrm{\overline{Weyl}}(a,b)&= 2 \cosh\left(\frac{a}{2}\right) \cos \left(\frac{b}{2}\right)
	-2i\sinh\left(\frac{a}{2}\right) \sin \left(\frac{b}{2}\right)\\
\chi_\mathrm{Dirac}(a,b)&= 4 \cosh\left(\frac{a}{2}\right) \cos \left(\frac{b}{2}\right).
\end{align}
\end{subequations}

Noting from Ref.~\cite{Gates:2009me} that the chiral multiplet is in the cis-representation, we see that the characters~(\ref{e:twistedchars}) obey the superfield equation

\begin{align}
\textrm{complex unconstrained} = \textrm{chiral}+\textrm{complex linear}.
\end{align}
Similarly, we comment that the results for the gauge-fixed vector multiplet and tensor multiplet given in the literature are consistent with their identification as real-linear multiplets. For example, the characters~(\ref{e:twistedchars}) obey the superfield equation
\begin{align}
\textrm{gauge-fixed vector multiplet} &= \textrm{real unconstrained}-\left(\textrm{chiral}+\overline{\textrm{chiral}}\right)\cr
	&=\textrm{real-linear}
\cr
&= \textrm{tensor multiplet} 
.
\end{align}
It is important to note that in this equation $\left(\textrm{chiral}+\overline{\textrm{chiral}}\right)$ is the real part of the chiral superfield, which can be identified with \emph{one} cis-representation, Eq.~(\ref{e:twistedcharscis}), as these are all real representations.

We represent all of this in Tab.~\ref{t:tc1}.
\begin{table}[!h]
\centering
\caption{Twisted characters of composite superfields are the sum of the twisted characters of their base superfields.}
$\begin{array}{c|ccccc}
	&\mathrm{\chi ral}~(4|4)
	&\mathbb R\textrm{-lin.}~(4|4) 
		&\mathbb R~(8|8) 
	&\mathbb C\textrm{-lin.}~(12|12)
	&\mathbb C~(16|16)
	\\
\hline
n_c+n_t	&1
	&1
		&2		
	&3	
	&4	
	\\
n_c-n_t	&1
	&-1
		&0 	 	
	&-1	
	&0	
\end{array}$
\label{t:tc1}
\end{table}

\noindent The super-dimensions of these representations are indicated as $(d_b|d_f)$.

Just as there are two inequivalent $(4|4)$-dimensional representations (chiral and real-linear), there is a second $(8|8)$-dimensional one: conformal supergravity.
Together with the equations
\begin{align}
\textrm{minimal SG} &= \textrm{conformal SG}+\textrm{chiral compensator}\cr
\textrm{non-minimal SG} &= \textrm{conformal SG}+\textrm{complex-linear compensator}
\end{align}
we obtain Tab.~\ref{t:tc2}.
\begin{table}[!h]
\centering
\caption{The twisted characters for the compensator of a SUGRA representation can be recovered by subtracting off the base cSG twisted characters.}
$\begin{array}{c|cccccccc}
&\mathrm{cSG}~(8|8)
&\mathrm{mSG}~(12|12) 
&\not\!\!{\mathrm{m}}\mathrm{SG}~(20|20) \\
\hline
n_c + n_t	&2	&3	&5\\
n_c - n_t	&-2 	&-1	&-3
\end{array}$
\label{t:tc2}
\end{table}

\noindent This completes the list of all known off-shell 4D, $\mathcal N=1$ representations whose adinkras have been been reported to date~\cite{Gates:2009me,Gates:2011aa}.

\section{Conclusion}\label{s:conc}
$~~~~$ In this paper, the SUSY isomer numbers and adinkras for mSG, {\nmSG}, and cSG were explicitly derived. It was found that these numbers compose the characters of the $Spin(4)_R$ that remains after reduction to the 0-brane, and that adinkras are graphical depictions of these characters. The base multiplet for supergravity is cSG, with SUSY isomer numbers $(n_c,n_t) = (0,2)$. The mSG multiplet is cSG plus a chiral compensator superfield with SUSY isomer numbers $(n_c,n_t) = (1,0)$. We indeed found the isomer numbers for mSG to be the sum of the isomer numbers for cSG and the chiral superfield: $(n_c,n_t) = (0,2) + (1,0) = (1,2)$. The same additive character behavior holds true for \nmSG, which is cSG added to a complex linear compensator superfield, and all other multiplets investigated so far~\cite{Gates:2009me,Gates:2011aa}. This additive behavior was easily seen in the graphical adinkra depictions of these multiplets.

We also unveiled a simple procedure for finding SUSY isomer numbers. This utilizes the cis- and trans-valise adinkra pictures and the 0-brane transformation laws for the multiplet whose isomer numbers are sought. The procedure is to force the 0-brane transformation laws to fit into either the cis- or trans-valise adinkra, leading to constraints on the possible linear combinations of the fields in the multiplet that define the nodes in the irreducible adinkras. Interestingly, the SUSY isomer numbers for \emph{all} representations studied to date~\footnote{In this discussion, we do not consider the complex unconstrained superfield as it is \emph{exactly} the sum of a chiral and complex linear superfield.} fall within the four cases shown in Tab.~\ref{t:MConjecture}, where $k = n_c + n_t = d/4$ is the total number of adinkras in a representation with $d$ bosons and $d$ fermions.
 \begin{table}[!h]
\centering
\caption{The four isomer number cases for 4D, $\mathcal{N}=1$ off-shell SUSY that have been found to date. The number of adinkras is $k =d/4$ for representations with $d$ bosons and $d$ fermions. }
\begin{tabular}{r|c|c|c|c}
 	case	 & I & II & III & IV \\
 	\hline 
 	$n_c$ & 0 & 1 & $k$ & $k-1$ \\
 	\hline
 	$n_t$ & $k$ &  $k-1$ & 0 & 1
\end{tabular}\label{t:MConjecture}
 \end{table}
Table~\ref{t:MConjecture} is the evidence for a selection rule that any adinkras with SUSY isomer numbers other than these cannot extend to a 4D, $\mathcal{N}=1$ representation. It will be interesting to see if this selection rule holds as we continue our investigation of \emph{all} 4D, $\mathcal{N}=1$ off-shell representations. Selection rules such as these are what we look for in our quest to study off-shell representation theory with adinkras, with the ultimate goal being to use such selection rules to find \emph{new} off-shell representations.

From the results of this paper, we see that cSG satisfies case I and mSG and \nmSG ~both satisfy case II. The three irreducible 4D, $\mathcal{N}=1$ off-shell cases each have $k=1$ and so satisfy two cases simultaneously. These are the chiral (II = III), vector (I = IV), and tensor (I = IV) multiplets.  The 4D, $\mathcal{N}=1$ off-shell cases investigated in Ref.~\cite{Gates:2011aa}, the real scalar superfield and complex linear superfield, both satisfy case II, though the real scalar superfield simultaneously satisfies case~IV since $k=2$. For $k>2$, cases I, II, III, and IV are all distinct and it will be interesting to see the pattern that emerges as we investigate higher superhelicity systems with larger $k$.

Ongoing work that will be part of a future publication has already produced case~I for the $k=3$ \newSG ~and case~IV for the $k=5$, 4D, $\mathcal{N}=1$ gravitino-matter multiplet. The gravitino-matter multiplet is therefore the only known $k>2$ instance of \emph{either} cases III or~IV for which the larger number is the cis number. The roles of cis and trans between the gravitino-matter multiplet and all other known $k>2$ multiplets are reversed!

 Moving forward, we plan to continue to investigate adinkranization of higher superhelicity off-shell systems and systems with extended supersymmetry. If the selection rule evident from Tab.~\ref{t:MConjecture} holds for increasing $k$, one of the isomer numbers will remain at zero or one and the other will increase. Now the questions seem to be, is it the cis or the trans number that will increase for these higher $d$ systems, what is the pattern, and what new information is encoded by these characters?

\vspace{.05in}
 \begin{center}
 \parbox{5in}{{\it ``Our nation must come together to unite.''}\,\,-\,\, George W. Bush}
 \end{center}  
 
\section*{Acknowledgements}
$~~~~$ 
This research has been supported in part by NSF Grant PHY-09-68854, the J.~S. Toll 
Professorship endowment and the UMCP Center for String \& Particle Theory.  
WDL3 is supported by {\sc fondecyt} grant number 11100425 and gratefully acknowledges the University of Maryland Physics Department and the CSPT for hospitality.
KS thanks V.G.J. Rodgers for laying the groundwork 
for the indispensable \emph{Mathematica} code used throughout this work and all 
genomics works of which KS has been a part.   Additional acknowledgement is given 
by the undergraduate students (JP, SR, and AR) for the hospitality  of the University of 
Maryland and in particular of the CSPT, as well as recognition for their participation 
in 2012 SSTPRS (Student Summer Theoretical Physics Research Session).  We also 
thank Keith Burghardt, Konstantinos Koutrolikos, Thomas Rimlinger, and Mathew Calkins
for discussions. KS would like to thank N. De Leon and L. Wozniewski for discussions on chemistry analogies. Most adinkras were drawn with the aid of \emph{Adinkramat} \copyright ~2008 by G. Landweber.

\appendix
\section{Definitions and Conventions}\label{a:conv}
$~~~~$  We will use the real representation of the $\gamma$ matrices as in Refs.~\cite{Gates:2009me, Gates:2011aa}
\begin{align}
  (\gamma^0)_{a}^{~b} = &\left(
\begin{array}{cccc}
 0 & 1 & 0 & 0 \\
 -1 & 0 & 0 & 0 \\
 0 & 0 & 0 & -1 \\
 0 & 0 & 1 & 0
\end{array}
\right) ~~~,~~~(\gamma^1)_a^{~b} = \left(
\begin{array}{cccc}
 0 & 1 & 0 & 0 \\
 1 & 0 & 0 & 0 \\
 0 & 0 & 0 & 1 \\
 0 & 0 & 1 & 0
\end{array}
\right)\cr
  (\gamma^2)_{a}^{~b} = & \left(
\begin{array}{cccc}
 0 & 0 & 0 & -1 \\
 0 & 0 & 1 & 0 \\
 0 & 1 & 0 & 0 \\
 -1 & 0 & 0 & 0
\end{array}
\right)~~~,~~~(\gamma^3)_a^{~b} = \left(
\begin{array}{cccc}
 1 & 0 & 0 & 0 \\
 0 & -1 & 0 & 0 \\
 0 & 0 & 1 & 0 \\
 0 & 0 & 0 & -1
\end{array}
\right)\\
  (\gamma^5)_{a}^{~b} \equiv & i (\gamma^0 \gamma^1 \gamma^2 \gamma^3)_{a}^{~b} = \left(
\begin{array}{cccc}
 0 & 0 & 0 & i \\
 0 & 0 & -i & 0 \\
 0 & i & 0 & 0 \\
 -i & 0 & 0 & 0
\end{array}
\right)~~~.
\end{align} 
We also use the following conventions for the totally antisymmetric Levi-Civita tensor and the $SO(1,3)$ generators for spinors
\begin{align}
 \epsilon_{0123} = & -\epsilon^{0123} = 1~~~ \mbox{and totally anti-symmetric}, \cr
   \sigma^{\mu\nu} \equiv & \frac{i}{2}(\gamma^\mu\gamma^\nu - \gamma^\nu\gamma^\mu)~~~.
\end{align}
Einstein summation convention is assumed throughout, for example
\begin{align}
   A^{\mu}A_{\mu} \equiv \sum_{\mu=0}^3 A^{\mu}A_{\mu} = A^0A_0 + A^1A_1 + A^2A_2 + A^3A_3~~~.
\end{align}
All lower case Greek indices $\mu,\nu,\alpha,\beta,\dots$ are space-time indices and are raised and lowered with the Minkowski metric
\begin{equation}
   \eta_{\mu\nu} = \eta_{\nu\mu} \left( \begin{array}{cccc}
       -1 & 0 & 0 & 0 \\
       0 & 1 & 0 & 0 \\
       0 & 0 & 1 & 0 \\
       0 & 0 & 0 & 1
   \end{array}\right)~~~,~~~\eta^{\mu\nu} = \eta^{\nu\mu} = \left( \begin{array}{cccc}
       -1 & 0 & 0 & 0 \\
       0 & 1 & 0 & 0 \\
       0 & 0 & 1 & 0 \\
       0 & 0 & 0 & 1
   \end{array} \right)~~~,
\end{equation}
as
\begin{align}
A_\mu = \eta_{\mu\nu}A^{\nu}\ ,\ A^{\mu} = \eta^{\mu\nu}A_{\nu}~~~.
\end{align}
Lower case Latin $a,b,c,\dots$ are fermionic indices and are raised and lowered with the spinor metric 
\begin{align}
 C_{ab} =& - C_{ba} = \left(\begin{array}{cccc}
          0 & -1 & 0 & 0 \\
          1 & 0 & 0 & 0 \\
          0 & 0 & 0 & 1 \\
          0 & 0 & -1 & 0
 \end{array}
 \right)~~,~~\ C^{ab} = - C^{ba} =\left(\begin{array}{cccc}
          0 & -1 & 0 & 0 \\
          1 & 0 & 0 & 0 \\
          0 & 0 & 0 & 1 \\
          0 & 0 & -1 & 0
 \end{array}
 \right)
\end{align}
according to the \emph{northwest-southeast} rules:
\begin{align}
   \psi^a = C^{ab}\psi_b \ ,\ \psi_a = \psi^b C_{ba}~~~.
\end{align}
Symmetrization and anti-symmetrization are defined as follows without any normalization:
\begin{align}
   (\g_{(\m})_a{}^b \psi_{ \n ) b} \equiv (\g_{\m})_a{}^b \psi_{ \n  b} + (\g_{\n})_a{}^b \psi_{ \m b}~~~,~~~(\gamma_{[\mu})_{a}^{~b} \psi_{\nu] b} \equiv (\gamma_{\mu})_{a}^{~b} \psi_{\nu b} - (\gamma_{\nu})_{a}^{~b} \psi_{\mu b}~~~.
\end{align}
We use the following conventions for the Riemann and Ricci tensors, Ricci scalar, and Christoffel symbols:
\begin{subequations}
\begin{align}
   R^{\a}{}_{\m\b\n} = & \partial_\n \G^{\a}{}_{\m\b} - \partial_\b \G^{\a}{}_{\m\n} + \G^\r{}_{\m\b} \G^{\a}_{\n\r} - \G^{\r}{}_{\m\n} \G^\a{}_{\b\r} \\
   R_{\mu\nu} = R^{\a}{}_{\m\a\n} =&  \partial_\nu\Gamma^{\alpha}_{~\mu\alpha}-\partial_\alpha\Gamma^{\alpha}_{~\mu\nu}  + \Gamma^{\alpha}_{~\mu\beta}\Gamma^\beta_{\nu\alpha}-\Gamma^\alpha_{~\mu\nu}\Gamma^\beta_{~\alpha\beta} \\
   R = & g^{\mu\nu}R_{\m\n} \\
   \Gamma^\mu_{~\alpha\beta} =& \frac{1}{2}g^{\mu\nu}(\partial_{(\beta}g_{\alpha)\nu}  - \partial_\nu g_{\alpha\beta})~~~.
\end{align}
\end{subequations}
We linearize with
\begin{equation}
   g_{\mu\nu} =  \eta_{\mu\nu} + h_{\mu\nu}~~~,~~~g^{\mu\nu} = \eta^{\mu\nu} - h^{\mu\nu}
\end{equation}
after which indices are raised and lowered by the Minkowski metric $\eta_{\m\n}$ and $h_{\mu\nu}$ is referred to as the graviton.
For instance, we have the linearized Christoffel symbol
\begin{align}
 \Gamma^\mu{}_{\alpha\beta} = \frac{1}{2}\eta^{\mu\nu}(\partial_{(\beta}h_{\alpha)\nu} - \partial_\nu h_{\alpha\beta}) ~~~.
\end{align}
We define the d'Alembertian operator as
\begin{align}
 \square \equiv \eta^{\mu\nu} \partial_{\mu}\partial_{\nu}~~~
\end{align}
and $h$ denotes the trace of the graviton which is symmetric
\begin{align}
   h \equiv & \eta^{\mu\nu} h_{\mu\nu}\ ,\ h_{\mu\nu} = h_{\nu\mu}~~~.
\end{align}
Conventions for reducing 4D transformation laws to the 0-brane as well as all conventions for drawing adinkras from these transformation laws are as reviewed in Sec.~\ref{s:arev} and are the same as in Parts I and II~\cite{Gates:2009me,Gates:2011aa}. Another review of these rules can be found in Appendix A of Part II~\cite{Gates:2011aa}.

\section{Proof of Most General mSG Lagrangian, Transformation Laws, and Algebra in Majorana Components}\label{a:Lproof}
$~~~~$ The most general linear 4D, $\mathcal{N} = 1 $ minimal  SUGRA Lagrangian with component fields in a real Majorana representation as used in this paper takes the form
\begin{align}\label{eq:Lagrangian}
     \mathcal{L} = & h_0\left(-\frac{1}{2}\partial_\alpha h_{\mu\nu} \partial^\alpha h^{\mu\nu} + \frac{1}{2}\partial^\alpha h \partial_\alpha h - \partial^\alpha h \partial^\beta h_{\alpha\beta} + \partial^\mu h_{\mu\nu} \partial_\alpha h^{\alpha\nu}\right) + \cr
     &- s_0 \frac{1}{2} S^2 -p_0 \frac{1}{2} P^2 + a_0 \frac{1}{2} A_\mu A^\mu -f_0 \frac{1}{2} \psi_{\mu a} \epsilon^{\mu\nu\alpha\beta}(\gamma^5\gamma_\nu)^{ab} \partial_\alpha \psi_{\beta b}
\end{align}
where $a_0$, $s_0$, $p_0$, $h_0$, and $f_0$ are constants to be determined via supersymmetry. Variation of the Lagrangian with respect to $h_{\mu\nu}$ results in
\begin{align}
   \delta \mathcal{L} =& -2 \delta h^{\mu\nu}\left( R_{\mu\nu} - \frac{1}{2} \eta_{\mu\nu} R \right)
\end{align}
where $R_{\mu\nu}$ is the linearized Ricci tensor
\begin{align}
   R_{\mu\nu} =& -\partial_\alpha\Gamma^{\alpha}_{~\mu\nu} + \partial_\nu\Gamma^{\alpha}_{~\mu\alpha} -\Gamma^\alpha_{~\mu\nu}\Gamma^\beta_{~\alpha\beta} + \Gamma^{\alpha}_{~\mu\beta}\Gamma^\beta_{\nu\alpha} \cr
   = & -\frac{1}{2} \partial_\beta \partial_{(\mu} h_{\nu)}^{~\beta} + \frac{1}{2} \partial_\alpha \partial^\alpha h_{\mu\nu} + \frac{1}{2} \partial_\mu \partial_\nu h~~~. 
\end{align}
This confirms we are using the correct linearized supergravity Lagrangian as it produces Einstein's equations in vacuum for the linear theory.

In the following subsections, we will first impose this supersymmetry via a set of general transformation laws, then use closure of the algebra of these transformation laws to find a final solution for these constants. This will leave us, up to field redefinition symmetries present in the Lagrangian, the most general Lagrangian and set of supersymmetric transformation laws.

\subsection{Transformation Laws}
$~~~~$ The most general transformation laws that are an invariant of the Lagrangian take the form
\begin{subequations}\label{eq:D}
\begin{align}
   {\rm D}_a S =&  i s_1 (\sigma^{\mu\nu})_{a}^{~b}\partial_\mu \psi_{\nu b} \\
   {\rm D}_a P =& p_1 (\gamma^5\sigma^{\mu\nu})_{a}^{~b}\partial_\mu \psi_{\nu b} \\
   {\rm D}_a A_{\mu} =& i a_1 (\gamma^5 \gamma^\nu)_a^{~b} \partial_{[\nu} \psi_{\mu] b} + a_2 \epsilon_{\mu}^{~\nu\alpha\beta}(\gamma_\nu)_{a}^{~b} \partial_\alpha \psi_{\beta b}  \\
   {\rm D}_a h_{\mu\nu} = & h_1 (\gamma_\mu)_a^{~b}\psi_{\nu b} + h_1 (\gamma_\nu)_a^{~b}\psi_{\mu b} + h_2 \eta_{\mu\nu} (\gamma^\alpha)_{a}^{~b}\psi_{\alpha b}  \\
   {\rm D}_a \psi_{\mu b} =& i f_1 (\gamma_\mu)_{ab} S + f_2 (\gamma^5\gamma_\mu)_{ab} P + f_3 (\gamma^5)_{ab} A_\mu + i f_4 (\gamma^5\sigma_{\mu}^{~\nu})_{ab}A_\nu  + \cr
   &+f_5 (\sigma^{\alpha\beta})_{ab}\partial_\alpha h_{\beta\mu} + i f_6 C_{ab} \partial_{\alpha}h^{\alpha}_{~\mu} + i f_7 C_{ab} \partial_{\mu} h + f_8 (\sigma_\mu^{~\nu})_{ab} \partial_\nu h + \cr
   &+ f_9 (\sigma_\mu^{~\alpha})_{ab}\partial_\beta h^\beta_{~\alpha}~~~.
\end{align}
\end{subequations}
A few comments about the things that have led us to this present form. We have excluded all terms on the right hand sides of the transformation laws for $S$, $P$, and $A_\mu$ that do not obey the symmetry in Eq.~(\ref{e:psisymmetry}).
It is also notable that the spin connection shows up in the $f_5$ term in the transformation laws for $\psi_{\mu b}$, and is equivalent to
\begin{align}
   \omega_{\mu\alpha\beta} = g_{\alpha\nu} e^{\underline{\rho}}_{~\beta} \partial_\mu e_{\underline{\rho}}^{~\nu} +  g_{\alpha\nu}\Gamma^{\nu}_{~\mu\beta} = \frac{1}{2} \partial_{[\beta}h_{\alpha]\mu}
\end{align}
where the linear frame fields are defined as
\begin{align}
e^{\underline{\mu}}_{~\mu} \equiv & \delta^{\underline{\mu}}_{~\mu} + \frac{1}{2} f^{\underline{\mu}}_{~\mu} \\
     e_{\underline{\mu}}^{~\mu} \equiv & \delta_{\underline{\mu}}^{~\mu} - \frac{1}{2} f_{\underline{\mu}}^{~\mu} 
\end{align}
and defined to satisfy
\begin{align}
    h_{\mu\nu} \equiv & \eta_{\underline{\mu}\underline{\alpha}}\delta^{\underline{\alpha}}_{~\nu}f^{\underline{\mu}}_{~\mu} = \eta_{\underline{\nu}\underline{\alpha}}\delta^{\underline{\alpha}}_{~\mu}f^{\underline{\mu}}_{~\nu}
    \\
    h^{\mu\nu} \equiv & \eta^{\underline{\mu}\underline{\alpha}}\delta_{\underline{\alpha}}^{~\nu}f_{\underline{\mu}}^{~\mu} = \eta^{\underline{\nu}\underline{\alpha}}\delta_{\underline{\alpha}}^{~\mu}f_{\underline{\mu}}^{~\nu}
\end{align}
so that we have the linear relations
\begin{align}
g_{\mu\nu} \equiv &e^{\underline{\mu}}_{~\mu} e^{\underline{\nu}}_{~\nu}   \eta_{\underline{\mu}\underline{\nu}} = \eta_{\mu\nu} + h_{\mu\nu} + \mathcal{O}(h^2) \\
   g^{\mu\nu} \equiv   &e_{\underline{\mu}}^{~\mu} e_{\underline{\nu}}^{~\nu}   \eta^{\underline{\mu}\underline{\nu}} = \eta^{\mu\nu} - h^{\mu\nu} + \mathcal{O}(h^2) \\
    g_{\mu\nu}g^{\nu\alpha} = &(\eta_{\mu\nu} + h_{\mu\nu})(\eta^{\nu\alpha} - h^{\nu\alpha}) + \mathcal{O}(h^2) =  \delta_\mu^{~\alpha} + \mathcal{O}(h^2) ~~~.
\end{align}

Enforcing the supersymmetry on the Lagrangian~(\ref{eq:Lagrangian}) such that
\begin{equation}
    {\rm D}_a \mathcal{L} = 0 + \mbox{total derivatives}~~~
\end{equation}
leads to the following solution for the $f_i$ in terms of $s_1$, $p_1$, $a_0$, $s_0$, $p_0$, $h_0$, $f_0$, $a_1$, $a_2$, $h_1$, and $h_2$  
\begin{align}
      f_1 = - \frac{s_0}{2 f_0} s_1 ~~~&,~~~f_2 = - \frac{p_0}{2 f_0}p_1 ~~~,~~~f_4 = -\frac{a_0}{2 f_0} a_1~~~, \cr
      f_3 = -\frac{a_0}{f_0} \left( a_2 - \frac{1}{2} a_1 \right)~~~&,~~~f_5 = -2 \frac{h_0}{f_0}h_1~~~,~~~h_2 = f_6 = f_8 = f_9 = 0 ~. 
\end{align}
The fact that $f_7$ is yet unconstrained is not surprising as this term encodes the gauge symmetry in Eq.~(\ref{e:psisymmetry}). For now, we shall keep it unknown.

This gives us, for transformation laws that are a symmetry of the Lagrangian~(\ref{eq:Lagrangian})
\begin{subequations}\label{e:D2}
\begin{align}
   {\rm D}_a S =&  i s_1 (\sigma^{\mu\nu})_{a}^{~b}\partial_\mu \psi_{\nu b} \\
   {\rm D}_a P =& p_1 (\gamma^5\sigma^{\mu\nu})_{a}^{~b}\partial_\mu \psi_{\nu b} \\
   {\rm D}_a A_{\mu} =& i a_1 (\gamma^5 \gamma^\nu)_a^{~b} \partial_{[\nu} \psi_{\mu] b} + a_2 \epsilon_{\mu}^{~\nu\alpha\beta}(\gamma_\nu)_{a}^{~b} \partial_\alpha \psi_{\beta b}  \\
   {\rm D}_a h_{\mu\nu} = & h_1 (\gamma_\mu)_a^{~b}\psi_{\nu b} + h_1 (\gamma_\nu)_a^{~b}\psi_{\mu b} \\ 
   {\rm D}_a \psi_{\mu b} =&  - i\frac{s_0}{2 f_0} s_1 (\gamma_\mu)_{ab} S - \frac{p_0}{2 f_0}p_1(\gamma^5\gamma_\mu)_{ab} P -\frac{a_0}{f_0} \left( a_2 - \frac{1}{2} a_1 \right) (\gamma^5)_{ab} A_\mu + \cr 
   &-i\frac{a_0}{2 f_0} a_1 (\gamma^5\sigma_{\mu}^{~\nu})_{ab}A_\nu 
   -2 \frac{h_0}{f_0}h_1 (\sigma^{\alpha\beta})_{ab}\partial_\alpha h_{\beta\mu} + i f_7 C_{ab} \partial_{\mu} h 
\end{align}
\end{subequations}

\subsection{Algebra}
$~~~~$ Next, we wish to finish solving for the leftover constants in the transformation laws~(\ref{e:D2}) so they satisfy the algebra:
\begin{align}
   \{ {\rm D}_a, {\rm D}_b \} S = 2 i (\gamma^\mu)_{ab} \partial_\mu S~~~&,~~~  \{ {\rm D}_a, {\rm D}_b \} P = 2 i (\gamma^\mu)_{ab} \partial_\mu P,\cr
   \{ {\rm D}_a, {\rm D}_b \} A_\nu = & 2 i (\gamma^\mu)_{ab} \partial_\mu A_\nu~~~, \\
 \{ {\rm D}_a, {\rm D}_b \} h_{\mu\nu} = & 2 i (\gamma^\alpha)_{ab} \partial_\alpha h_{\mu\nu} + 2i\partial_{(\mu} (V_{\nu)})_{ab},~~~\\
  \{  {\rm D}_a, {\rm D}_b \} \psi_{\mu c} =& 2 i (\gamma^\alpha)_{ab} \partial_\alpha \psi_{\mu c} + 2 i \partial_\mu \varphi_{abc}~~~.
\end{align}
with the gauge freedom encoded by
\begin{align}
  (V_{\nu})_{ab} =& b_1 (\gamma^\alpha)_{ab} h_{\nu\alpha} + b_2 (\gamma_\nu)_{ab} h \\
  \varphi_{abc} = & c_1 (\gamma^\alpha)_{ab} \psi_{\alpha c} + c_2 (\gamma^\alpha)_{c (a} \psi_{|\alpha | b)} + \left[c_3 (\gamma^5\gamma^\sigma)_{c(a}(\gamma^5)_{b)}^d + c_4 C_{c(a} (\gamma^\sigma)_{b)}^d \right] \psi_{\sigma d}
  \label{e:psiclosurea}
\end{align}
with  $b_i$ and $c_i$ new constants to be solved for. We in fact started from a much more complicated gauge invariant term than $\varphi_{abc}$, but here only summarize the part of the proof for the terms that did not vanish.

By direct, brute force calculation, we have found that the most general set of parameters which satisfy closure as above and Lagrangian invariance are
\begin{align}
    s_0 =& \frac{2 f_0}{3 s_1^2}~~~,~~~p_0 = \frac{2 f_0}{3 p_1^2}~~~,~~~a_0 = \frac{2 f_0}{3 a_1^2}~~~,~~~h_0 = \frac{f_0}{4 h_1^2}~~~, \cr
 a_2 =& - \frac{1}{2} a_1~~~,~~~f_1 = - \frac{1}{3 s_1}~~~,~~~f_2 = -\frac{1}{3 p_1} ~~~,~~~f_3 = \frac{2}{3 a_1} ~~~,\cr
 f_4 =& -\frac{1}{3 a_1}~~~, ~~~f_5 = -\frac{1}{2 h_1}~~~,~~~b_1 = -\frac{1}{2}~~~,~~~c_1 = -1~~~,\cr
 c_2 =& c_3 = c_4 = \frac{1}{4}~~~,~~~f_7  = b_2 = f_6 = h_2 = f_8 = f_9 = 0~~~.
 \label{e:parameters}
\end{align}
There are still has five free parameters, $s_1, p_1, a_1, h_1$, and $f_0$, which encode the left over normalizations in the Lagrangian, an overall rescaling of the Lagrangian, and the binary symmetries of the Lagrangian, i.e. $h_{\mu\nu} \to - h_{\mu\nu}$, etc. Notice, that closure has now forced $h_7 = b_2 = 0$.

The final, most general Lagrangian, transformation laws, and algebra are:
\begin{align}\label{eq:LagrangianFinal}
     \mathcal{L} = & \frac{f_0}{4 h_1^2}\left(-\frac{1}{2}\partial_\alpha h_{\mu\nu} \partial^\alpha h^{\mu\nu} + \frac{1}{2}\partial^\alpha h \partial_\alpha h - \partial^\alpha h \partial^\beta h_{\alpha\beta} + \partial^\mu h_{\mu\nu} \partial_\alpha h^{\alpha\nu}\right) + \cr
     &- \frac{2 f_0}{3 s_1^2} \frac{1}{2} S^2 -\frac{2 f_0}{3 p_1^2} \frac{1}{2} P^2 + \frac{2 f_0}{3 a_1^2} \frac{1}{2} A_\mu A^\mu -f_0 \frac{1}{2} \psi_{\mu a} \epsilon^{\mu\nu\alpha\beta}(\gamma^5\gamma_\nu)^{ab} \partial_\alpha \psi_{\beta b}
\end{align}
and
\begin{subequations}\label{eq:Dfinal}
\begin{align}
   {\rm D}_a S =&  i s_1 (\sigma^{\mu\nu})_{a}^{~b}\partial_\mu \psi_{\nu b} \\
   {\rm D}_a P =& p_1 (\gamma^5\sigma^{\mu\nu})_{a}^{~b}\partial_\mu \psi_{\nu b} \\
   {\rm D}_a A_{\mu} =& i a_1 (\gamma^5 \gamma^\nu)_a^{~b} \partial_{[\nu} \psi_{\mu] b} -\frac{1}{2}a_1 \epsilon_{\mu}^{~\nu\alpha\beta}(\gamma_\nu)_{a}^{~b} \partial_\alpha \psi_{\beta b}  \\
   {\rm D}_a h_{\mu\nu} = & h_1 (\gamma_\mu)_a^{~b}\psi_{\nu b} + h_1 (\gamma_\nu)_a^{~b}\psi_{\mu b}   \\
   {\rm D}_a \psi_{\mu b} =&  - \frac{i}{3 s_1} (\gamma_\mu)_{ab} S -\frac{1}{3 p_1} (\gamma^5\gamma_\mu)_{ab} P + \frac{2}{3 a_1} (\gamma^5)_{ab} A_\mu -\frac{i}{3 a_1} (\gamma^5\sigma_{\mu}^{~\nu})_{ab}A_\nu  + \cr
   & -\frac{1}{2 h_1} (\sigma^{\alpha\beta})_{ab}\partial_\alpha h_{\beta\mu} 
\end{align}
\end{subequations}
which satisfy the algebra in Eq.~(\ref{e:Algebra}). With no loss of generality, we therefore make the following choices for the final five parameters
\begin{align}
a_1 = s_1 = p_1 = f_0 = 2 h_1  = 1
\end{align}
which puts the Lagrangian and transformation laws into the forms used throughout the paper, i.e., Eqs.~(\ref{eq:LParChoice}) and~(\ref{eq:Dfinalnumeric}). Also, various gamma matrix identities take us from Eqs.~(\ref{e:psiclosurea}) and~(\ref{e:parameters}) to Eq.~(\ref{e:varphi}) for the closure term on the gravitino.

\bibliographystyle{utphys}
\bibliography{Bibliography}

\end{document}